\documentclass[11pt]{article}
\usepackage{graphicx,color}
\usepackage{caption}
\usepackage{setspace}
\begin{document}
\captionsetup{font={small,stretch=1.1}}
\title{Exploring the 3D Ising gauge-Higgs model in exact Coulomb gauge and with a
gauge-invariant substitute for Landau gauge}
\author{Michael Grady\\
Department of Physics\\ State University of New York at Fredonia\\
Fredonia NY 14063 USA\\ grady@fredonia.edu}
\date{\today}
\maketitle
\thispagestyle{empty}
\begin{abstract}
The Z2 gauge-Higgs model in three dimensions has two different types 
of phase transition, confinement-deconfinement and Higgs magnetization.
Here they are explored through two order parameters, 
the Coulomb magnetization which is a local order parameter for confinement, and 
a replica-based gauge-invariant order parameter which tracks the Higgs 
transition in a way similar to Landau-gauge magnetization. Minimal Coulomb
gauge is set exactly, using the polynomial-time minimum-weight matching algorithm
of Edmonds. This is a tremendous speed improvement over relaxation/annealing  methods and
completely eliminates the systematic error. The replica-overlap is
also an improvement over relaxation methods for setting Landau gauge, in that it has
an easily controllable and measurable systematic error. These simulations show the
phase transitions not ending at the first-order endpoint but bifurcating near there and
continuing even through the expected analyticity region.  The specific 
heat critical exponents, $\alpha$, are highly negative, explaining the lack 
of strong signals in the energy.  Nevertheless, energy moments are seen to follow curves consistent
with the predictions from the order parameters, showing these to be true thermal transitions, albeit weak ones.
These results 
cast further doubt on Fradkin-Shenker analyticity.  They also suggest one or more new 
yet to be explored phases in gauge-Higgs models in the bifurcated region.

\end{abstract}


\linespread{1.2}
\newpage
\section{Introduction}
The Ising gauge-Higgs theory in three dimensions is the simplest nontrivial model
in this very important class.  Therefore, in investigating the fundamentals of these systems,
it makes sense first to understand this system fully.  A somewhat unappreciated but very important issue is
whether the two basic kinds of phase transitions in these models, confinement-deconfinement 
and Higgs magnetization, are
symmetry breaking.  This is important because, as emphasized by Landau \cite{ll}, a line of symmetry-breaking transitions cannot end except at an edge of a phase diagram. For an exact symmetry, the magnetization
is exactly zero in the unbroken phase. A function which is exactly zero in a finite region can
only become nonzero at a singular point. Thus one cannot move from the symmetric phase to the
symmetry-broken phase without passing through a singularity.  One argument that they are not
symmetry breaking is that the symmetries present in this model are local gauge symmetries, and Elitzur's theorem
does not allow local symmetries to break spontaneously\cite{elitzur}.  This, incidentally, 
would mean that the Higgs mechanism
in the standard model has nothing to do with symmetry breaking, despite the usual textbook treatments.
However, this strict interpretation would also apply to the edges of the phase diagram, where one
has the familiar 3D Ising model on the Higgs axis and its dual, the 3D Ising gauge theory, on the gauge
axis.  There is no doubt that the 3D Ising model itself has a symmetry-breaking phase transition. However,
if the gauge field is retained at the Ising(Higgs) edge of the gauge-Higgs phase diagram and no gauge 
fixing is performed, then even here there is no longer an order parameter with
which to observe this transition.  The local gauge symmetry will force the magnetization to zero in both phases.
Nevertheless, a transition could still be seen in gauge-invariant 
energy quantities such as the specific
heat which diverges there. At $\beta = \infty$ all plaquettes are unity. If one fixes to a complete
axial gauge, which is also Landau gauge in this instance, then all of the links can be set to unity also. 
(In this work $\beta$ is the gauge coupling parameter and $\lambda$ the Higgs, defined below).
At this
point one is left with the 3D Ising model itself, with the global Higgs symmetry still present. 
It was there all
along but apparently part of the gauge symmetry.
This lesson shows that the local gauge symmetry
can effectively hide a global symmetry which is technically its subgroup.  Gauge fixing can remove this masking
of the symmetry breaking, by leaving an exact global symmetry unfixed.
The key point here is that Elitzur's theorem does not apply to
subgroups which affect an {\em infinite} number of spins, even though one could imagine building the global symmetry
operation one by one from an infinite number of local transformations.

The pure-gauge theory limit, $\lambda \rightarrow 0$, is an even more interesting example
of a hidden symmetry breaking.  Here there is no obvious global symmetry at all since the 
Higgs fields have completely decoupled.
However, the 3D Ising gauge theory is dual to the 3D Ising model. This means it has the same free energy so it 
has to
have a matching phase transition, with the same specific heat critical exponent. Since the Landau 
theory that explains
this critical behavior is dependent on symmetry breaking, it seems this must also be symmetry-breaking, even though
there is no obvious candidate.  Duality proves that, at least for this system, the 
confinement-deconfinement
transition of the pure-gauge theory is symmetry-breaking.  

As mentioned, one method of uncovering global symmetries that can break spontaneously and 
allow construction of order parameters
sensitive to these symmetry-breakings
is gauge fixing.  As seen below and in previous work, Landau gauge can be used to trace the Higgs magnetization 
transition, and Coulomb gauge the symmetry breaking transition associated with the gauge field, usually identified as confinement/deconfinement. Many gauges leave a remnant global symmetry which is subject to symmetry breaking. 
Some 
lattice researchers are dubious of any result that comes from gauge fixing, 
despite its almost universal acceptance in
continuum perturbation theory. It can be seen as similar to choosing a coordinate system 
in general relativity.
Some results are simply more easily addressed with the scaffolding of a coordinate system.
By allowing order parameters to be defined, gauge fixing can lift the veil imposed by the 
local gauge invariance to
see underlying structures. That is not to say there are not some practical 
limitations to the gauge-fixing approach and
hard-to-solve technical issues. These will be addressed here through the introduction of two new techniques.
One is to use the exact and extremely fast Edmonds algorithm to set Coulomb gauge. This removes all 
problems of systematic error from approximate methods, which usually do not find an absolute minimum. The
second is to use a spin-glass inspired real-replica overlap order parameter introduced earlier\cite{megh} as
a gauge-invariant replacement for the Landau-gauge order parameter.

Roughly 500 separate heat-bath Monte Carlo simulations were performed, both to study the order parameter 
behaviors as well as to measure specific heats and higher moments of the internal energies (about 80\% of 
the effort was toward the latter).  Lattices from $20^3$ to
$64^3$ were studied.  Overall, roughly $10^{11}$ Monte Carlo sweeps were performed on three eight-threaded PC's 
over a period of three years.

The remainder of the paper is organized as follows.
After testing the techniques thoroughly in the Ising model and Ising gauge theory limits, the interior of the 
phase diagram is explored, first at $\beta=0.708$ which is the previously observed place of strongest first-order 
behavior\cite{ggrt}.  
Here both transitions are coincident on the self-dual line, and our order parameters both show a transition here.  In particular the 
replica order parameter is seen to jump in a way that exactly tracks  the jumps in average plaquette and Higgs energy.
This demonstrates dramatically the symmetry-breaking occurring at the first-order transition.  A symmetry-breaking first-order transition is described by a Landau theory with even-order terms in the free energy up to 6th order, with a negative fourth-order term\cite{cl}.  It cannot end at a critical point\cite{ll}, but rather at a tricritical point where it turns into one or more higher-order transitions which continue to an edge of the phase diagram.  This is in contrast to the liquid-gas type of first-order transition which is controlled by a third order Landau free-energy which explicitly breaks the symmetry.  With no exact symmetry being broken, the phase transition line in that case can end at a critical point.

 We then move on to
the region beyond the previously known first-order transition. Surprisingly
the transitions which are coincident in the first-order region, following the self-dual line, bifurcate once again in the 
low-$\beta$ region.
They are tracked well into the expected Fradkin-Shenker analyticity region, all the way to $\beta=0.05$. These
transitions are higher-order and have high values of correlation length exponent $\nu$ (of order 1.5)
resulting in {\em highly negative} specific heat exponents, $\alpha$, from the hyperscaling relation
\begin{equation}
\alpha= 2- d \nu   \label{hsc1}
\end{equation}
where $d$ is the spatial dimension.
 This means the specific heat does not have an infinite
singularity - only some high derivative does.  Such singular behavior is difficult to observe with numerical 
simulations, due to the lack of easily-measured quantities that diverge with lattice size, but one can still check if scaling is consistent with the $\alpha$-exponent  
predicted from the order-parameter analysis. We find this is true in every case examined.  Critical exponents and couplings independently determined from energy quantities all agree with those determined from the order parameters. Thus one can see the effect of these transitions on the energy, signifying weak singularities. Observation of energy effects and the 
fact that one of our order parameters is gauge invariant means that these transitions cannot be dismissed
as gauge artifacts.  A similar bifurcation was previously observed by Caudy and Greensite\cite{cg}. They
observed, for the 4D SU(2)  theory, Landau-gauge and Coulomb-gauge magnetization transitions following separate tracks in the same area, on the strong-coupling
side of the first-order region.  They interpreted this as a sign of ``gauge ambiguity"  casting doubt on the reality of
these transitions.  However, there is no reason the transitions must track the self-dual line at strong coupling - after all they do not do so at weak coupling.  The transitions start out separate and have different order parameters so there should be two transitions, perhaps sometimes coincident, but also apparently sometimes not.

The presence of such transitions is in obvious conflict with the presence of an analyticity region
connecting the confinement and Higgs phases. A widely-accepted proof of the existence such a region
was given by Fradkin and Shenker(FS)\cite{fs}
based on previous work by Osterwalder and Seiler(OS)\cite{os}, utilizing cluster expansion techniques pioneered by
Glimm Jaffee and Spencer\cite{gjs}.  However, this has always been in conflict with the much simpler analyticity
argument of Landau that a symmetry-breaking phase transition cannot end in the middle of a phase diagram.  This 
is why the issue of symmetry breaking is so important and why many insist that there is no spontaneous
symmetry breaking in gauge theories. What is being presented here is the alternative hypothesis, that
the gauge symmetry merely hides the underlying symmetry breaking of global symmetries, 
which can be uncovered by gauge fixing
or other means.  As pointed out in a previous work\cite{megh}, there is a possible flaw in the Fradkin-Shenker approach.
Although there is little doubt that the cluster expansion converges, the coefficients themselves contain
infinite-lattice quantities that could be singular functions of the temperature. This possibility is not
addressed in their paper.  Seiler has presented a similar work using a ``polymer expansion" which 
does not apparently have
this feature\cite{poly,seilernew}.  However, the polymer expansion result only guarantees analyticity at low $\beta $ and 
low $\lambda $
or high-$\lambda $ for a given $\beta $, which seems to allow for possible phase transitions at 
intermediate $\lambda $.   An abbreviated version of our critique from \cite{megh} is given here in the penultimate section for 
clarity.

\section{Z2 gauge-Higgs theory}

The Z2 gauge-Higgs theory in three dimensions has the action
\begin{equation}
S=-\beta \sum _p  U_p  -\lambda \sum _{\vec{r},\mu } 
\phi (\vec{r})U_\mu (\vec{r})\phi (\vec{r}+\hat{\mu})\label{eqn1}
\end{equation}
where $U_\mu (\vec{r})$ are Z2 gauge fields on links, $U_p$ is the elementary plaquette made from
the product of four links around a 1x1 box, and the $\phi$ are site-based Z2
Higgs fields.  $\beta \propto 1/g^2$ is the
inverse gauge coupling and $\lambda$ is the 
Higgs coupling.
The phase diagram as heretofore known\cite{stack,ggrt,tkps} is shown in Fig.~1. On the right axis is the 3D Ising 
transition.
\begin{figure}[b!]\centering
                    \includegraphics[width=3.2in,  clip]{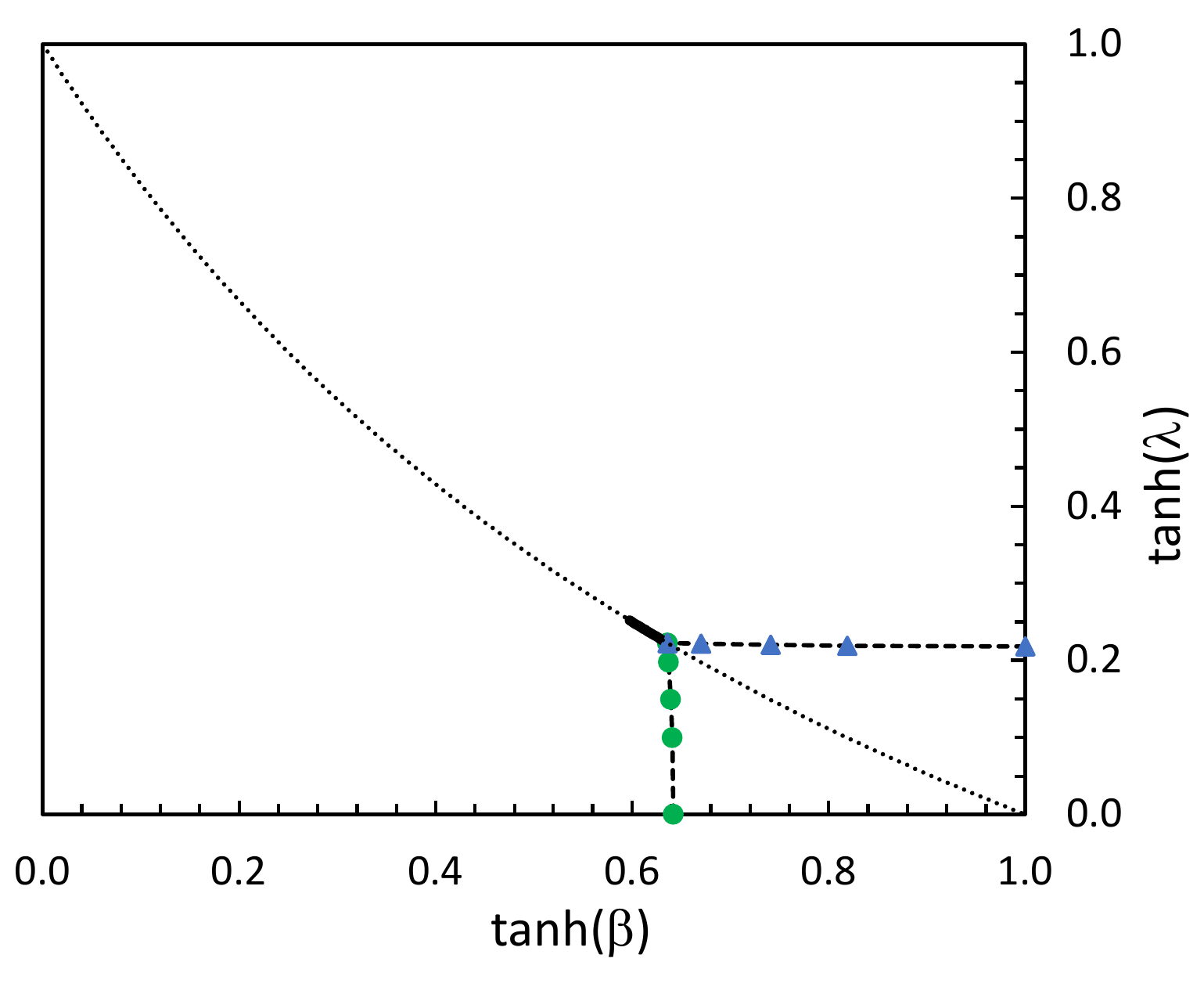}
                                  \caption{The phase diagram for the 3D Ising gauge-Higgs theory established from previous Monte Carlo studies.  The Ising (i.e.\ Higgs, at right) and Ising-gauge critical point (bottom) are endpoints of second-order lines that run along dual-related paths.  They likely become first-order exactly at the place where they join, and then run up the self-dual line for a short distance, whence the transition is seen to terminate. Second-order transition lines are dashed, 
first-order solid. The self-dual line, for which every point is its own dual is shown as a dotted line. Datapoints are from Ref. \cite{tkps}.}
          \label{fig1}
       \end{figure}
 It enters
the phase diagram with little change initially, but eventually becomes first-order. On the lower 
axis is the 3D Ising
gauge theory. It also enters the diagram with little change at first.
The two phase transitions
then merge and continue as first-order, running up the self-dual line (before this the two paths are linked by
duality, so it is not a surprise that the behaviors match).  The initial study showed them becoming first-order before the merger, but the  later studies indicate it is likely they only become first-order at the point of merger. The first-order transition runs for a while up the self-dual line and then
comes to an end at (0.689(1),0.2575(5))\cite{tkps}.
It is generally believed that this is an ordinary critical endpoint, similar to the liquid-gas transition.
Such an ending of the transition is needed for consistency with the  
the Fradkin-Shenker theorem.  However, such
a critical endpoint requires a 
non-symmetry-breaking transition.  Since the initial
transitions on the axes are symmetry breaking (the Higgs explicitly 
and the gauge through duality) the question becomes
at what point does it become non-symmetry-breaking? A purist would 
probably argue as soon as the transitions enter the
diagram they become non-symmetry-breaking, but this is a little hard to swallow 
considering how little changed the transitions initially
appear from those on the axes. Recall that the Landau description of second-order phase transitions
relies on the spontaneous breaking of an exact symmetry.
In particular, for the Higgs case near the Ising transition, one only has a few negative plaquettes at high-beta.
This makes for a very few negative links in Landau gauge. So, correspondingly, there are very few anti-ferromagnetic
Higgs interactions.  The Ising model is known to keep its ferromagnetic transition when a small
fraction of interactions become antiferromagnetic, so the transition must enter 
the phase diagram with little change.
However, it is even harder to picture how a transition could change from symmetry-breaking to
non-symmetry-breaking further into the phase diagram.
This change
would itself be a new type of phase transition along a transition line which 
would appear to have no analogue in
statistical mechanics. It would not be consistent with Landau theory in which 
symmetry-breaking transitions have only even-order terms in the effective potential and
non-symmetry-breaking transitions have odd-order terms.  The only way for odd-order terms to
arise is {\em explicit} symmetry breaking which is not present here.

Another way out of this conundrum has been suggested by Caudy and Greensite\cite{cg}.
They put forward the notion that a transition
could at some point become ``non-thermal." In other words, the order parameter could completely
decouple from the energy at some point. 
This would allow for the order parameter to be singular
in a region where the energy is not (the FS theorem only concerns free-energy singularities). 
This would also require a new type of phase transition --
thermal to non-thermal.  However, phase transitions only occur for energetic reasons involving
energy/entropy balance.  In other words, there would be no phase transition if the order parameter
were not related to the energy, and correlations always go both ways.  An easy way to see this
is to examine the correlations between order parameter and either the average plaquette 
or the Higgs energy.
Simulations show these to be non-zero in the symmetry-broken phase.  They must be since this quantity
is easily interpreted as the derivative of order parameter with respect to $\beta$ or $\lambda$, and
the order parameter certainly does change as these are varied. But the same correlation can also be
interpreted as the derivative of average plaquette or Higgs energy with respect to an external field
coupled to the order parameter. Thus the energy must ``care about" such an external field which shows 
it cannot be oblivious to the order parameter.
Because it is outside the main
scope of this paper, a detailed examination of
non-thermal transitions is given in Appendix A.  We find that non-thermal transitions can exist, but only in the extremely limited circumstance that the specific heat exponent $\alpha$ is a non-positive even integer. In this case the specific heat can become ``accidently" non-singular at a place where the order parameter is singular.  It is shown that the 2D Ising-gauge theory actually has such a non-thermal transition, but is probably an exceptional case.  In 3D, only one of the six phase transition regions examined in this paper possibly meets this condition, and even that case is unlikely.

\section{Landau and Coulomb-gauge order parameters}
There are two different kinds of phase transition in gauge-Higgs models. 
One starts out as the spin transition - a magnetization transition - in our case the 3D Ising model.
The other is the confinement-deconfinement
gauge transition. For the 3D Z2 case,
these are related by duality, but in other cases they generally are not. As seen in Fig.~1 these transitions
track separately along paths related by duality and eventually
join together. Subsequently they follow the self-dual line, as a single combined transition.  Since these are
separate in part of the phase diagram, they presumably have different order parameters. These can
be uncovered by transforming to Landau and Coulomb gauge respectively. Landau gauge seeks as many positive
links as possible.  Because negative plaquettes are gauge invariant features, there will always be some
negative links associated with these that cannot be removed by a gauge transformation, but only moved around.
At weak gauge coupling (high $\beta$), where negative plaquettes are rare, most links will be positive in Landau gauge.
The Higgs interactions associated with positive links are ferromagnetic.  For fixed gauge field, the 
Higgs field essentially sees the gauge background as a spin glass, with a handful of antiferromagnetic links
thrown into a sea of ferromagnetic links.  Such a system still has a ferromagnet to paramagnet phase transition.
The Higgs field itself is the order parameter, and the remnant global Z2 symmetry is the symmetry that breaks
spontaneously in the magnetized phase.  (Note that for Abelian theories this global symmetry only affects 
the Higgs field).  Setting the Landau gauge presents practical problems, however. 
Generally relaxation or simulated-annealing methods are used, which often do not
find absolute maxima.  Typically procedures such as ``best of ten" annealings are used, but one is never
sure of the quality of the solution and the systematic error in the magnetization or its moments that 
could arise from bad maximizations.  A rare ``truly bad" maximization can severely affect results on moments, giving a false signal of tunneling or onset of randomness.
The quality of the procedure also varies with the couplings, so may
work well in some regions of the phase diagram but not others.  Simulated annealing was used at the start of
this study as a ``warm up" and most of the phase transitions observed here were seen with it, but detailed
characterizations, such as determining critical exponents, were not easily obtained due to the 
slow speed of the technique
and the presence of systematic errors.  Instead a new technique introduced in \cite{megh} is used as a replacement
for the Landau-gauge magnetization.  It also has the advantage of being a gauge-invariant order parameter.
This helps to allay any concerns that things seen in a fixed gauge may be unphysical.  This order parameter uses
a second ``replica" Higgs field which is equilibrated to the gauge field before being used.  The original Higgs
field is part of the Monte Carlo simulation, so already equilibrated through detailed balance.  The number
of equilibration sweeps for the replica field is similar to that needed at the start of any Monte Carlo simulation. 
It varies with couplings
and lattice size, but is easily studied, and measured quantities are seen to approach asymptotic values
exponentially in the number of equilibration sweeps. So although this procedure does have a systematic error,
that error is easily studied and can be reduced below the random error simply by equilibrating long enough.
The order parameter is the product of the Higgs field with the replica Higgs field, averaged over the lattice.
This two-real-replica overlap is sometimes used as an order parameter for spin glasses\cite{2rr}. It will show
magnetization in a system with either spin-glass order (frozen pattern random over space) or 
ferromagnetic order (or even anti-ferromagnetic order).  For the cases studied here, the transitions
were also observed
in the same places with the Landau-gauge order parameter, so it is clear that it is  a ferromagnetic 
to paramagnetic transition being seen.  The order parameter is defined as
\begin{equation}
q=\sum _{\vec{r}}\phi _{R}(\vec{r}) \phi (\vec{r}) /V .
\end{equation}
where $V$ is the number of lattice sites. It
is gauge invariant because the 
replica Higgs, $\phi _{R}$, also transforms under the local Z2 symmetry.  
However, there is a new global Z2 in which
the replica is flipped but the regular Higgs field is not, or vice versa.  This is 
the symmetry that is spontaneously broken when the $q$ order parameter becomes non-zero. Below, the method is
first tested for the $\beta \rightarrow \infty$ case, the already extremely well-studied 3D Ising model.  
If one insists
on not fixing the gauge, even for this case the Higgs field averages to zero due to Elitzur's 
theorem and standard methods
cannot be used to study the transition. Of course it can still be seen in the diverging specific heat. 
As shown below, the $q$ order parameter can successfully be used to characterize this transition without
gauge fixing.  In this limit, Landau and full-axial gauge coincide.  If the gauge is fixed in this way,
all of the links become unity and one then has the 3D Ising model explicitly, with the Higgs field itself again
as order parameter. This is a clear case of the local symmetry hiding a symmetry-breaking transition, which
can be uncovered either by gauge fixing or using the replica-overlap order parameter.  In the following, the square root of the absolute value of this 
order parameter, denoted $m$, is used in analyses, since that has the dimensions of the normal magnetization and behaves more like it. However, $q$ itself has similar scaling behavior, so this step is not absolutely required.

The other type of transition in these models, what is normally considered confinement-deconfinement but is 
actually more general than that, may be uncovered by fixing to minimal Coulomb gauge.  Although confinement-deconfinement
is a proper characterization for the pure-gauge theory, as soon as the Higgs coupling is turned on confinement
is no longer absolute. This is because the Higgs field can screen the potential. Nevertheless, the gauge
transition continues to exist. As seen below, this is because the symmetry being broken is still an exact one.
For the 3D Z2 theory, Coulomb gauge seeks as many positive links as possible in two of three lattice directions,
say the first two.  The third direction links are ignored in the gauge-fixing procedure.  This leaves
a larger remnant symmetry than Landau gauge does.  One can perform a partially-global Z2 transformation
on each 1-2 lattice layer separately.  Such a transformation only affects third-direction links so does not interfere with the gauge condition. All of the
third direction links attached to a layer are flipped if a -1 is applied to the layer.  In Coulomb gauge,
the $L$-fold order parameter is the average third-direction link, averaged separately over 2D slices (here $L$ is the linear lattice size in the third direction).  
For fixed one and two-direction links,  third
direction links see a mostly ferromagnetic self-interaction  at weak coupling from the plaquette interaction
(because most plaquettes have all positive
1- or 2-direction links). The third direction links will, therefore, magnetize.
It has been shown that in the pure-gauge theory, this magnetization is a local order parameter for
deconfinement.  Specifically, in the Coulomb-magnetized phase the potential energy of two quark sources has been proven to be finite 
at infinite separation, showing that phase to be deconfined\cite{zw}.  
Setting the Coulomb gauge has many of the same
practical difficulties as the Landau gauge - problems with local maxima and systematic error as well as 
heavy computational effort.  However, the 3D Z2 gauge-Higgs system is an exception. For it minimal Coulomb gauge
can be set exactly with a graph-theory derived method described
in the next section.  The algorithm is not only exact but surprisingly fast, allowing high-precision determination
of the Coulomb magnetization and its moments. As a warm-up for this method, it is first used 
on the pure-gauge theory.
Because this is dual to the 3D Ising model, one knows it has a phase transition with a precisely
predicted location and accurately determined critical exponent $\nu$.  
This allows a robust test of the Coulomb-gauge method, which indeed 
finds the transition very close to the expected coupling.
This also gives 
a highly nontrivial test of the concept of symmetry-hiding in a gauge theory.  Due to the duality connection
to the 3D Ising model, one knows there must be a hidden symmetry-breaking phase transition here even though the only explicit symmetry is a local one.  It is shown
that Coulomb gauge-fixing is able to uncover such a transition and provide a symmetry-breaking order parameter.

\section{Setting the exact Coulomb gauge using Edmonds' algorithm}

Edmonds' algorithm\cite{edmonds} is a sophisticated graph-theory technique which finds a 
minimum-weight perfect matching to a general graph. It was recognized by Bieche et.\ al.\cite{bieche}
that this could be used to find the exact ground-state energy 
for the 2D Ising spin-glass by exploiting the hidden gauge invariance in that system. 
The ground-state energy is found by using this algorithm to set an exact Landau gauge
(minimize the number of negative links, i.e. antiferromagnetic interactions).
Unfortunately Landau gauge in 3D does not map into a graph-matching problem, but
Coulomb gauge does, because it seeks to minimize negative links in only two of the 
three dimensions. It can be set on 2D layers independently. Not only is the
gauge set exactly (absolute minimum found) but the algorithm is thousands of times
faster than simulated annealing, which also usually does not find the absolute minimum.
This makes possible high-statistics studies of the confinement-deconfinement 
transition using the Coulomb-magnetization order parameter.  The main disadvantage of this
approach is that it does not work with periodic boundary conditions.  This is
inconvenient but tolerable if large lattices are used.

Consider setting the Coulomb gauge on a 2D surface, using Z2 gauge transformations at 
lattice sites. The objective is to make as many positive links as possible. For instance 
if a site has three or four 1 or 2-direction negative links touching it, then a gauge transformation(GT) at that site 
which flips
all of the links touching it will reduce the number of negative links. To proceed beyond
this obvious step is difficult. One method is simulated annealing(SA), where a ``fake energy"
consisting of the total number of negative links is constructed and a Monte Carlo
simulation with a slowly decreasing temperature is performed, with the hopes of finding
the ground state.  However, in practice, one usually ends up only at a local
minimum. If the SA is repeated a number of times one may be satisfied 
that one is close to the global minimum.  However, without knowing how close one actually is, it
is difficult to estimate the size of any systematic error that may arise from the imperfect
algorithm.  Thus, although Coulomb-gauge methods are quite intriguing, in that they
provide a local order parameter for confinement, the possibility of systematic errors and
the difficulty of characterizing them has limited their usefulness.  

The gauge-invariant plaquettes form a scaffold on which the Coulomb gauge-configurations
can be built. Negative plaquettes require at least one negative link. A valid gauge
configuration can be built by joining pairs of negative plaquettes with strings of
negative links. The string is most easily pictured as running
along dual lattice links, with the negative links on the original lattice perpendicular to these.
Along the string, the negative links 
occur in pairs within plaquettes, resulting in positive plaquettes except at the
string ends.  The minimal Coulomb gauge will be achieved if the sum of these string
lengths is a minimum.  Finding such a minimum is exactly the weighted perfect graph-matching
problem.  The dual lattice sites at the center of negative plaquettes are taken
to be the vertices of the graph. The edges are minimum-length paths between each vertex
and all other vertices. A perfect graph matching is a set of edges that touches 
each vertex exactly once.  If we count the string lengths as weights of the corresponding edges,
then a minimum-weight perfect matching will produce a gauge configuration with the 
smallest possible number of negative links, i.e. the minimal Coulomb gauge.  Generally there
will be more than one minimum-weight solution, since any non-straight 
string has a number of possible 
equal-length  paths, when using the taxicab metric. 

If one tries to retain
periodic boundary conditions then a number of related problems occur.  One can choose
to connect two vertices by a path that goes through a boundary or one that does not.
These might even have the same length. The configurations will differ by the values
of gauge-invariant Polyakov loops.  There is also no guarantee that a given solution
will match the Polyakov loops of the original gauge configuration one started with,
and thus may not be gauge equivalent.  Of course this is not acceptable as the whole
point of the exercise is to find a gauge transformation from the original configuration.
Clearly the algorithm has no way of matching the original Polyakov loops. Opening the
boundary in the 1 and 2 directions removes the Polyakov loops (PBC can be retained
in the third direction).  In practice a combination of fixed and open boundary conditions
works best.  Links pointing out from the boundary do not exist (open BC). Links
lying within the boundary are set to unity. This fixes a problem caused by negative links 
on the boundary as they would make alternate endpoints for strings.  One could also hope that the
combination of open and fixed might partially compensate the boundary effect,
in that the fixed boundary link is cold and the open one hot. Note that the third-direction
links on boundaries are all active. This boundary condition,
specifically tailored to the Edmonds algorithm could be called OFA (open, fixed, active).

The exact Coulomb gauge is built up layer by layer, initially ignoring the third-direction
links which will be put in later.  First the locations of all negative plaquettes on a 1-2 layer are found.
Say there are $P$ negative plaquettes. Then a $P\times P$ cost matrix is constructed
from the minimum taxicab distances between pairs of negative plaquettes.  This is then input to the
Edmonds minimum-weight perfect matching algorithm.  This rather remarkable 
polynomial-scaling algorithm
employs Edmonds' ``blossom" algorithm
for generating a maximum cardinality matching along with a primal-dual linear programming
approach which handles the weights.  Meeting the complimentary slackness conditions
of the linear program and its dual proves that the resulting solution is optimal,
i.e an exact minimum-weight solution has been found.  A FORTRAN implementation by G. Kazakidis, published 
by Burkard and Derigs\cite{burkard-derigs} was used.  Faster implementations exist\cite{otheralgs}, 
but the older version was 
completely adequate. Generally there are hundreds to at most thousands of vertices on
the lattices used here (up to 64x64 sheets); however, these algorithms can easily handle vertices in
the millions. On the $50^3$ lattice the actual algorithm time, including pre- and post-processing, is about the 
same as a single Monte Carlo sweep. The output of the algorithm is a list of matched
vertices.  The Coulomb gauge configuration is then constructed by placing negative links 
on shortest paths between matched vertices (choosing one if more than one equivalent paths exist
- more on this later).  Then the entire matching procedure is repeated for the other layers,
64 total in the case of a $64^3$ lattice.  This is a solution for only the 1 and 2 directions.
To get the third direction and the proper associated Higgs field, one must discover the 
actual gauge transformation that connects the original gauge configuration to the new one
that has just been generated. This is done by going through a serpentine path on each layer. The 
gauge transformation on the first site (0,0) is arbitrarily set to unity, $J(0,0)=1$, where $J(i,j)$ is the 
gauge transformation scalar. The pair $(i,j)$ are
Cartesian lattice coordinates on the given 1-2 plane. Moving one step in
the 1-direction, set $J(1,0)=J(0,0)U_1 (0,0)U'_1 (0,0)$.
Here $U_k (i,j)$ is the original gauge field link and $U'_k (i,j)$ is the new one just generated.  
Using an obvious extension of the above formula, one
continues in the 1-direction to the end of the lattice, then up in the 2-direction one link, 
and continuing back along the 1 direction until the left boundary is hit again.  After moving up
once again in the 2-direction the process is repeated until the entire layer is covered.  The 
process is then repeated for each layer. In the end one has the complete gauge transformation $J(i,j)$
linking the two configurations.  Then this is then used in the normal way to transform the entire
gauge configuration (including the third-direction links), as well as the Higgs field. It is a good idea
then to check that all of the links in the 1 and 2-directions of the transformed gauge configuration
match the corresponding link in $U'$.  This verifies that the procedure described 
does generate a gauge-equivalent
configuration in the Coulomb gauge, as theory indicates it must.  This is where trouble would be detected
if one tried to hang onto periodic boundary conditions.  It is non-trivial because there are many
2-direction links not on the serpentine path used to discover the gauge transformation. No configuration
generated in this study has failed this test, which is a fairly robust check of program reliability, and
a reassuring verification of the theoretical underpinning.  

The algorithm was compared to a fairly gentle SA algorithm 
in the vicinity of one of the critical points. It was found that the best of ten procedure usually did not find the minimal solution, but generally was within 1\%.  In a longer test on one gauge configuration SA finally found an optimal solution after several hundred tries.  The  exact algorithm could help in testing or tuning SA or other approximate methods, for use in other models for which exact methods do not exist.  However, the behavior of these methods is often model and coupling dependent, so this might be of limited use.

Finally one must decide what to do, if anything, about the degeneracy of the minimum-weight solution. This occurs
because for matched vertices that do not lie on-axis, there are multiple equal-length paths. Two approaches
were tried. One is to construct these paths with a random algorithm so any path is equally probable. Then
one simply defines the Coulomb magnetization for a configuration to be the average over the degenerate
solutions. Only one is chosen randomly for a given configuration, but in principle all will be visited over 
a very long
run. The other approach is to attempt to use the degeneracy to enhance the Coulomb magnetization further. For
a fixed set of 1 and 2-direction links, the third direction links behave as spins interacting with 
each other within
a layer through ferromagnetic interactions if the other two links in a plaquette match or antiferromagnetic
if they do not.  The point of going to Coulomb gauge in the first place was to increase the 
number of 1 and 2-direction
positive links as much as possible, in order to increase the chances of ferromagnetic interactions.  Then one
can hope to see a magnetization transition at weak coupling.  However, one can also obtain a 
ferromagnetic interaction
from the rarer negative links if they match on adjacent layers.  So by prioritizing link overlaps one can 
devise a criterion to choose among degenerate solutions.  The maximum overlap may still be degenerate, 
but the degree
is much smaller.  Essentially this maximum-overlap gauge derived from the Coulomb gauge is a new gauge. 
It is only
relevant to the discrete gauge groups because continuous theories do not have the same degree of degeneracy. 
Note
that biasing toward more ferromagnetism does not break the residual symmetry.  
This ``enhancement" was implemented
with a zero-temperature Monte Carlo relaxation in which changes to equivalent Coulomb gauges 
that increase overlap
are always accepted, those that decreased it were rejected, and neutral changes were accepted 50\% of the time.
This 
typically achieved a stable solution in a few hundred sweeps.  
In each coupling region, the number of sweeps was
adjusted so that at least 95\% of configurations reached a stable solution by the halfway point. 
Typically there is
a 1-2\% increase in overlap. Another run of the relaxation will typically give a different solution with
an increase that differs of order 1\% (i.e. 1\% of the 1-2\%). This was small enough that only one relaxation
for each configuration was performed.  However, although initial indications showed some magnetization enhancement,
a longer study showed only a very minor change in magnetization.   Both methods 
were tested for the pure-gauge theory.
The conclusion is that both work, and give very similar answers for critical
points and critical exponents, although raw numbers for magnetizations are a bit different.  It was concluded
that there is no real advantage to performing the extra overlap-enhancement step, which is fairly time consuming.
Considering that better statistics can be obtained with the faster and simpler random-path algorithm, that 
is what was employed for all of the studies below.

\section{Pure-gauge theory}
In order to test the Coulomb gauge method outlined above, we first applied it to the pure-gauge theory for which
precise results are known.  Using $\lambda _c = 0.2216595(26)$ from \cite{fl} for the 3D Ising model, one
predicts from duality a phase transition at $\beta _c= -0.5\ln (\tanh (\lambda _c))= 0.7614027$. The critical exponent $\nu$
should match that of the Ising model, given by Monte Carlo as $0.63002(10)$\cite{mh}. The other critical exponents $\gamma$ and $\beta '$ depend on the dimensionality of
the space the order parameter is averaged over, through the other hyperscaling relation 
\begin{equation}
\gamma /\nu + 2 \beta '/\nu =d  . \label{hsc2}
\end{equation}
In this paper we will use the notation $\beta '$ to refer to the magnetic critical exponent in order to eliminate any confusion with the inverse coupling parameter $\beta $.
 Since our
Coulomb order parameter is  averaged only over two-dimensional layers of the 3D lattice, $d=2$ for the above
formula, whereas $d=3$ for the ordinary Ising model magnetization. Therefore one cannot expect our  $\beta '$ and $\gamma$ exponents to match the Ising values.

With the exception of one case detailed later, all lattices for all studies in this paper are first equilibrated with 100,000 sweeps.  Multiple tests were done to show this is sufficient.  Errors in measured quantities are from binned fluctuations.  For the pure-gauge theory, for each run of the $40^3$ and $50^3$ lattices $2\times10^6$ sweeps
were performed, with the Coulomb gauge transformation and measurements performed once every 20 sweeps.  For the 
$64^3$ lattice half this number of sweeps were performed.  Boundary conditions were periodic in the third direction and 
a mixture of fixed and open in the first two, as described above, to meet the requirements of the Edmonds algorithm. In addition, quantities such as magnetization were only measured in the interior region of the 2D slices, at least
five spaces from the boundary.  This is not absolutely necessary, but seems to help make sense of the smaller lattice
results.  As a result, the effective linear size of the lattice for finite-size scaling purposes is  $L_{\rm{eff}}=L-10$ instead of $L$, for the $L^3$ lattice.   The separate layers are averaged after taking the absolute value of the within-layer average, similar to the standard method of averaging magnetizations over separate lattices.  This is necessary because each layer has its
own symmetry, and the magnetizations on separate layers are independent, except for a single Polyakov-loop constraint. Error bars for measured quantities are from binned fluctuations. 

To locate the critical point we searched for a Binder cumulant crossing, defined here as 
\begin{equation}
U= 1-<\! M^4\! >/(3<\! M^2 \! >^2 ) .  
\end{equation}
Here $M$ is the Coulomb magnetization, which is simply the average of third-direction pointing links over each 1-2 lattice layer.  At leading order,
finite-size scaling theory predicts U for all lattices has a universal value at the critical point, whereas in the magnetized
phase $U\rightarrow 2/3$ as $L\rightarrow \infty$ and $U \rightarrow 0$ as $L\rightarrow \infty$ in the unmagnetized phase.  This leads to a crossing of U for different lattice sizes at the infinite lattice critical point. In practice, next-to-leading terms spoil this a bit, giving a slightly shifting crossing as the lattice size is increased.  With enough different lattice sizes, this shift can itself be quantified and used to extrapolate precisely to the true infinite-lattice critical point.  In the current study, we will be satisfied to use only the leading-order scaling analysis, since measurements are taken in most cases on three lattice sizes, sometimes only two of which are used in fits.  Ignoring sub-leading scaling contributions may introduce possible systematic errors in critical exponents due to finite-lattice effects, 
estimated at 5-10\% (see below).  A follow-up study using more sophisticated finite-size scaling methods is planned.  The pure-gauge Binder cumulant is shown in Fig.~2a.  A clear crossing is observed at $\beta=0.7605$, although within errors the largest 
lattices could be crossing at 0.761.  This is close to the expected value of 0.7614 for the infinite-lattice critical point. 
According to leading finite-size scaling theory, U plotted against the scaled coupling variable 
\begin{equation}
(1/\beta _c -1/\beta)*L_{\rm{eff}}^{1/\nu} \label{sc} \label{scalevar}
\end{equation}
should give a universal function - i.e. for the correct $\beta _c$ and $\nu$, the data from different lattice sizes should collapse onto a single plot. This method can be used as a more systematic method of finding $\beta _c$ and also $\nu$.
Such a ``collapse plot" is given in Fig.~2b, which also shows the scaled magnetization $<|M|>L_{\rm{eff}}^{\beta '/\nu }$ vs. the scaling variable, which should also show a collapse for the correct magnetic scaling exponent, $\beta '$ (the optimization method used here for the critical exponents is described below).
\begin{figure}[b!]\centering
                    \includegraphics[width=0.47\textwidth,  clip]{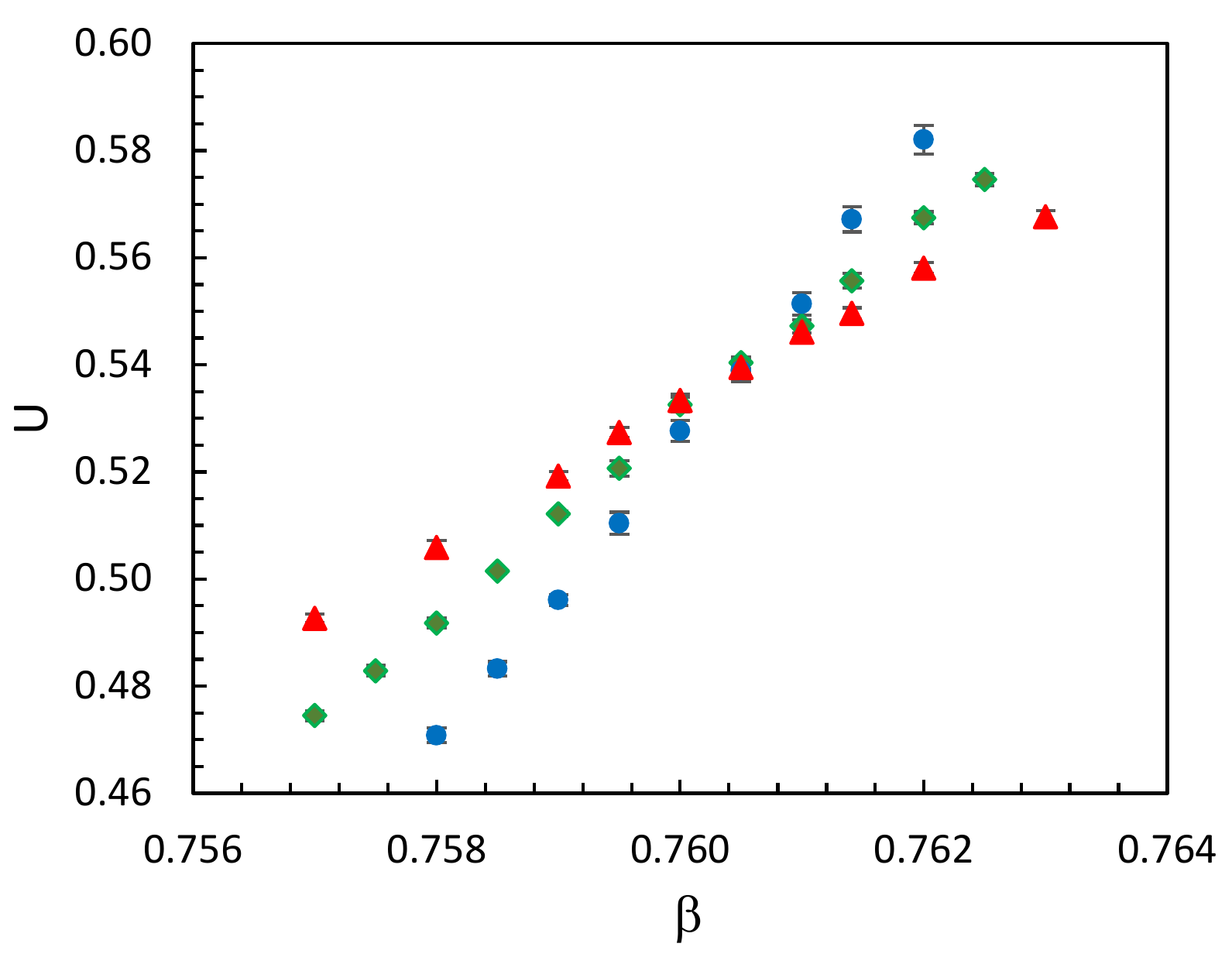}
                         \includegraphics[width=0.515\textwidth, clip]{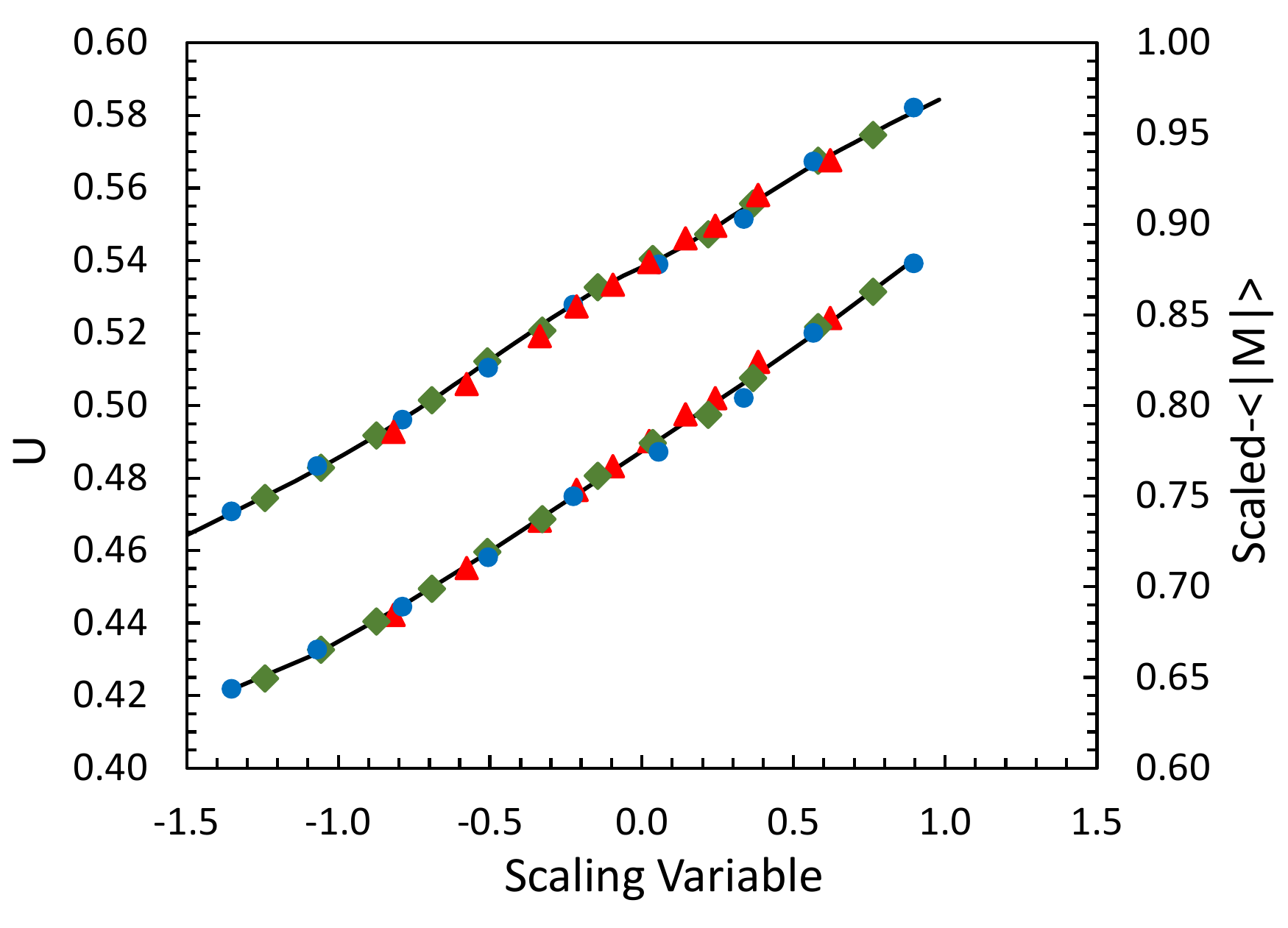}
				\includegraphics[width=0.47\textwidth,  clip]{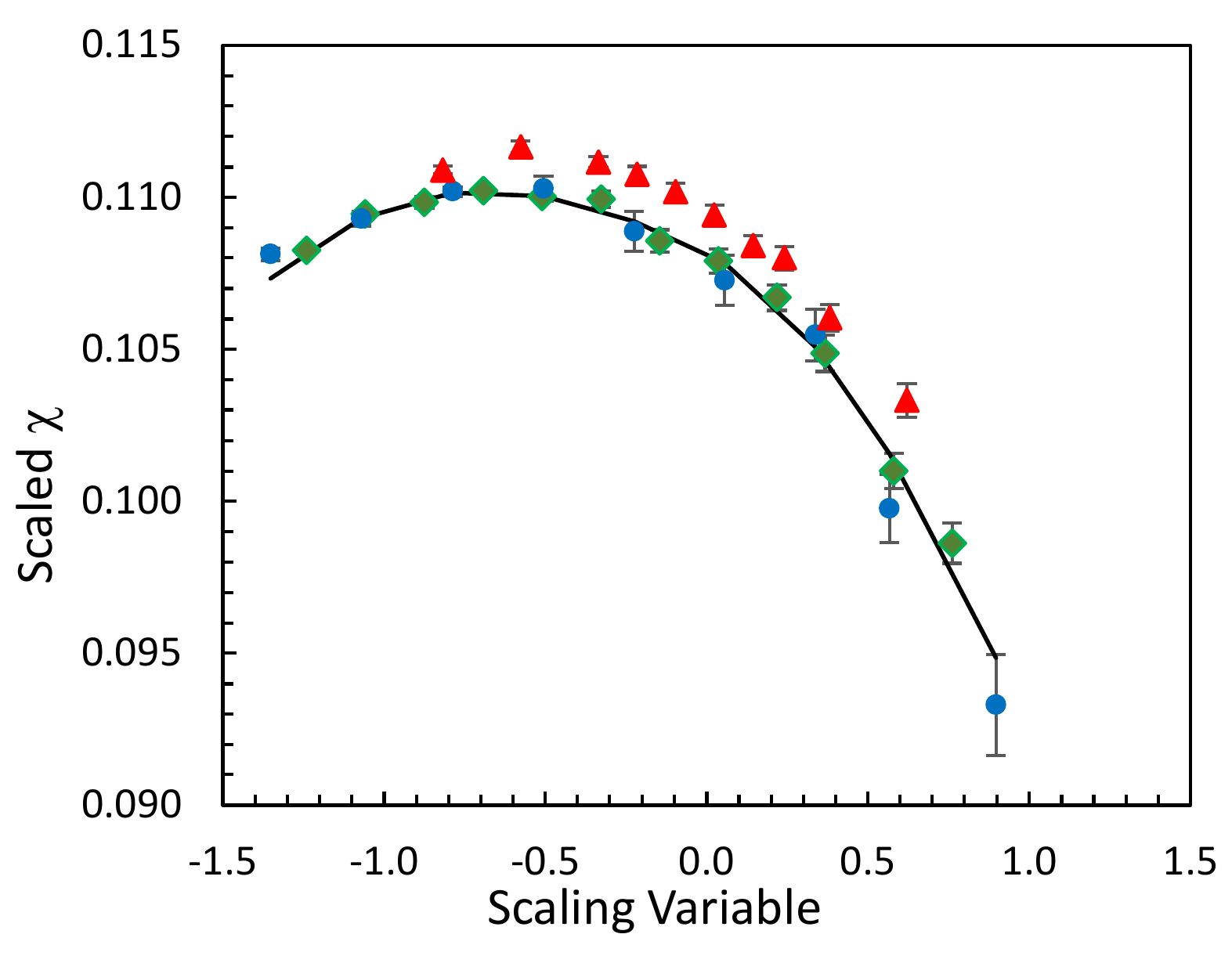}
                                  \caption{Pure-gauge theory Binder cumulant crossing (a), scaling-collapse curves for Binder cumulant (upper curve) and Coulomb magnetization (lower curve - right scale) (b), and scaling-collapse curve for susceptibility (c). Triangles are $40^3$, diamonds $50^3$, and circles $64^3$ lattice. Error ranges for $<|M|>$ are about one-half the size of plotted points. } 
          \label{fig2}
       \end{figure}

Another method of locating a phase transition is to look for a peak in the susceptibility, growing with lattice size.  The position of the peak can be taken as a ``finite lattice critical point," which is $0.759$ for the $64^3$ lattice  
The $\chi$-peak shifts with lattice size, expected to eventually approach the infinite lattice critical point.  A scaling plot for the scaled susceptibility $\chi L_{\rm{eff}}^{-\gamma /\nu }$ is given in Fig.~2c.  Here 
\begin{equation}
\chi= (<M^2>-<|M|>^2)*L_{\rm{eff}}^2 .  
\end{equation}

Our method of determining the three critical exponents and $\beta_c$ is to do a single non-linear least-squares fit to all three quantities,
using various parameterized smooth functions such as power laws to represent the universal functions.  For the pure-gauge case only the $50^3$ and $64^3$ data could be reasonably fit this way, showing the $40^3$ data having too large a finite-size correction from next-to-leading terms (especially apparent in $\chi$, but also present in the others).  The fit with 31 degrees of freedom gives a $\chi ^2/$d.f. goodness of fit of 0.8. The critical point and exponents were determined as 
$\beta _c =.7604(1)$, $\nu =0.69(3)$, $\gamma /\nu =1.871(4)$ and $\beta ' /\nu = 0.061(8)$.  The quantity
$d_{\rm{eff}}\equiv  \gamma /\nu + 2\beta /\nu$, which should equal 2, is measured as 1.990(16).  Errors quoted are determined by forcing the quantity larger or smaller than the optimum value, allowing all other parameters to vary freely, until the best fit has a $\chi^2$ increased by unity over the overall best fit.  The value of $\nu$ is about 
$10\%$ off the accepted value of 0.63.  This could be just a $3\sigma$ random error, but likely is at least half systematic.
Generally to reduce finite lattice size errors in exponents to under 5\% one needs to use a large number of different lattice sizes and fit to the next-order corrections.  In this study, which is simply about the existence of phase transitions, we will be satisfied with this level of error.  The Coulomb-gauge method has, therefore, successfully found the pure-gauge theory phase transition, finding the critical point within 0.13\% and $\nu$ within 10\%.  The other exponents are unique to this order parameter due to the two-dimensionality, so have no comparison values, but are seen to satisfy the $d=2$ hyperscaling relation (\ref{hsc2}) to under 1\%.

\section{Pure-Higgs Theory}
\begin{figure}[b!]\centering
                    \includegraphics[width=0.48\textwidth,  clip]{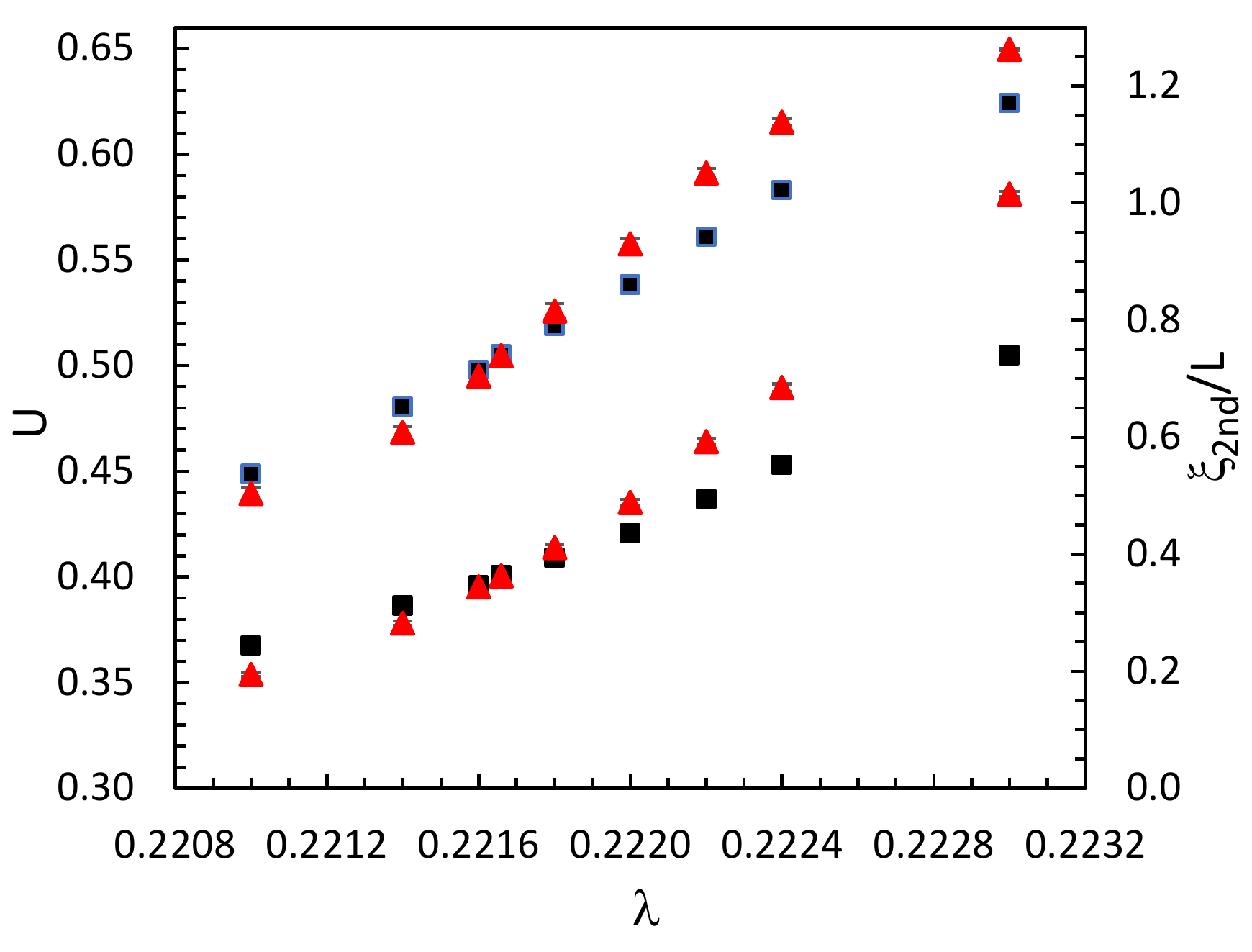}
                         \includegraphics[width=0.48\textwidth,  clip]{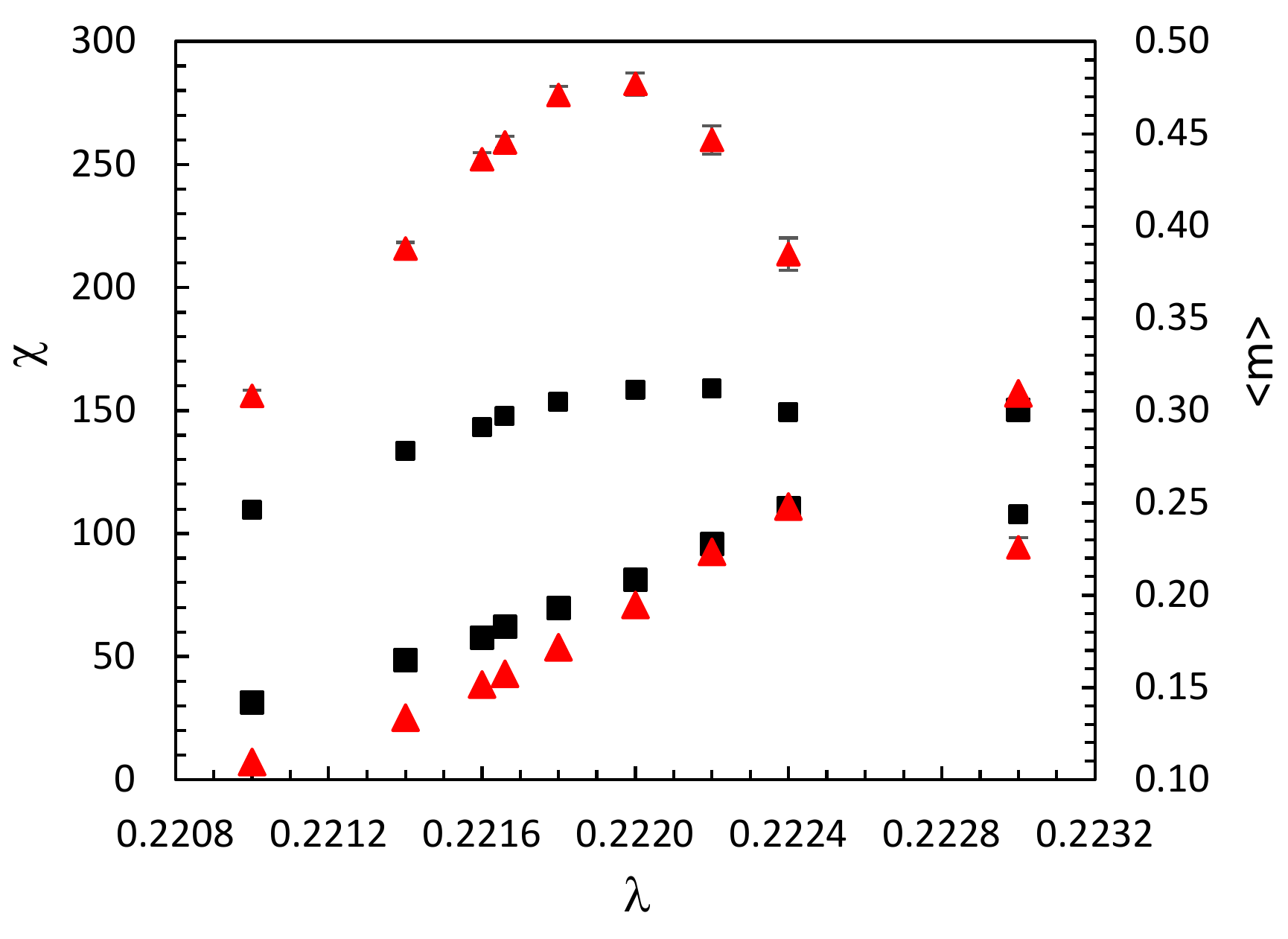}
                                  \caption{Binder cumulant and $\xi _{\rm{2nd}}/L$ (lower graph, right scale) crossings (a), magnetization (larger symbols, right scale) and susceptibility (b) for the pure-spin (Ising) model. Squares are $30^3$ lattice. Error ranges for $<m>$ are about one-fifth the size of plotted points.}
          \label{fig3}
       \end{figure}
The replica-overlap order parameter was already successfully demonstrated for the 4d SU(2) gauge-Higgs theory\cite{megh}.  Here we will test it on the pure-spin theory ($\beta=\infty$).  Because the order parameter is gauge invariant, transforming to axial gauge is not necessary. All plaquettes are unity, but negative links may still be present. As a
matter of convenience, however, all gauge links were set to unity for these pure-spin simulations.  In general, this method  requires a second ``replica"
Higgs field equilibrated to the gauge field.  For the pure-spin case the gauge field does not react to the Higgs field so no re-equilibration to follow changes in the gauge field is necessary.  In this case one simply has two independent Higgs fields.  Due to translational invariance, overlaps will be random, but in the magnetized phase, each Higgs will have a dominant phase, so
the net-overlap will be non-zero.  For instance, suppose the first Higgs has 60\% positive links and 40\% negative.  The ordinary magnetization would be $0.6-0.4=0.2$.  At this coupling the replica Higgs will also be 60/40 but could be either positive or negative.  Correlations between the two in this case are random, since they are completely independent, so the absolute value of the overlap, 
$|q|$, would average $0.6^2 +0.4^2-2(0.6)(0.4)=0.04$.  As mentioned we actually use $m\equiv \sqrt{|q|}$ in our analysis to
better mimic the ordinary magnetization, which here gives exactly the expected value of 0.2.  Aside from some extra noise,
the replica overlap should give the same results as the ordinary magnetization.  The advantage is that it will still work the same after a random gauge transformation which would render the ordinary magnetization useless (it would average to zero at all couplings).  
\begin{figure}[bt!]\centering
                    \includegraphics[width=0.48\textwidth,  clip]{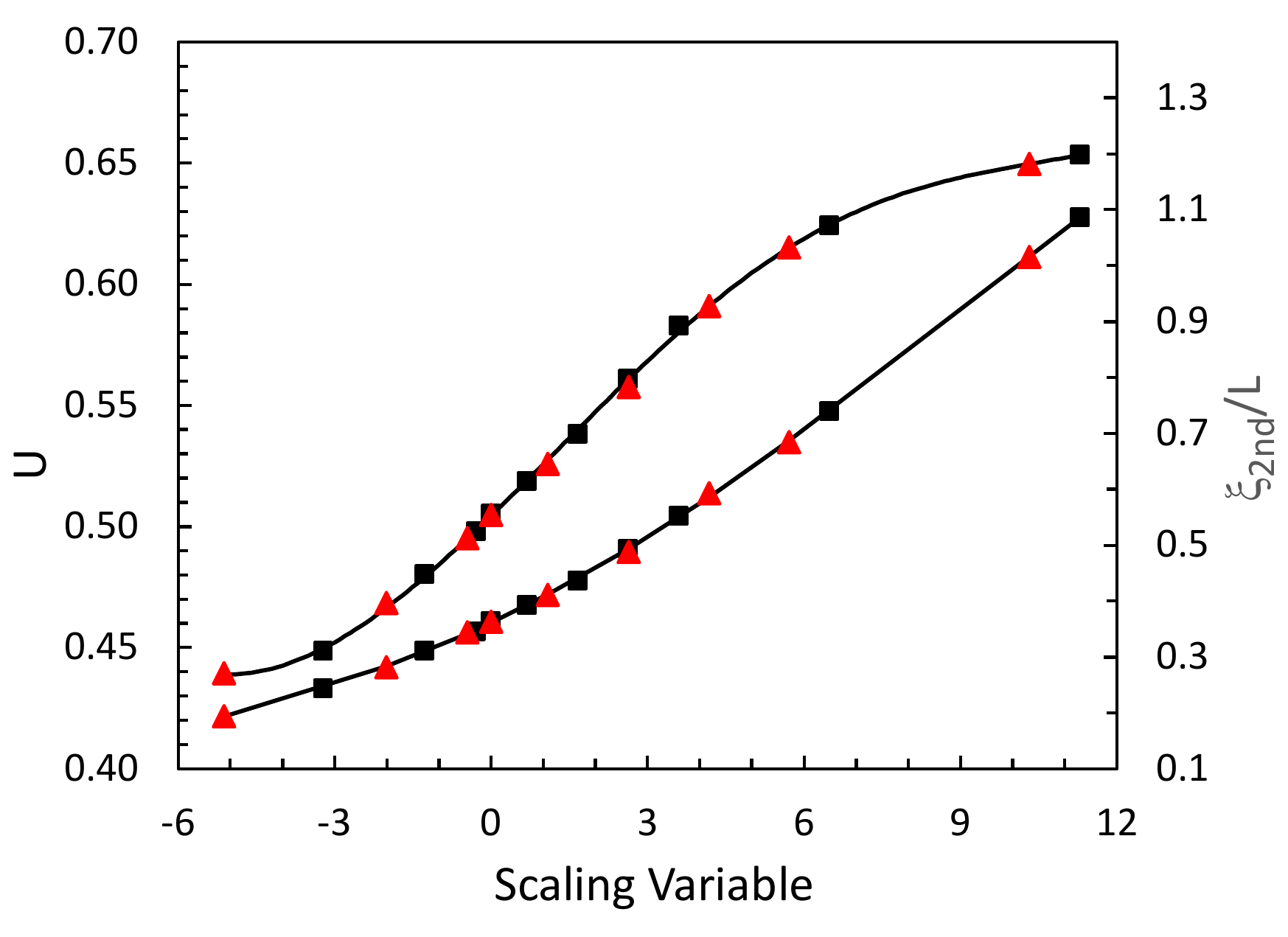}
                         \includegraphics[width=0.48\textwidth,  clip]{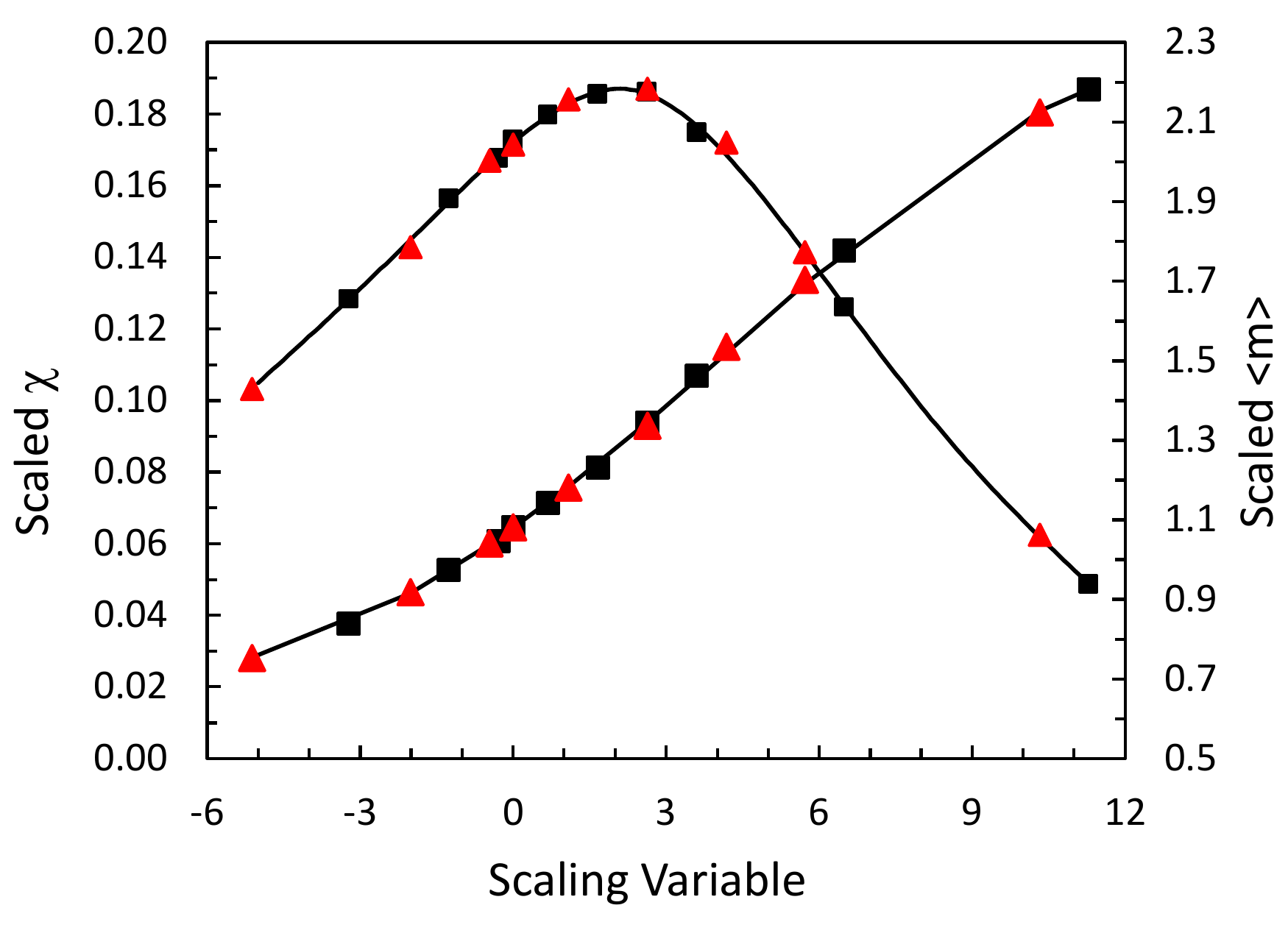}
                                  \caption{Pure-spin scaling collapse graphs for Binder cumulant and $\xi _{\rm{2nd}}/L$ (lower curve, right scale) (a).  Collapse graphs for magnetization (larger symbols, right scale) and susceptibility (b).}
          \label{fig4}
       \end{figure}

To be sure our techniques were working, the pure-spin theory was studied with this order parameter.  Because this case was expected to be straightforward, only two lattice sizes, $30^3$ and $40^3$, were used.  For $30^3$, $2.5\times 10^6$ sweeps with measurements every five were used, and 40\% fewer for the $40^3$ lattice.  Note, of course, that these simulations use ordinary periodic boundary conditions.  Fig.~3a shows the Binder cumulant, $U$ vs. coupling, as well as another quantity with a similar behavior, the second moment correlation length\cite{cooper} $\xi _{\rm{2nd}}$ divided by the linear lattice size, $L$. Like $U$ this quantity also has a common finite value for all lattices at the critical point.  It diverges with lattice size in the magnetized region and vanishes with lattice size in the random phase, thus it has the same type of crossing behavior as U.  In many cases it is practically superior, because U has an upper limit of 2/3, but $\xi _{\rm{2nd}}/L$ is divergent.  The calculation of $\xi _{\rm{2nd}}$ \cite{cooper} involves Fourier methods, and has the advantage that there is no fitting or human choices that need to be made.   

The expected critical point is at 0.2216595(26)\cite{fl}.  The data for both $U$ and 
$\xi _{\rm{2nd}}/L$ are consistent with crossing here (one simulation was run at 0.22166). The magnetization
 ($m=\sqrt{|q|}$) and susceptibility have typical behavior (Fig.~3b).
Scaling collapse fits for the four quantities are shown in Fig.~4a,b, where the scaling variable is Eq.~\ref{scalevar} but with 
$\lambda$ and $\lambda _c$ in place of $\beta$ and $\beta _c$. Here we are using PBC so $L _{\rm{eff}}=L$.  The overall fit to all four quantities has 47 degrees of freedom and $\chi^2$/d.f.$=0.7$.  The critical exponents determined all agree well with expectations.  We obtain $\lambda _c=.221659(11)$,
$\nu = 0.621(9)$, $\gamma /\nu =1.98(3)$ and $\beta ' /\nu = 0.523(9)$, compared to conventionally determined values by a more sophisticated Monte Carlos of $\nu=0.63002(10)\cite{mh} $, $\gamma /\nu =1.9828(57)\cite{fl} $ and  $\beta ' /\nu = 0.518(7)$\cite{fl} .  We see
that the replica overlap order parameter is quite successful and performs similarly to the ordinary magnetization in the 
pure-spin model.

\section{First-order region}
\begin{figure}[b!]\centering
                    \includegraphics[width=0.48\textwidth,  clip]{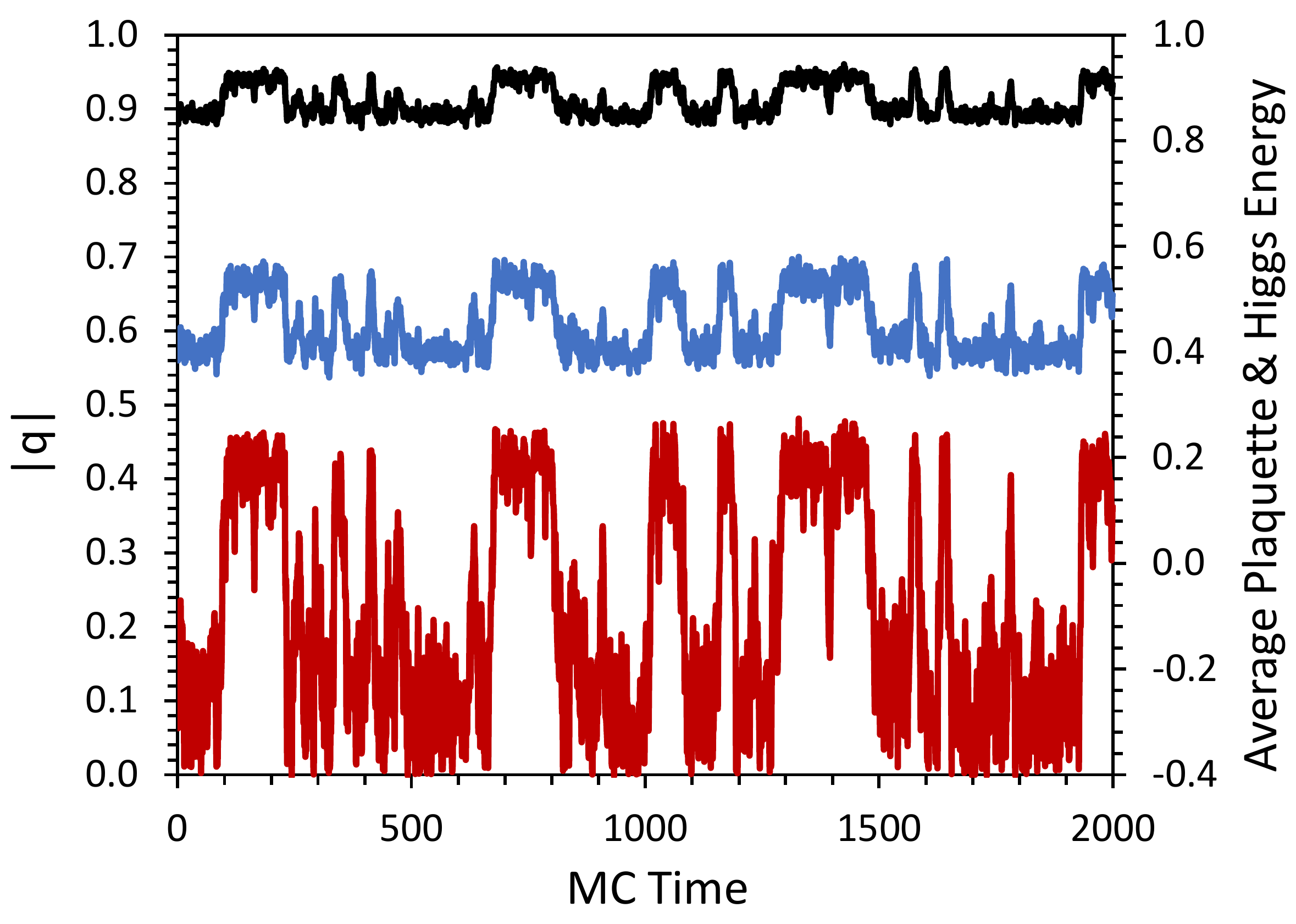}
                                  \caption{Time series for run in first-order region at $\beta = 0.708$ on the self-dual line. Replica order parameter is shown in lower trace (left scale), Higgs energy in middle (right scale) and average plaquette at top (right scale). The replica order parameter is seen to exactly follow first-order jumps in energy quantities}
          \label{fig5}
       \end{figure}
We next explore the region previously observed to have a first-order phase transition running along the self-dual line.  The strongest first-order transition has been identified as occurring near ($\beta=0.708$, $\lambda=0.2476$)\cite{ggrt}. We first explored this point with the replica-overlap order parameter.  As with all such runs, an equilibration study was first performed to determine the 
number of equilibration sweeps needed to reduce the remaining systematic 
error from equilibration to a level significantly below the random error.  A typical study will be detailed below.   In the first-order region the errors were dominated by the varying times spent in the two different phases for runs of the length performed,  so a systematic study of equilibration could not easily be performed.  However, no differences could be observed between 50,000, 80,000,  and 100,000 equilibration sweeps, so 50,000 was deemed adequate here (perhaps much more than adequate).  The precise values of magnetizations are not really needed here - only the observation of two distinct levels.  Fig.~5 shows a time series for a run on a $30^3$ lattice. This was a 400,000 sweep run with measurements performed every 200 sweeps. The figure shows the replica-overlap, $|q|$, tracking exactly the jumps in the plaquette and Higgs energy.  Histograms of magnetization and Higgs energy are shown in Fig.~6ab, both of which show a clear bimodal distribution characteristic of a first-order transition (plaquette histogram, not shown, is very similar to Higgs energy).  Moreover, if the histograms are each divided in the middle to identify phases, the magnetization and Higgs energy agree on
which phase the system is in (lower or upper) 99\% of the time.  One could hardly ask for a clearer confirmation that this transition is symmetry-breaking, and that the replica overlap order parameter can see it clearly.  This one piece of evidence is, by itself, enough to nullify the long-standing hypothesis that the endpoint of the first-order line is an ordinary critical point.  As is well known, symmetry-breaking first-order transitions cannot just end in the middle of a phase diagram\cite{ll}. The endpoint of a symmetry-breaking first-order transition is a tricritical point, where the transition shifts from first-order to higher order.  In the Landau theory this is where the fourth-order term in effective free-energy switches sign from negative to positive\cite{cl}.  The symmetry-breaking transition line or lines must continue to the edge of the phase diagram.
\begin{figure}[t!]\centering
                    \includegraphics[width=0.48\textwidth,  clip]{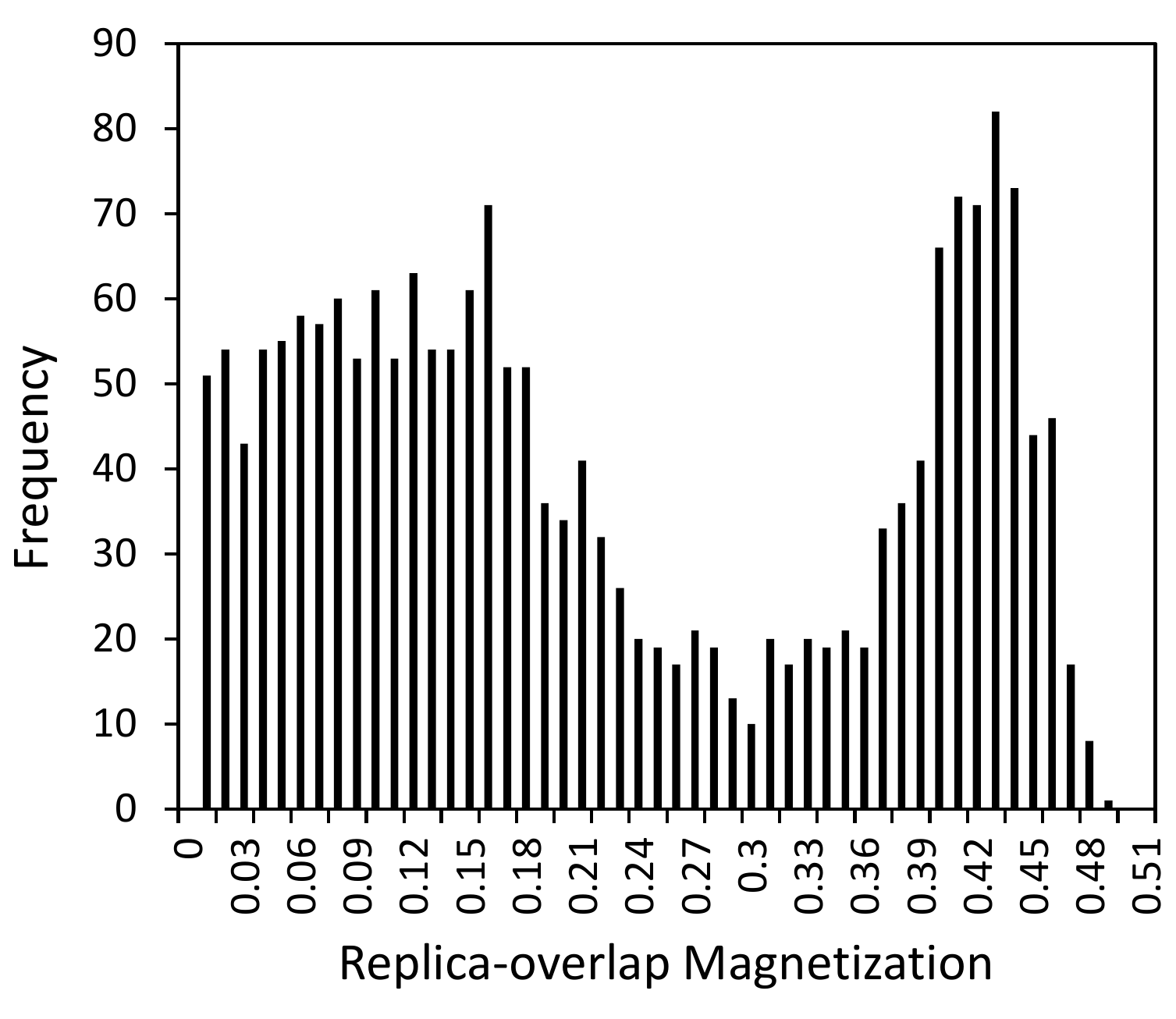}
                         \includegraphics[width=0.48\textwidth,  clip]{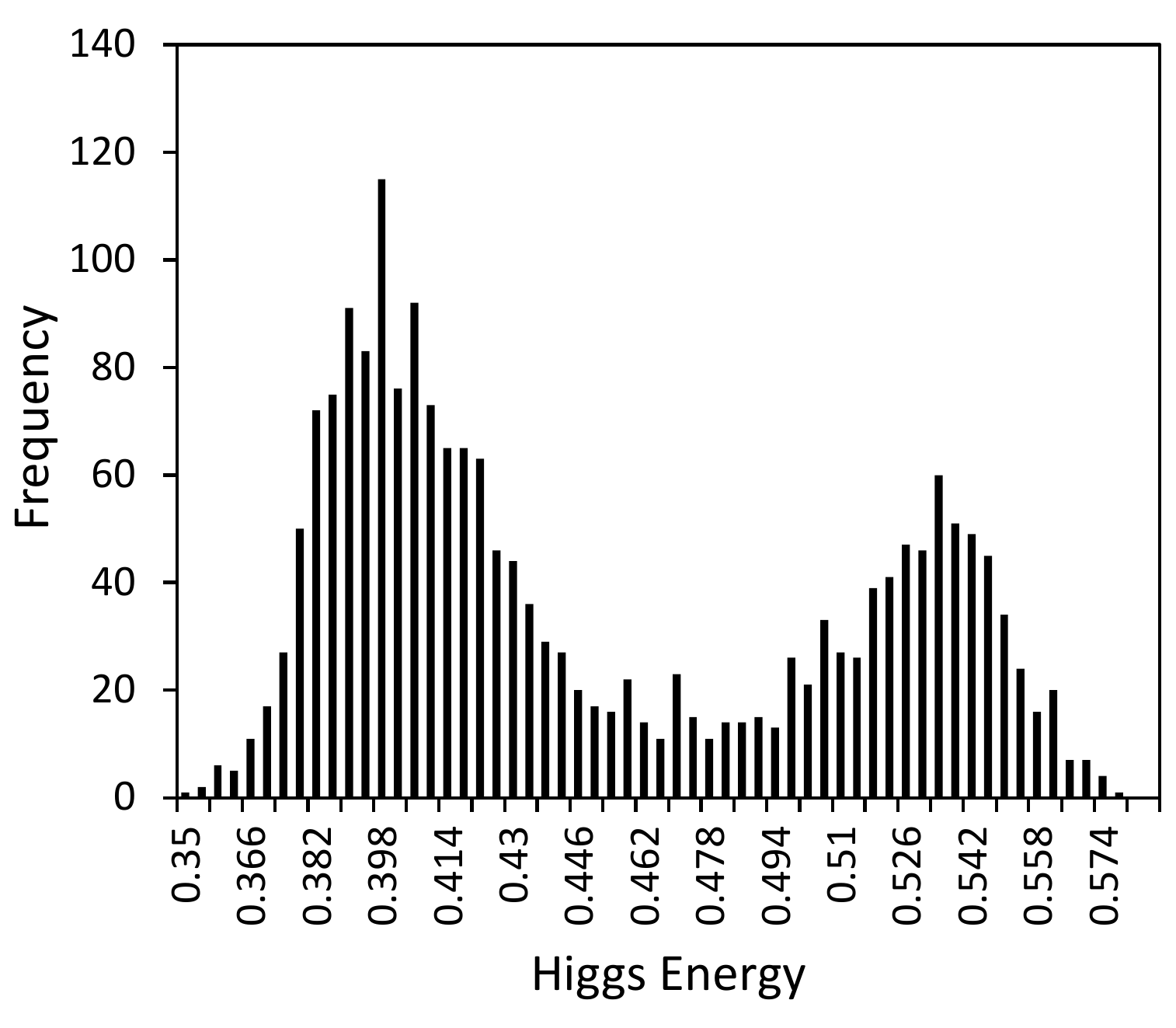}
                                  \caption{Histograms for replica order parameter (a)  and Higgs energy (b).}
          \label{fig6}
       \end{figure}

The question arises what the Coulomb-gauge magnetization shows here.  It too sees  a transition very close to the self-dual line, so probably coincident. However, there is a surprise here due to the open boundary condition.  The transition no longer appears first-order!  Energy distributions throughout the critical region are unimodal.  In addition, the specific heat (plaquette second moment) measured in the open boundary condition runs on $30^3$, $40^3$, and $50^3$ lattices shows a peak growing much too slowly to match first-order.  The peak-heights fit a finite-size scaling law of $L^{\alpha /\nu}$ with $\alpha /\nu =1.39$(12).  For a first-order transition it should scale as $L^3$, following the spatial dimension of three. Indeed for our PBC runs, scaling of both plaquette and Higgs energy second moments averaged to an exponent of 
$2.84(11)$, fulfilling that expectation. The exponent on the OFA lattice is clearly nowhere near this. There is a qualitative difference between the open boundary and PBC cases, in that, for a lattice to tunnel between phases, two percolating boundaries are needed for PBC but only one for OBC (or OFA).  This could account for the different behavior.  Of course in the infinite lattice limit  the boundary condition should not matter.  Because the open boundary is in some sense ``larger than infinity" this result suggests that the first-order nature of the phase transition even where it appears strong could be a finite-lattice artifact.  Because this issue is not of crucial importance to the main thrust of this paper, it will be left for a future study.  Very large lattices may be needed for a definitive conclusion. 
\begin{figure}[t!]\centering
                    \includegraphics[width=0.475\textwidth,  clip]{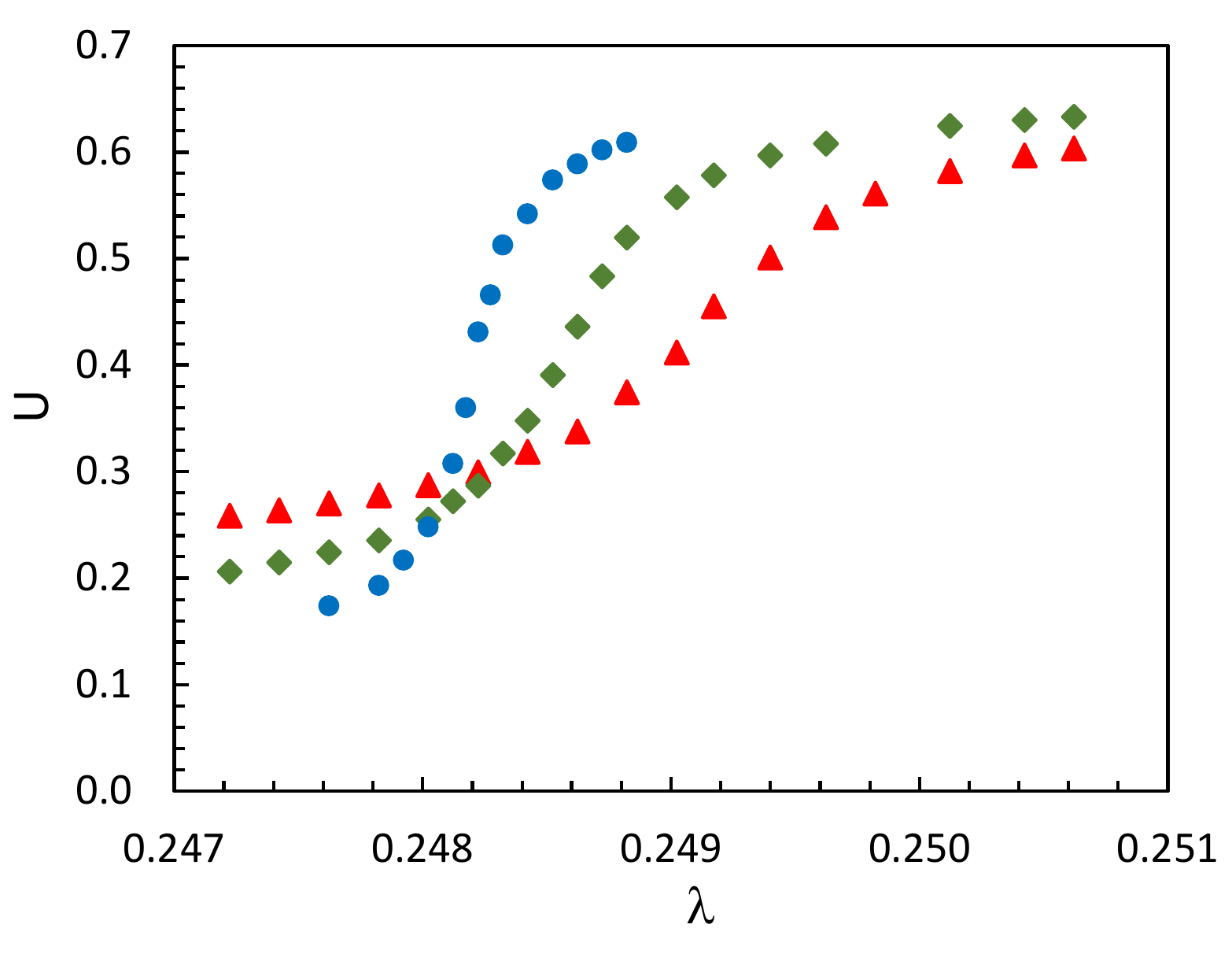}
                         \includegraphics[width=0.515\textwidth,  clip]{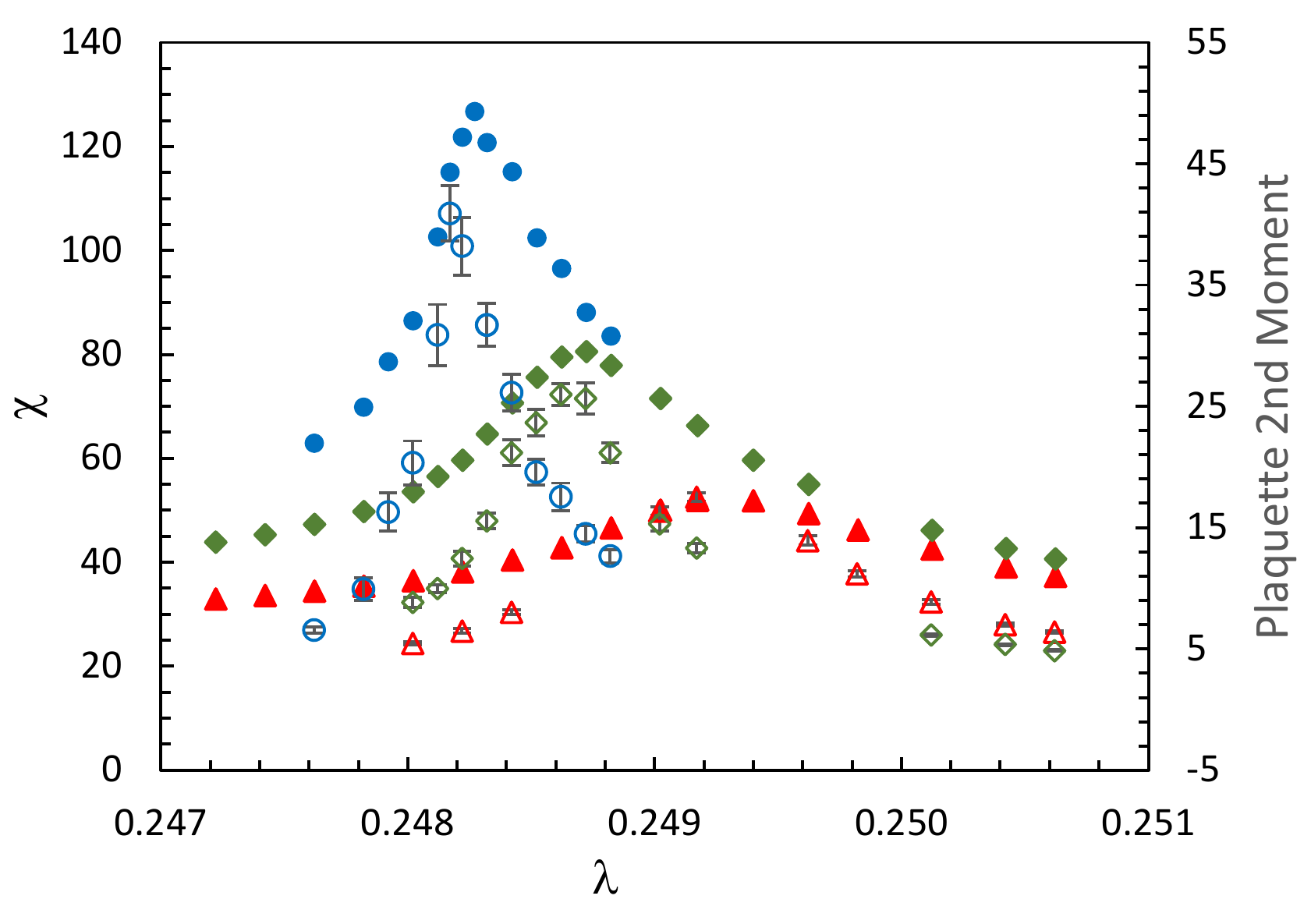}
                        \includegraphics[width=0.48\textwidth,  clip]{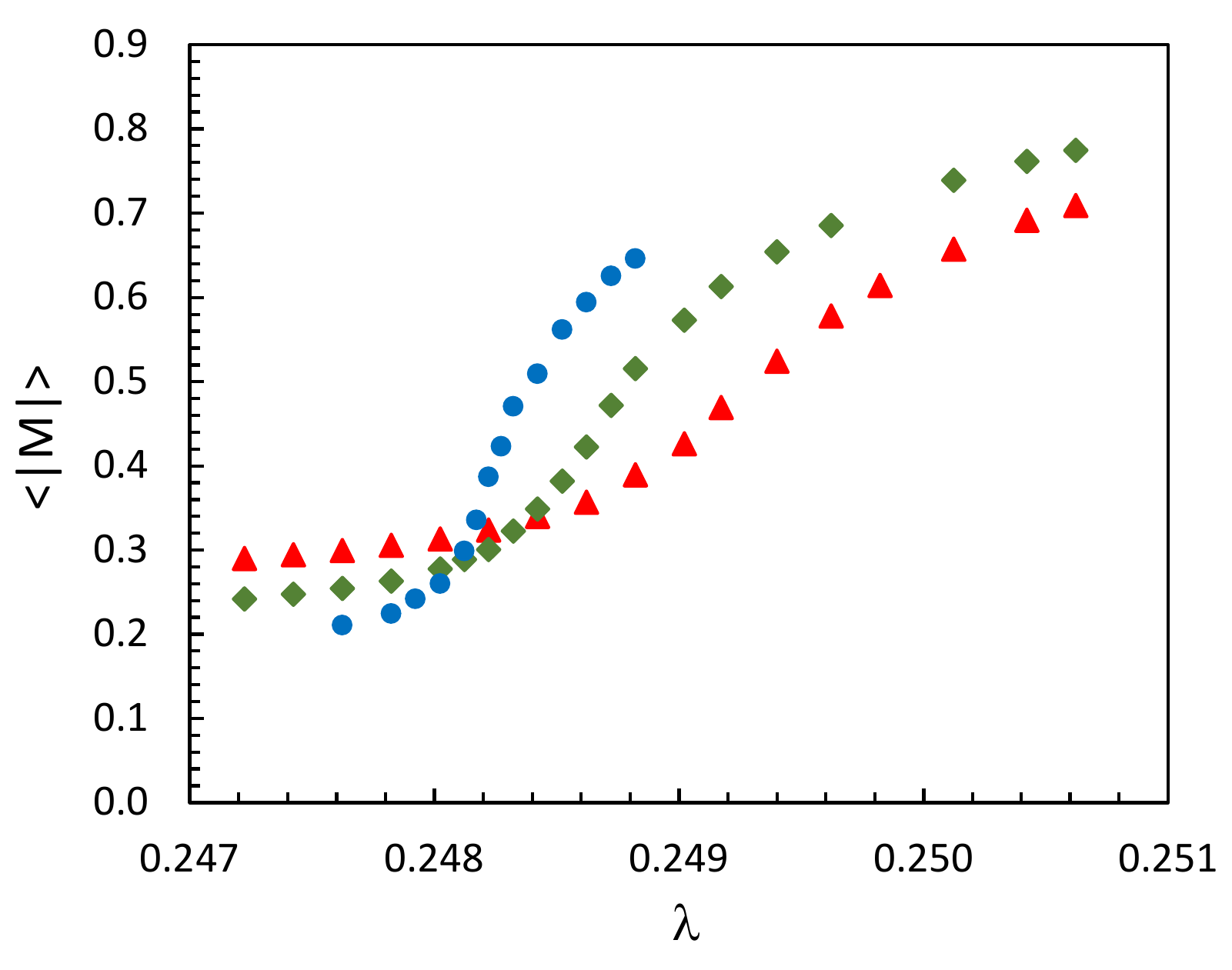}
                                  \caption{Binder cumulant crossing (a),  susceptibility (b), and Coulomb magnetization (c) for $\beta =0.708$. Error ranges are about one-half the size of plotted points for Binder cumulant, one-third for susceptibility, and one-quarter for magnetization. Open symbols on (b) show the plaquette second moment (specific heat), using right scale.}
          \label{fig7}
       \end{figure}
\begin{figure}[bth!]\centering
                    \includegraphics[width=0.51\textwidth,  clip]{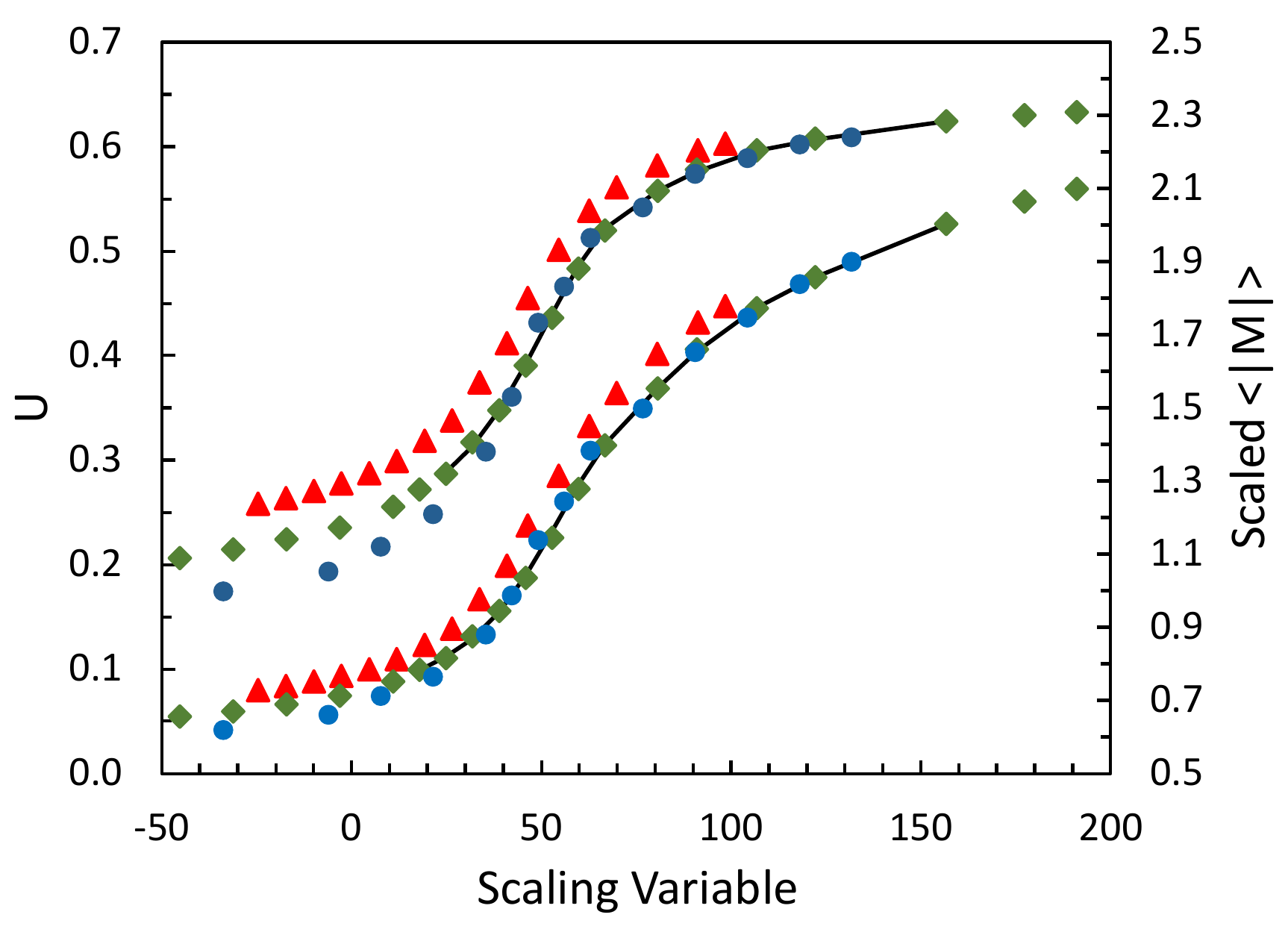}
                         \includegraphics[width=0.475\textwidth,  clip]{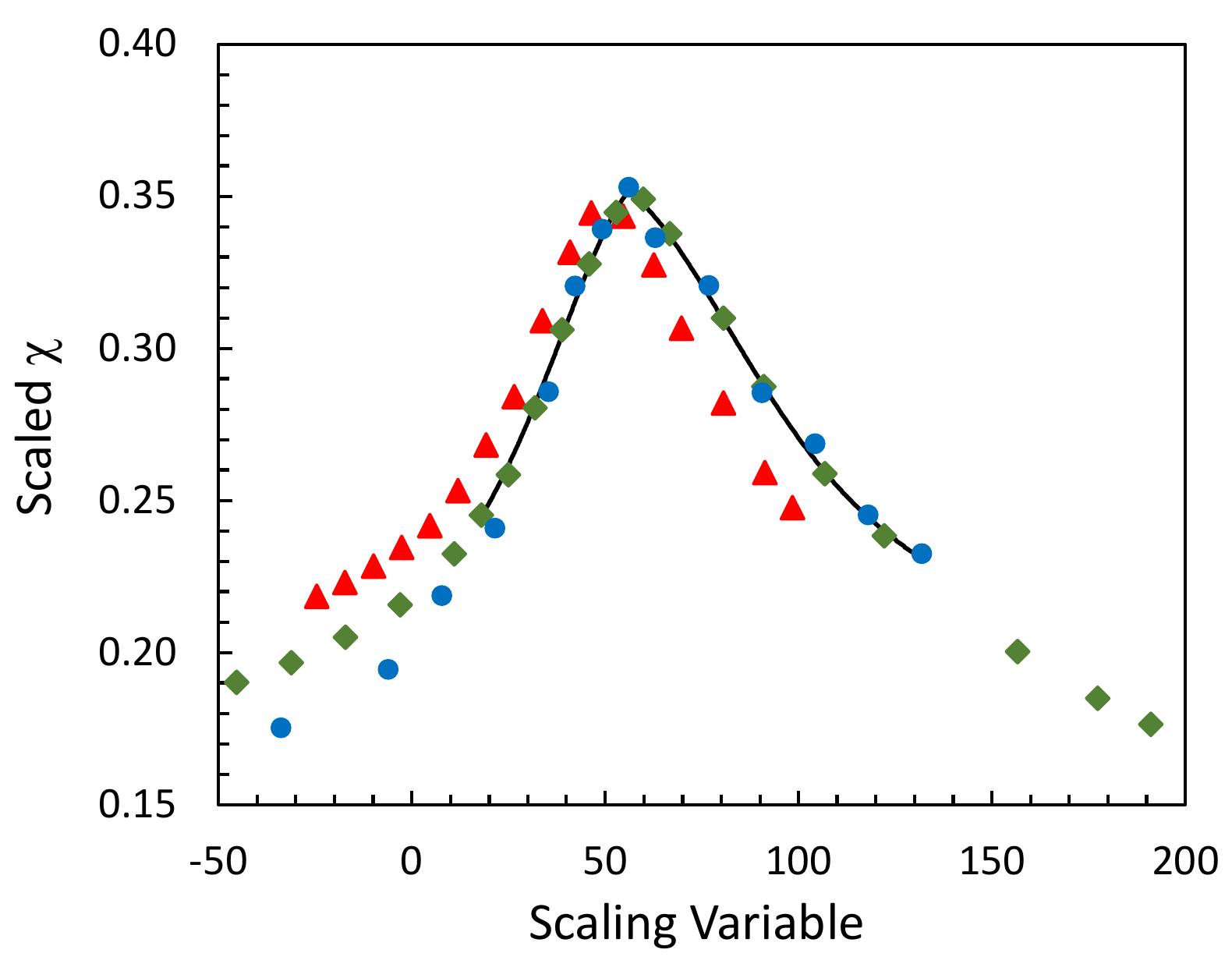}
                                  \caption{Scaling collapse graphs for $\beta =0.708$. Binder cumulant (left graph), and scaled Coulomb magnetization (a), and scaled susceptibility (b).}
          \label{fig8}
       \end{figure} 

The Coulomb-gauge order-parameter analysis is shown in Figs.~7,8.  For each coupling $2.5\times 10^6$ sweeps were performed on the $40^3$ lattice, $1.25\times 10^6$ on the $50^3$ and $6.25\times 10^5$ on the $64^3$ lattice. The Coulomb gauge-transformation and measurements were performed every five Monte Carlo sweeps.  This large number of measurements was possible due to the high speed of the Edmonds algorithm, which took a time roughly equal to 1-5 
Monte Carlo sweeps, depending on the coupling and lattice size. Fig.~7a plots the Binder cumulant. This shows a clear crossing, but the smaller lattice crosses at a somewhat different point showing some noticeable higher-order scaling effects.  The $50^3$ and $64^3$ cross around $\lambda=0.2480$, slightly above the expected value of 0.24762 from self-duality.  However, considering the shift seen from the $40^3$ lattice, it is reasonable to expect a finite-size shift of this magnitude.  The susceptibility (Fig.~7b) shows a rising peak, also shifting toward the expected critical point (the specific heat shows a similar shift). The magnetization (Fig.~7c) also shows a crossing.  Magnetizations are not required to cross and often do not. If they do cross, the crossing is always above the critical point.  Nevertheless, a growing magnetization with lattice size is a sure indicator of an ordered phase. The evidence here for an order-disorder phase transition is very strong. Separations of $U$ between $50^3$ and $64^3$ lattices are of order $50\sigma$ both above and below the transition.  Scaling collapse plots are shown in Fig.~8a,b.  Only the larger two lattices were used.  The $40^3$ data could be made to agree fairly well  if a different $\lambda _c$ were to be used for it, but we elected not to do this.   An oddity of this system that did not occur at any of the other couplings studied was that no reasonable fit to the data below the phase transition could be found.  This is presumably because the crossing takes place so low on the sigmoid curve for this case.   The U plot here is nearly horizontal making it nearly impossible for a horizontal shift to bring the curves together.  However, the $64^3$ data is more sloped, so larger lattices may be all that is needed to see scaling in this region.  The overall fit to the three datasets had 56 degrees of freedom and gave $\chi ^2$/d.f.=1.8.  This probably indicates a slight tension with higher-order scaling effects.  The results for critical point and scaling exponents were $\lambda _c=0.24787(4)$, $\nu =0.441(3)$, $\gamma /\nu = 1.47(2)$, $\beta '/\nu = 0.27(2)$ and 
$d_{\rm{eff}}=2.02(3)$.  In this case errors were determined by allowing $\chi ^2$ to increase by 1.8 rather than 1.0 to compensate for the less-than-perfect initial fit.  This $\nu$ value is consistent with that seen from the specific heat analysis. Using the hyperscaling relationship  (\ref{hsc1}) in the form $\nu = 2/((\alpha /\nu )+d)$, and using the value of 
$\alpha /\nu$ extracted from specific-heat scaling above, gives $\nu=0.456(13)$.  What is seen here is a fairly strong second-order transition, also clearly symmetry-breaking. It is strong in the sense that $\nu$ is well {\em below} the 3D Ising value of 0.63.  So, although the second-order nature was a surprise, there clearly is a thermal phase transition here, with the specific heat finite-size scaling and correlation-length critical exponent from the Coulomb order-parameter analysis
in agreement.  These observations show that the transition here is both symmetry-breaking and thermal.

\section{Transitions at $\beta=0.5$}
We now turn to simulations at $\beta=0.5$, which is well beyond the first-order endpoint at $\beta=0.689$.  This is the region where no phase transition is believed to exist.  On the pure-spin axis the last singularity is believed to be the
roughening transition at $\lambda=0.409(4)$ \cite{rougheningising}, which has a dual point on the gauge axis at 
$\beta= 0.474$. For lower $\beta$ on the axis($\lambda = 0$), there is good evidence that the strong-coupling expansion converges,
and the behavior is analytic.  It is an interesting question what happens to the roughening singularity as one enters the phase diagram.  Fradkin and Shenker (FS) argue that the Higgs coupling actually helps convergence. The analyticity region they draw curves to the right, reflecting this.  This suggests that $\beta=0.5$ near the self-dual line may lie in the expected analytic region.  
Of course, here we are proposing the alternate hypothesis that transitions persist in this region, continuing all the way to 
($\beta =0$, $\lambda =\infty$).
\begin{figure}[t!]\centering
                    \includegraphics[width=0.48\textwidth,  clip]{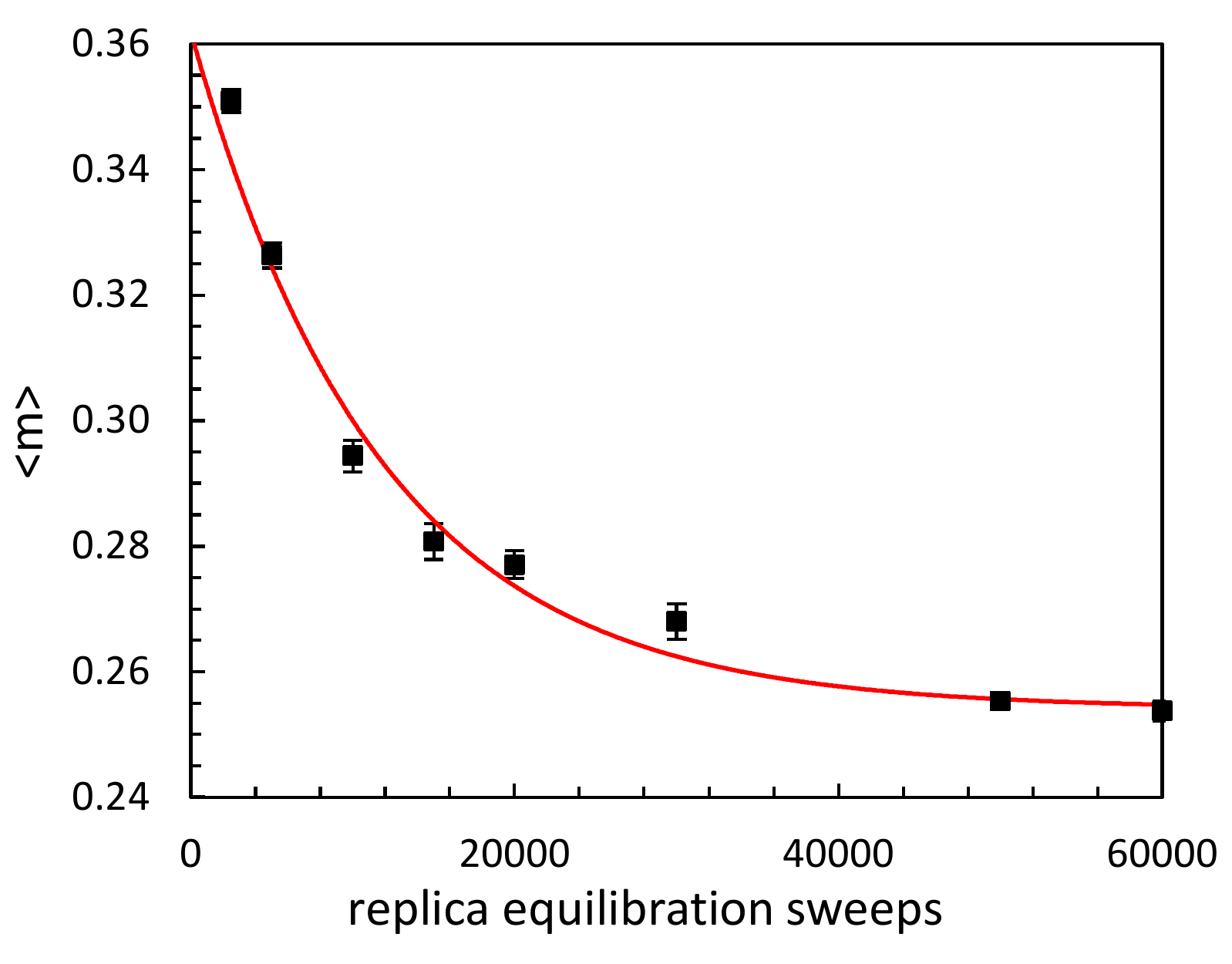}
                         \includegraphics[width=0.48\textwidth,  clip]{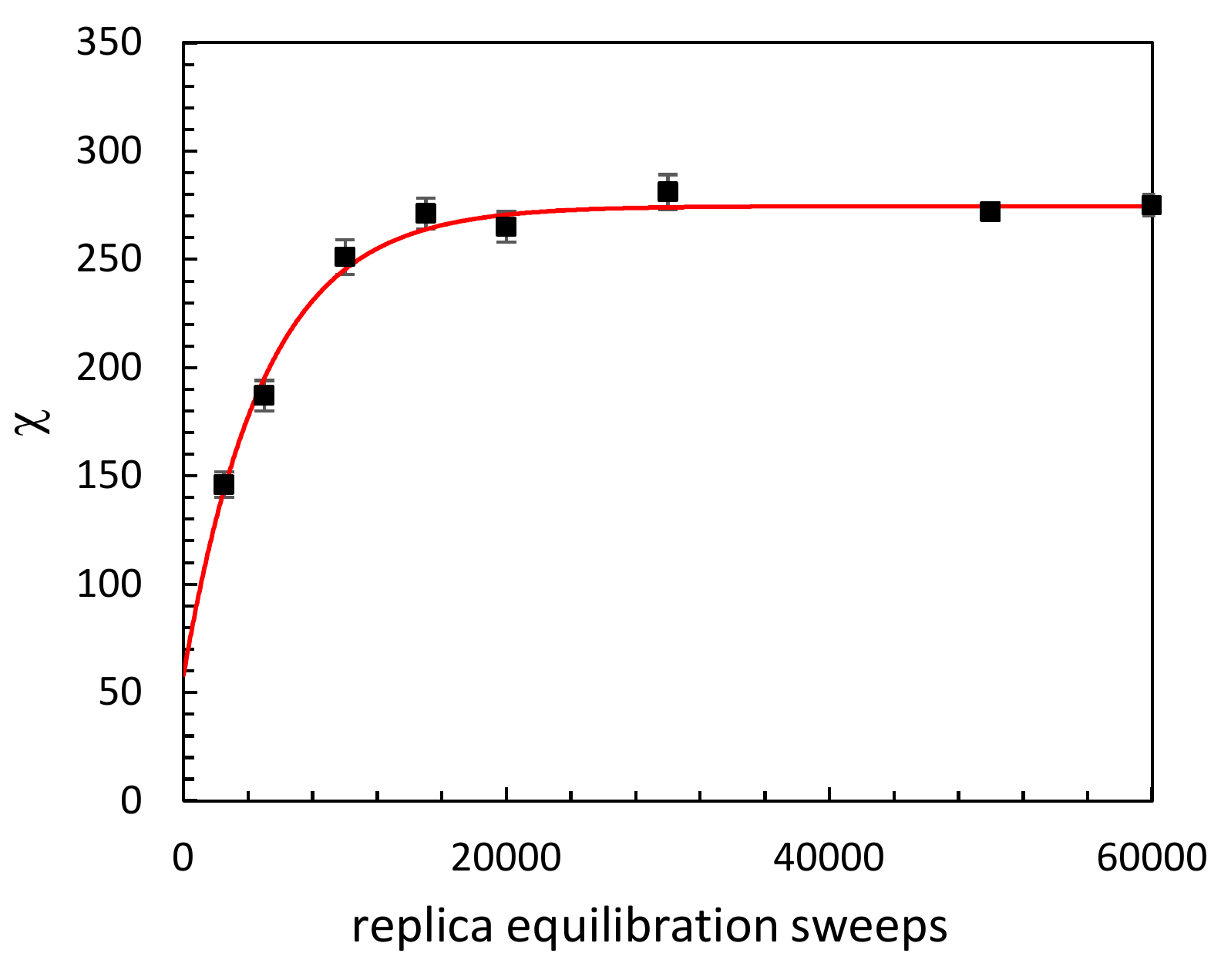}
                        \includegraphics[width=0.48\textwidth,  clip]{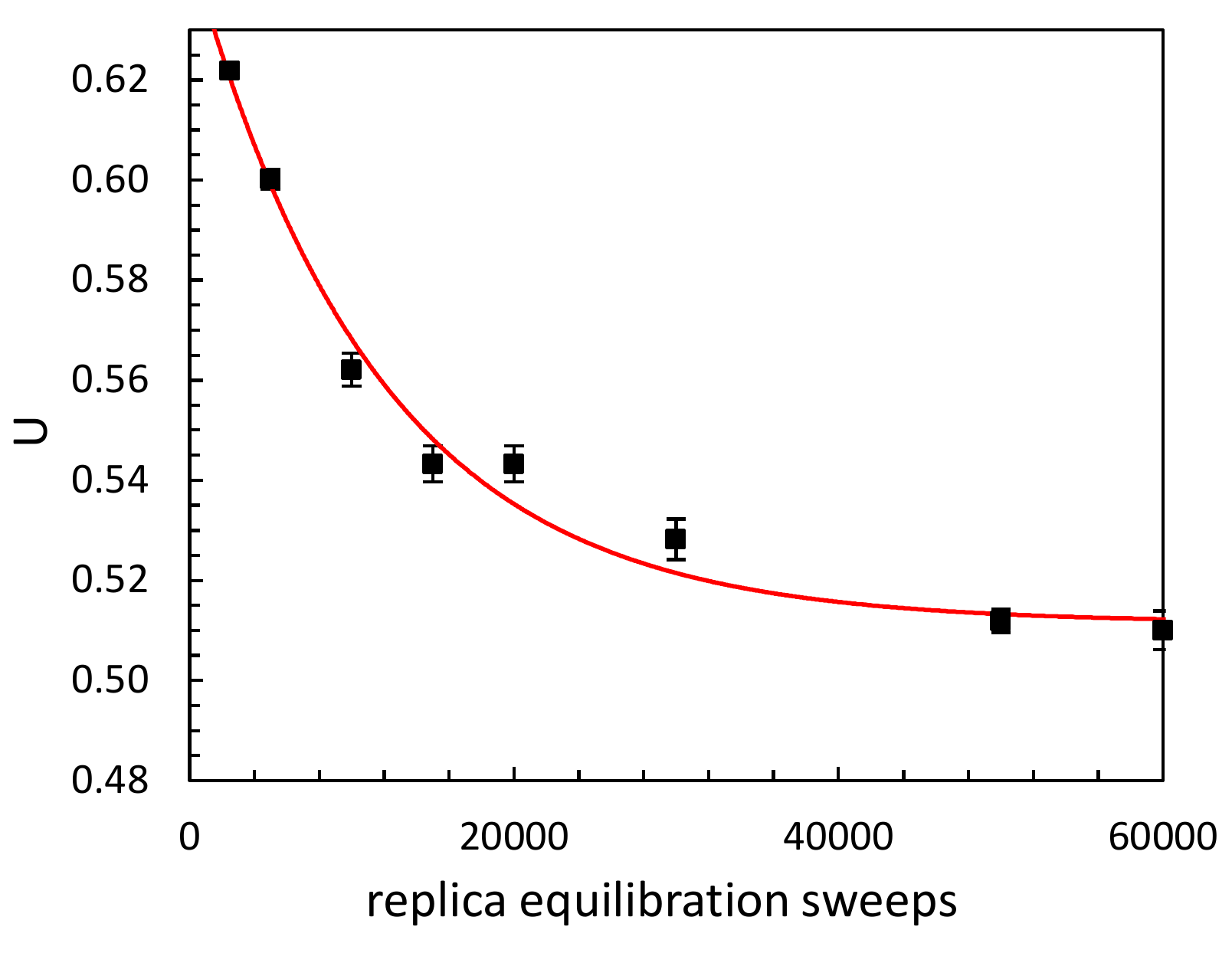}
                                  \caption{Replica equilibration for magnetization (a), susceptibility (b), and Binder cumulant (c) at $\beta=0.5$, $\lambda =0.34$ on a $30^3$ lattice.}
          \label{fig9}
       \end{figure}
\begin{figure}[bt!]\centering
                    \includegraphics[width=0.52\textwidth,  clip]{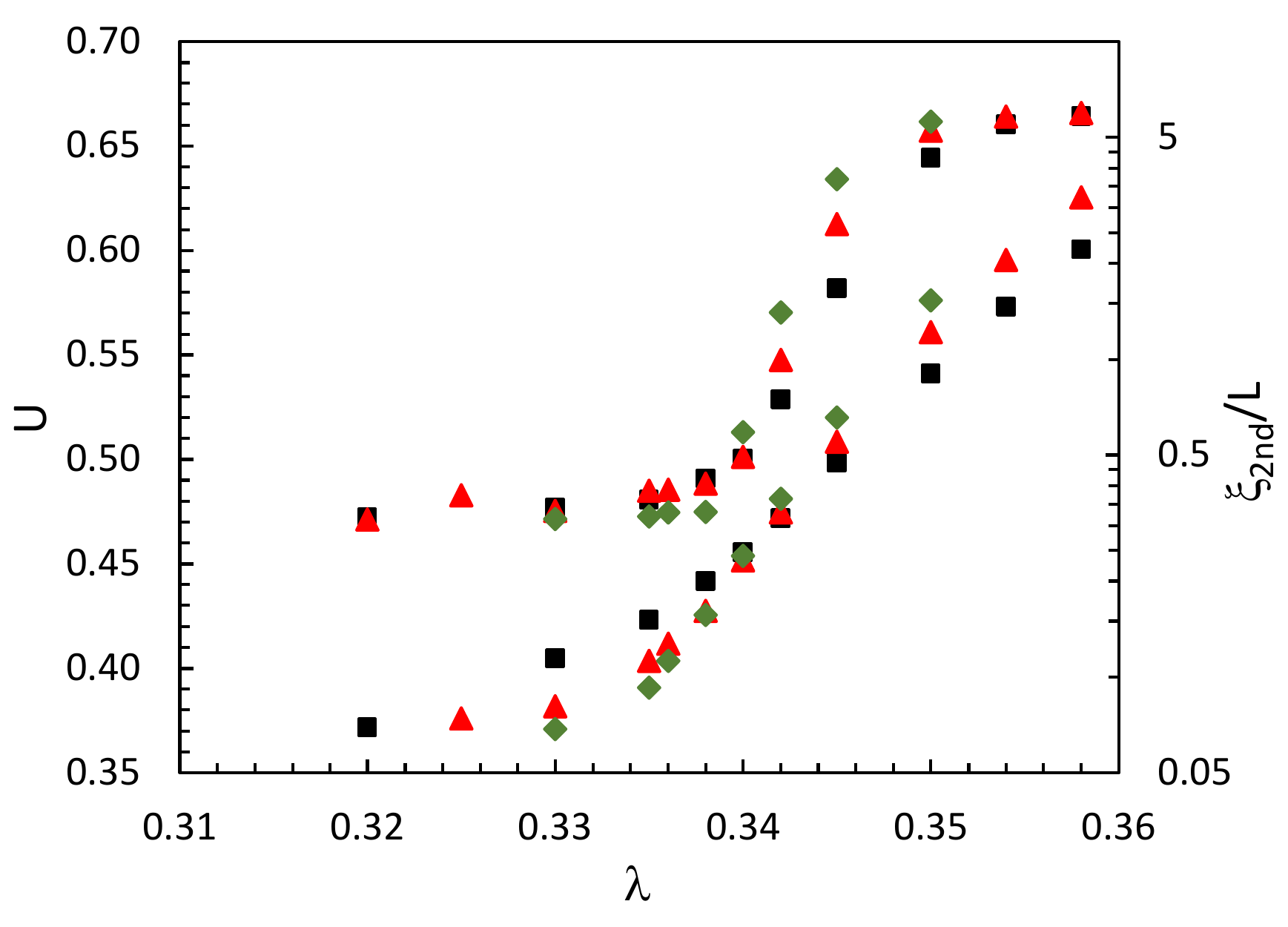}
                         \includegraphics[width=0.48\textwidth,  clip]{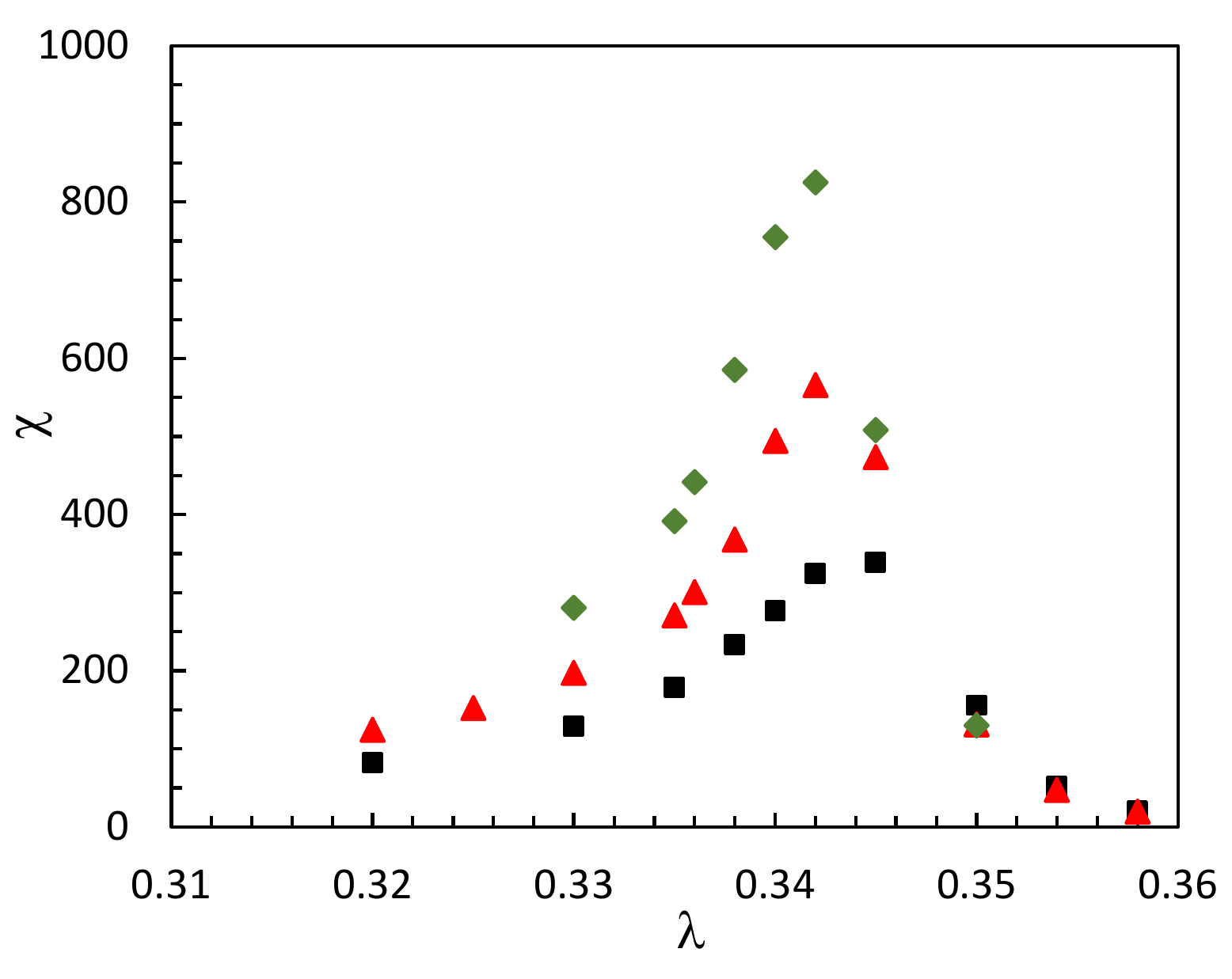}
                        \includegraphics[width=0.48\textwidth,  clip]{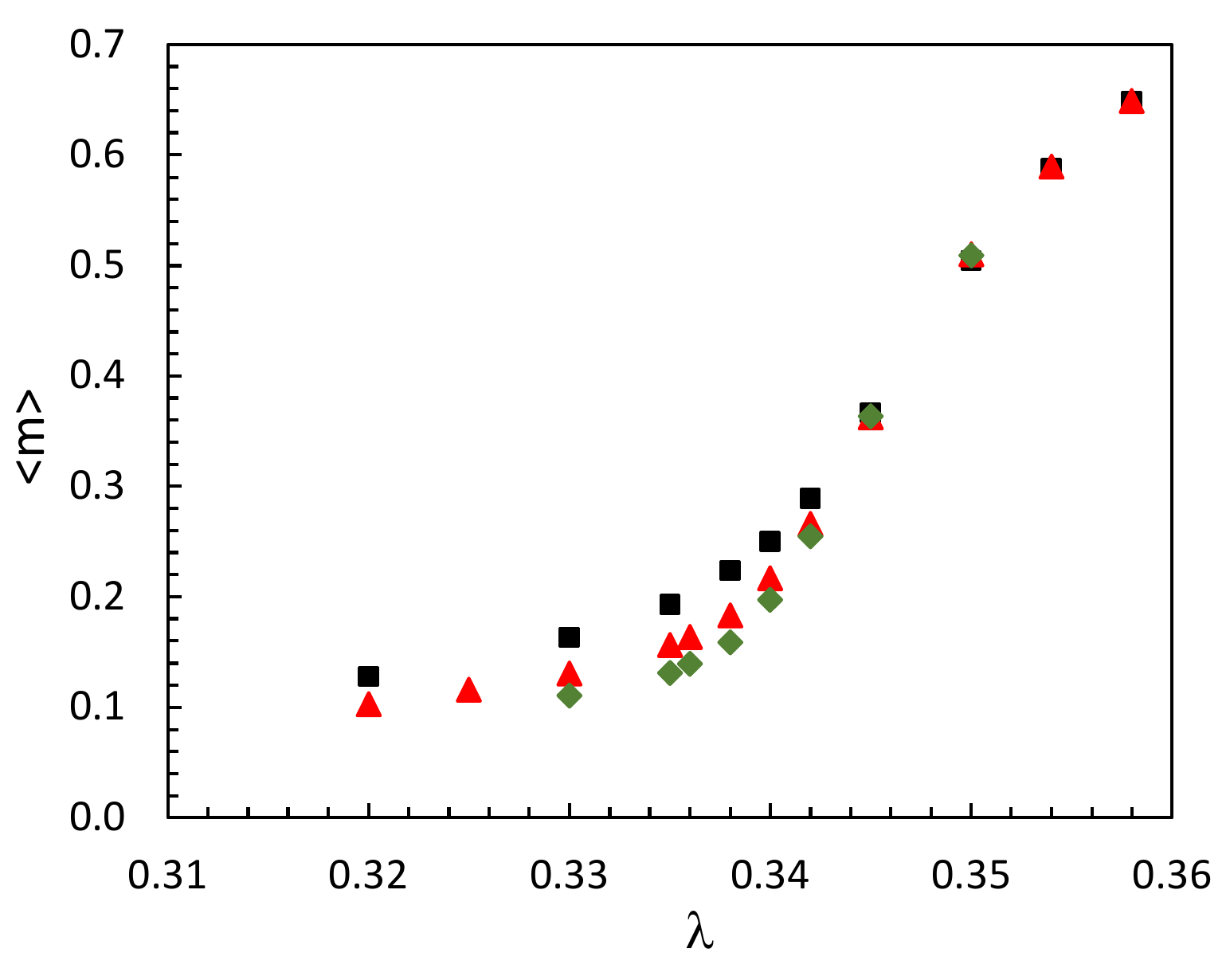}
                                  \caption{Crossing graphs for replica Binder cumulant and $\xi _{\rm{2nd}}/L$ (lower plot - note logarithmic scale) (a), susceptibility (b), and replica-overlap magnetization (c) at $\beta=0.5$.  Boxes are $30^3$, triangles $40^3$ and diamonds $50^3$ lattices. Error ranges for Binder cumulant are about one-half the size of plotted points, for
$\xi _{\rm{2nd}}$ one-eighth, for susceptibility one-quarter, and for magnetization one-tenth.}
          \label{fig10}
\end{figure}
\begin{figure}[bth!]\centering
                    \includegraphics[width=0.48\textwidth,  clip]{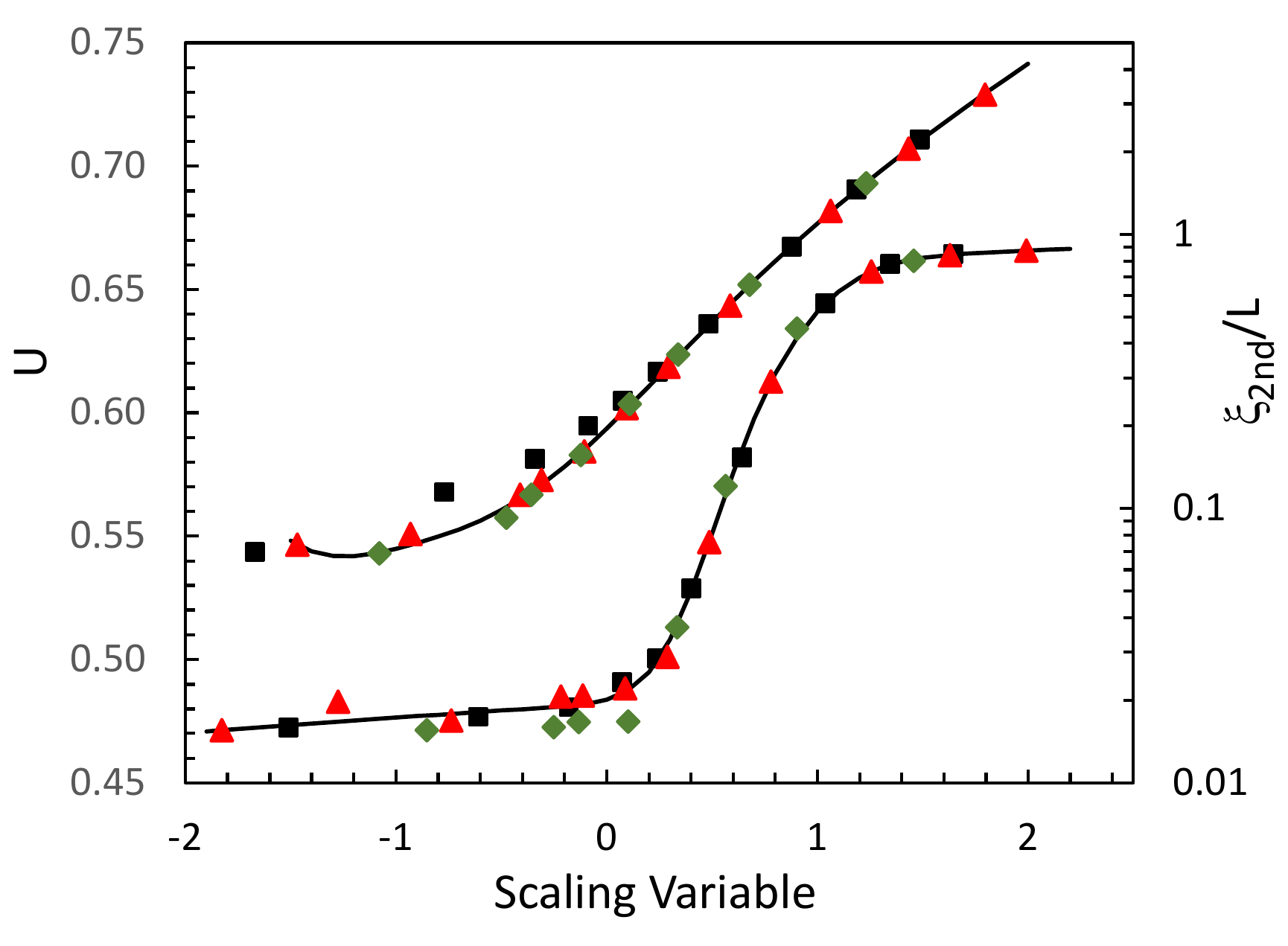}
                         \includegraphics[width=0.486\textwidth,  clip]{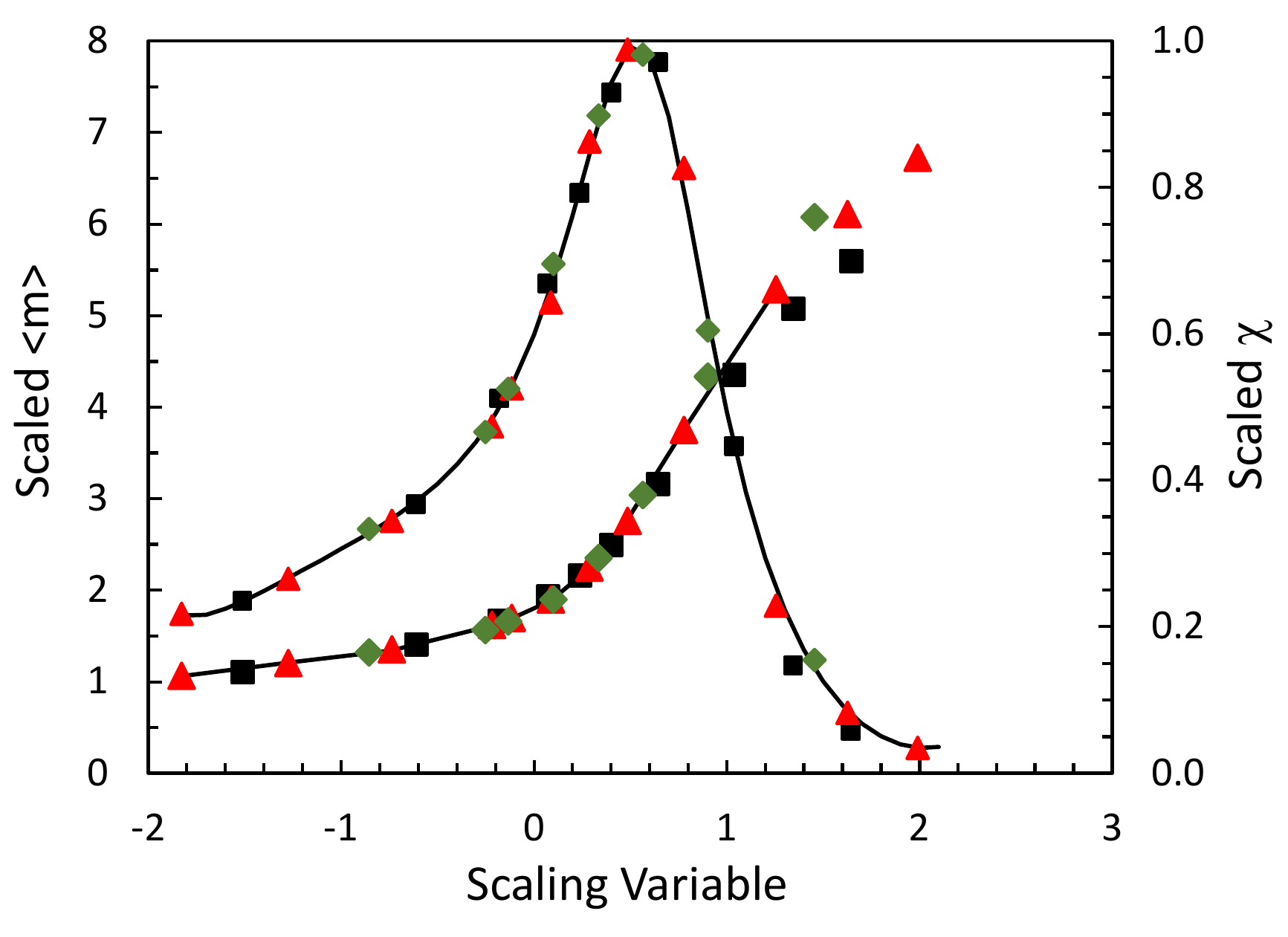}
                                   \caption{Scaling-collapse fits for Binder cumulant and $\xi _{\rm{2nd}}/L$ (this time upper curve - right logarithmic scale) (a), and for susceptibility and replica-overlap magnetization (larger symbols - left scale) (b), at 
$\beta=0.5$.} 
          \label{fig11}
\end{figure}

At $\beta= 0.5 $, the self-dual line has $\lambda = 0.38597$. At this coupling the replica-overlap magnetization appears to be well within the ordered phase, but the Coulomb-gauge magnetization is still in the random phase.  This suggests the transitions have split again, which would reflect the situation at the right end of the first-order line, where it clearly turns into two split higher-order lines. Recall also that Caudy and Greensite see a split transition between Coulomb and Landau-gauge magnetizations for the 4D SU(2) case\cite{cg} in the strong-coupling region.  For the replica-overlap runs, we  first show the requisite equilibration study.  In this paper the current Higgs field was taken to be the starting point of the replica-Higgs equilibration.  This has the advantage of always being in detailed balance with the gauge configuration. The equilibration is needed only to establish a completely independent Higgs field.  Equilibration studies were done at both $\lambda =0.34$, which is essentially the critical point and also at $\lambda =0.35$, in the magnetized phase.  These gave very similar results.  Also, both $30^3$ and $40^3$ were studied.  These did not have significantly different equilibration times.  Runs at 
$\lambda= 0.34$ are shown in Fig.~9a-c for various replica-Higgs equilibration times.  Single-exponential fits are shown.  These give a time constant of about 10,000 sweeps.  One sees that quantities have more or less settled at 50,000 sweeps.  The remaining systematic error here is slightly above the size of the random errors in our production runs.  Using the observed equilibration time to extrapolate, one can predict that 100,000 sweeps would reduce the systematic error
to less than 10\% of the random error.  This gives a bit of a safety factor in case the situation is worse at other couplings.  Even though no difference was seen between $30^3$ and $40^3$ in equilibration times, for the $50^3$ lattice 120,000 equilibration sweeps were used, for an additional safety factor.  

These simulations consisted of $10^6$ sweeps with replica equilibration and measurements performed every 200 sweeps for all three lattice sizes.
The Binder cumulant for the replica magnetization is shown in Fig.~10a, along with $\xi _{\rm{2nd}}/L$.  These both show a crossing near $\lambda =0.34$.  The separation is present, but not large, for the Binder cumulant on the disordered side ($2.5\sigma$) and very definite on the ordered side ($25\sigma$).  $\xi _{\rm{2nd}}/L$ shows good separation on both sides ($7\sigma$ and $40\sigma$). Really there is never any doubt about the existence of a disordered phase. It is proof of an ordered phase that demonstrates a phase transition to long-range order must exist. The susceptibility and magnetization (Fig.~10b,c) also show behavior consistent with a magnetization transition around $\lambda =0.34$.  The collapse fits (Fig.~11a,b) had 53 degrees of freedom with 
a $\chi ^2/$d.f.=2.2.  The $30^3$ data were excluded from $\xi _{\rm{2nd}}/L$ and magnetization fits. Results were 
$\lambda _c=0.3371(8)$, $\nu = 1.51(7)$, $\gamma /\nu = 1.72(4)$, $\beta ' /\nu =0.634(24)$, and $d_{\rm{eff}}=2.99(6)$.  The hyperscaling relationship predicts a specific heat exponent $\alpha=2-d\nu =-2.53(21)$.  The negative value means that the singularity expected in the specific heat is non-infinite, predicted as $|1/\lambda -1/\lambda _c |^{2.53}$. This means that only the third and higher derivatives of the specific heat (fifth and higher moments of the internal energy) have infinite singularities.  This makes it extremely difficult to detect a signal of non-analyticity in the energy functions from numerical simulations.  However, one can still look for consistent scaling behavior of energy moments, which we show below, along with what appear to be developing infinite singularities in higher derivatives.

\begin{figure}[tb]\centering
                    \includegraphics[width=0.515\textwidth,  clip]{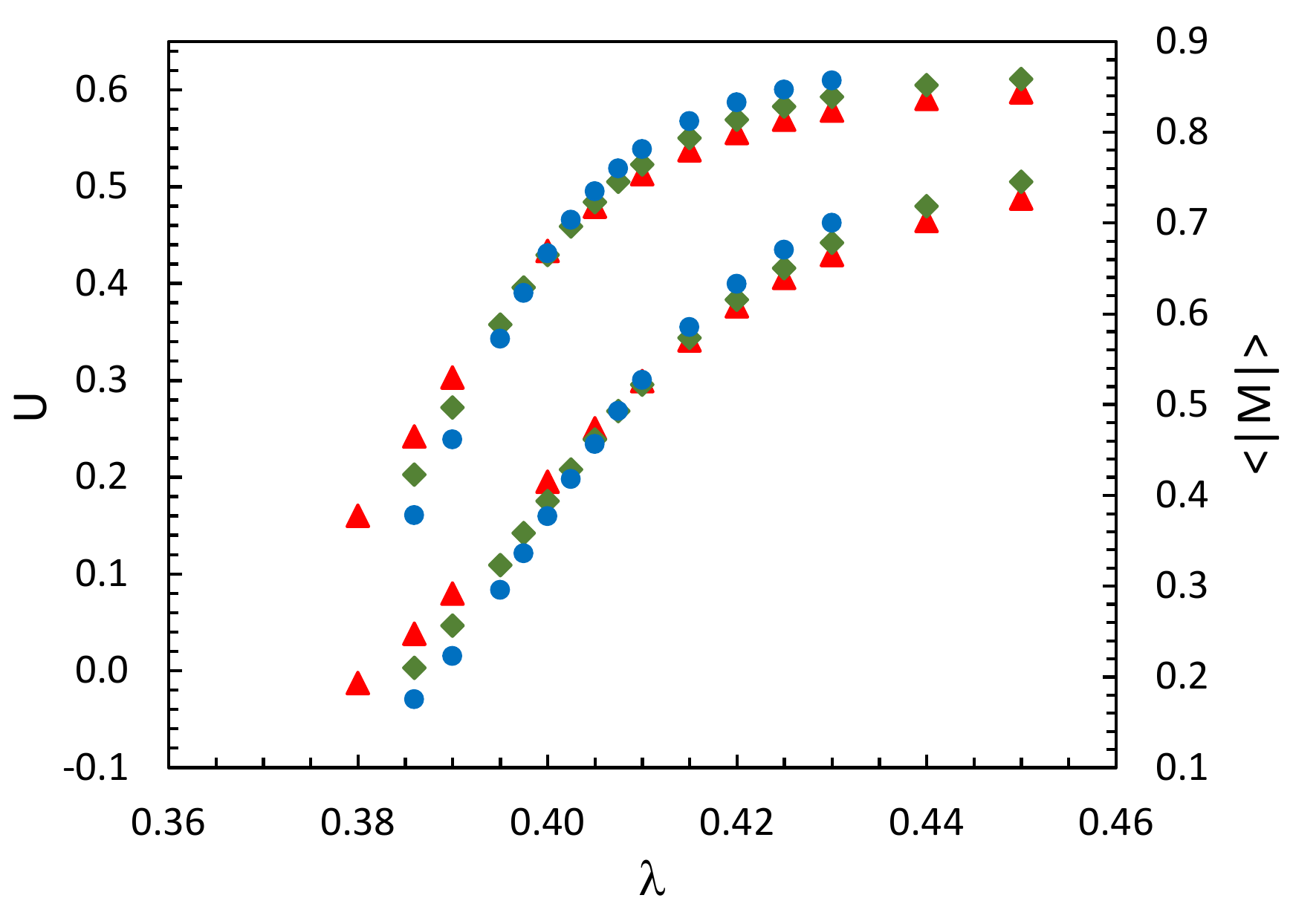}
                         \includegraphics[width=0.475\textwidth,  clip]{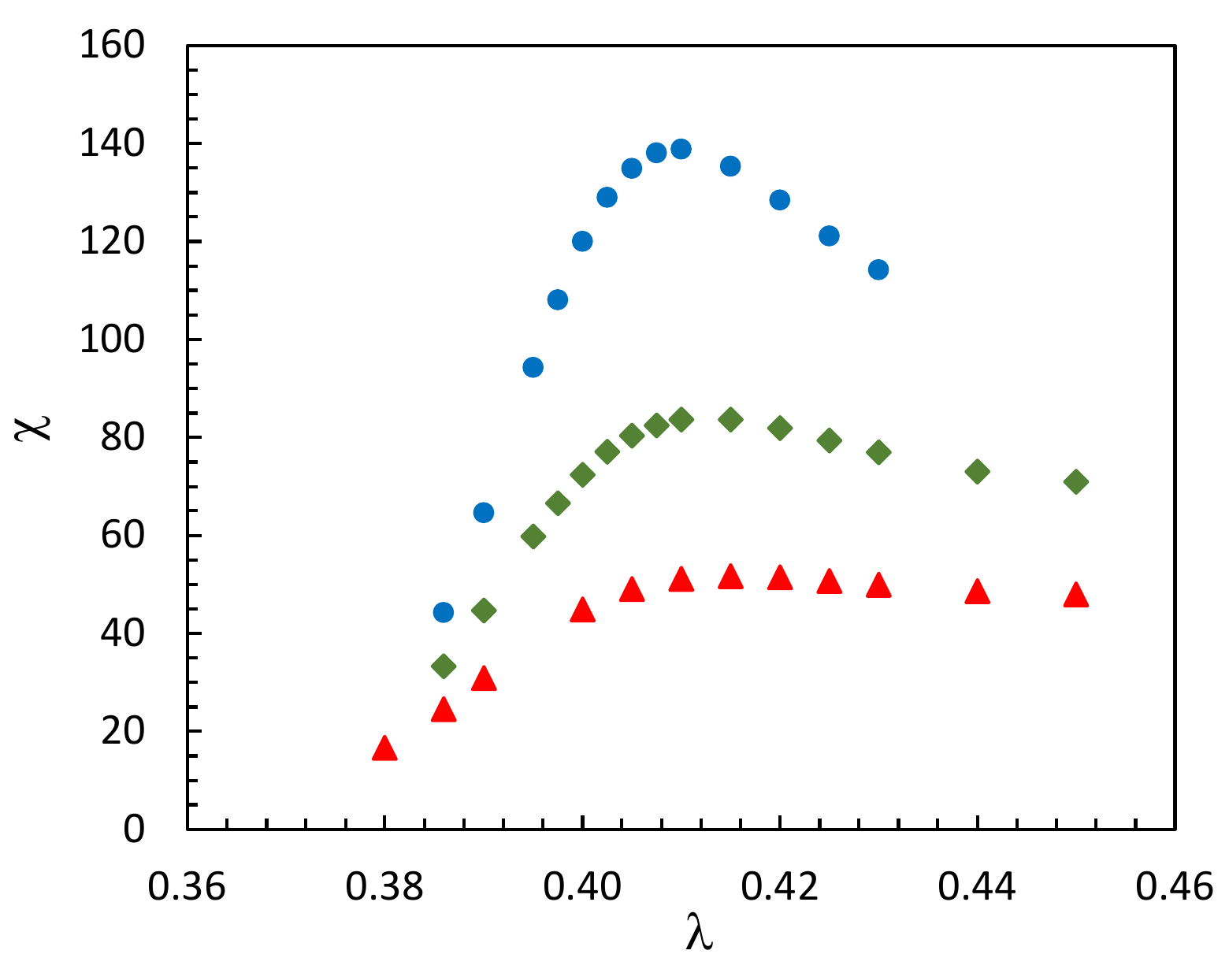}
                                   \caption{Binder cumulant crossing  and Coulomb magnetization (lower graph)  (a),  and susceptibility (b) for $\beta =0.5$. Circles are $64^3$ data. Error ranges in all graphs are of order 50-100 times smaller than plotted points.}
          \label{fig12}
       \end{figure}
We now consider the Coulomb magnetization.  This shows a phase transition above the self-dual line, near $\beta=0.4$. The fast Edmonds algorithm allowed gathering a great deal of data here, $2.5\times 10^6$ sweeps for the $40^3$ lattice, $2 \times 10^6$ for the $50^3$, and $10^6$ for the $64^3$ lattice, each with gauge transformations and measurements every five sweeps.  Each run took about a week on a standard PC.  In some ways, the data were ``too good" to match our fitting procedure based on only leading-order finite-size-scaling theory. This bodes well for a more precise future study involving more lattice sizes that will allow fits to higher-order scaling behavior and more accurate determinations of critical points.  Fig.~12a shows the $U$ crossing near $\lambda = 0.4$.  Errors in U were only of order $10^{-4}$, allowing the crossing to be confirmed by $200\sigma$ below and $400\sigma$ above (even higher confidence if multiple points are combined). The susceptibility shows a growing peak (Fig.~12b) with the $64^3$ data peaking at 
$\lambda= 0.41$.  Fig.~13ab shows the collapse fits.  The very low statistical errors, especially deep into the ordered phase made these difficult.  Only data from the largest two lattices were used and data for scaling variable $>1$ were excluded. Results of the overall fit to the three quantities gave
$\lambda _c = .3993(4)$, $\nu= 1.54(3)$, $\gamma /\nu = 1.674(5)$, $\beta ' / \nu =0.158(8)$, and
$d_{\rm{eff}} = 1.990(16)$.  The $\chi ^2 /$d.f. was a somewhat unsatisfactory $5.1$ despite very good-looking collapses.  This basically shows that these high-quality data contain additional information that could be extracted on higher-order scaling behavior.  A full treatment would require at least three more lattices of a larger size, however, so this was left for a follow-up study.  Note that the 
$\nu$ value is quite similar (equal within errors)  to that found in the lower transition with the replica order parameter.  Energy behavior at $\beta = 0.5$ will be examined later.
\begin{figure}[th!]\centering
                    \includegraphics[width=0.515\textwidth,  clip]{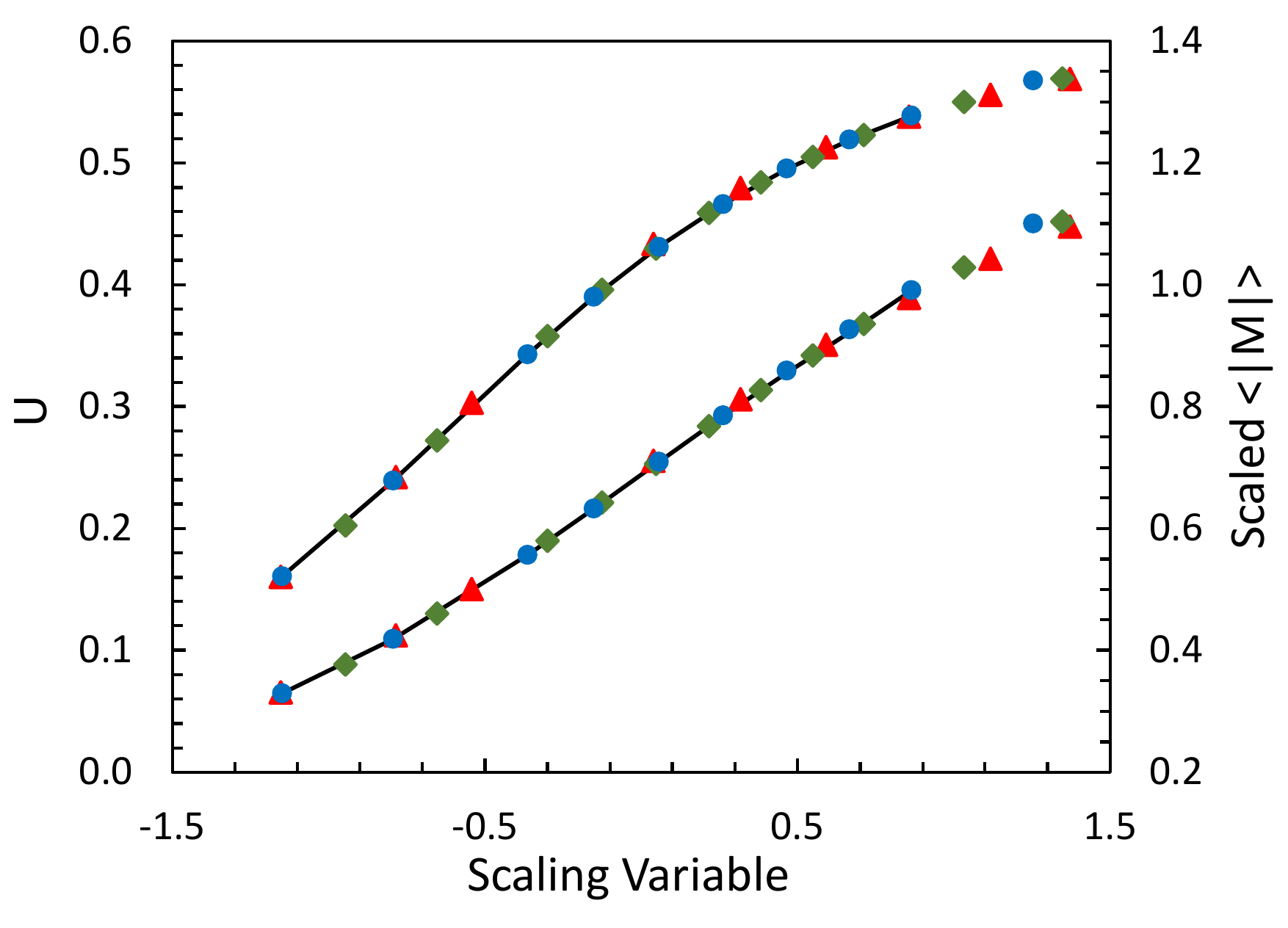}
                         \includegraphics[width=0.475\textwidth,  clip]{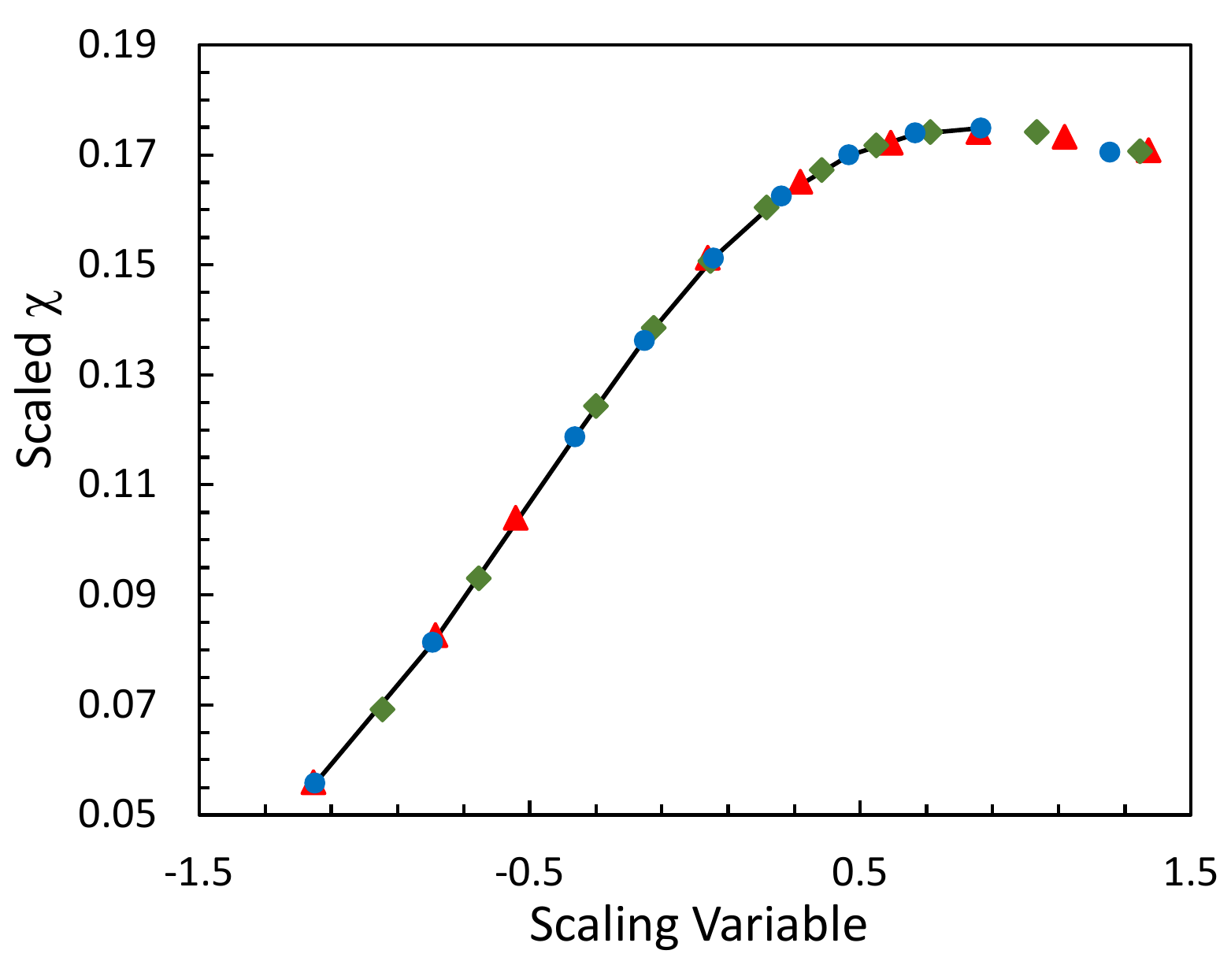}
                                  \caption{Scaling collapse graphs for Binder cumulant (left graph) and scaled Coulomb magnetization (a) and susceptibility (b) at $\beta =0.5$.}
          \label{fig13}
       \end{figure}
\section{Implications of duality}
The self-duality of this system leads to some interesting constraints and features. Basically everything in the upper
part of the phase diagram must be reflected in the lower, with plaquette and Higgs energies switching roles.  A useful check on algorithms is the duality-derived relationship\cite{savit}
\begin{equation}
U_{p}(\beta, \lambda) = (\cosh (2 \beta)-U_{H}(\beta ^* ,\lambda ^* ))/\sinh (2 \beta) \label{duality-relation}
\end{equation}
which agrees with all results in our dataset to which it applies, within statistical errors.  Here $U_p$ is the average plaquette, $U_H$ the average Higgs link energy, $\beta ^* = -0.5\ln (\tanh (\lambda ))$ and $\lambda ^* = -0.5\ln (\tanh (\beta ))$.  
On the other hand, it is not reasonable that the Coulomb-gauge
order parameter itself could be the direct dual of the Landau-gauge one, because of the layered-nature and larger remnant symmetry group of the former.  Indeed, the transformation of the $<\phi (\vec{r_1 }) \phi (\vec{r_2 })>$ correlation function in Landau gauge
to its dual can be performed explicitly \cite{savit}. The result is the average of a string-like operator, 
which involves fields 
lying all along a string connecting the two lattice points, rather than a two-point function involving only fields at the endpoints.  Therefore, there is a dual
correlation function, but not a local dual order parameter per se.  (Apparently all that is needed to produce a critical phase transition is a correlation length that becomes infinite - not necessarily an order parameter that breaks an explicit 
symmetry.) Presumably the same is true for the Coulomb gauge. So logically, there could actually be four phase transition lines: Landau, Coulomb, dual-Landau and dual-Coulomb.  Since the transitions we have found are not close to being equally distant from the self-dual line ($\lambda = 0.38597$ for $\beta = 0.5$) this possibility needs to be taken seriously.   Moving an equal distance up from the self-dual line as the Landau-replica critical point is below it, gives $\lambda =0.435$. Dual points are not vertically related, so the dual transformation line need not be exactly here, but it should be close.  However, the Coulomb transition was found at 0.3993(4),  almost four times closer to the self-dual line (this is shown graphically in Fig.~\ref{fig-final} below).  Another possible explanation of this situation could be the different boundary conditions being used for the Coulomb case. The duality transformation isn't exact for all boundary conditions, so duality predictions only need to hold in the infinite lattice limit.  To test this possibility simulations using Coulomb gauge in the PBC case were performed, using simulated annealing.  A 40-try algorithm with a 20,000 step logarithmic cooling schedule was used on $20^3$ and $30^3$ lattices (after studying convergence for lesser algorithms).  Binder cumulants and second-moment correlation lengths showed this system to definitely lie in the ordered phase at $\lambda = 0.41$, with crossing in the 0.395-0.405 range, agreeing with the OFA boundary condition result.  In addition, an OFA run using the Edmonds algorithm was performed at the exact self-dual point of the Landau-replica transition (0.5, 0.3371) which is (0.56214, 0.38597).  This proved to be well into the ordered phase.  Further simulations showed a transition for this $\beta$ at $\lambda =0.344(1)$, again much closer to the self-dual line which lies at $\lambda = 0.3371$ for this $\beta $.   These additional investigations would seem to indicate that in this region, the replica-Landau transition and Coulomb-magnetization transition are not dual to each other, a surprising additional twist to this assumed simple system which has surprised before.  
The data certainly seem to indicate, at least in this region, there appear to be four transition lines: Coulomb, Landau, dual-Coulomb and dual-Landau.

\section{Transitions at $\beta=0.05$} Finally we pushed deep into the strong-coupling region, at $\beta =0.05$.  If there is to be an analyticity region, certainly it is expected here.   Simulations were more difficult due to the noisier environment at strong coupling.  The Coulomb-gauge transition also required rather extreme $\lambda$ values to tame the wildly fluctuating plaquettes. 
The two transitions were found far apart, after a fair amount of searching.  They are extremely split away from the self-dual line, 
$\lambda= 1.498$, with the Higgs transition occurring around $\lambda = 0.54$ and the Coulomb-gauge magnetization not until $\lambda = 2.5$. At this latter point the plaquette is of order $0.95$ and Higgs energy nearly $0.99$.  However, this is not as extreme as it may look since the transition in the pure Ising-gauge theory itself takes place when the plaquette is around $0.94$.  In this region the Monte Carlo moved very slowly through configuration space and millions of sweeps were needed to obtain sensible results.  
At $\lambda = 0.54$, by contrast, the average plaquette is only $0.116$ and Higgs energy is $0.52$.  Here noise was the main problem, especially in energy measurements.
\begin{figure}[bth!]\centering
                    \includegraphics[width=0.48\textwidth,  clip]{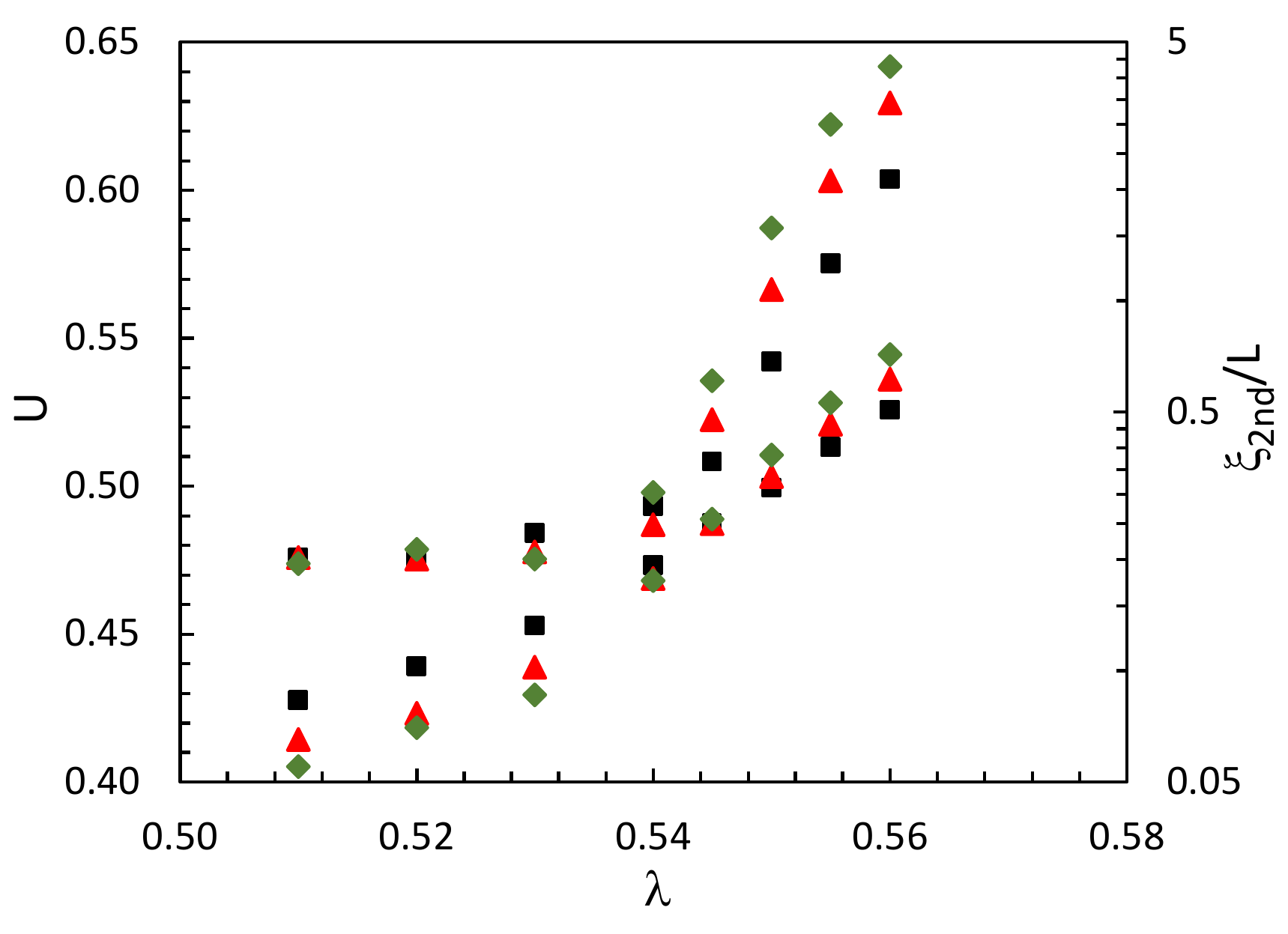}
                         \includegraphics[width=0.48\textwidth,  clip]{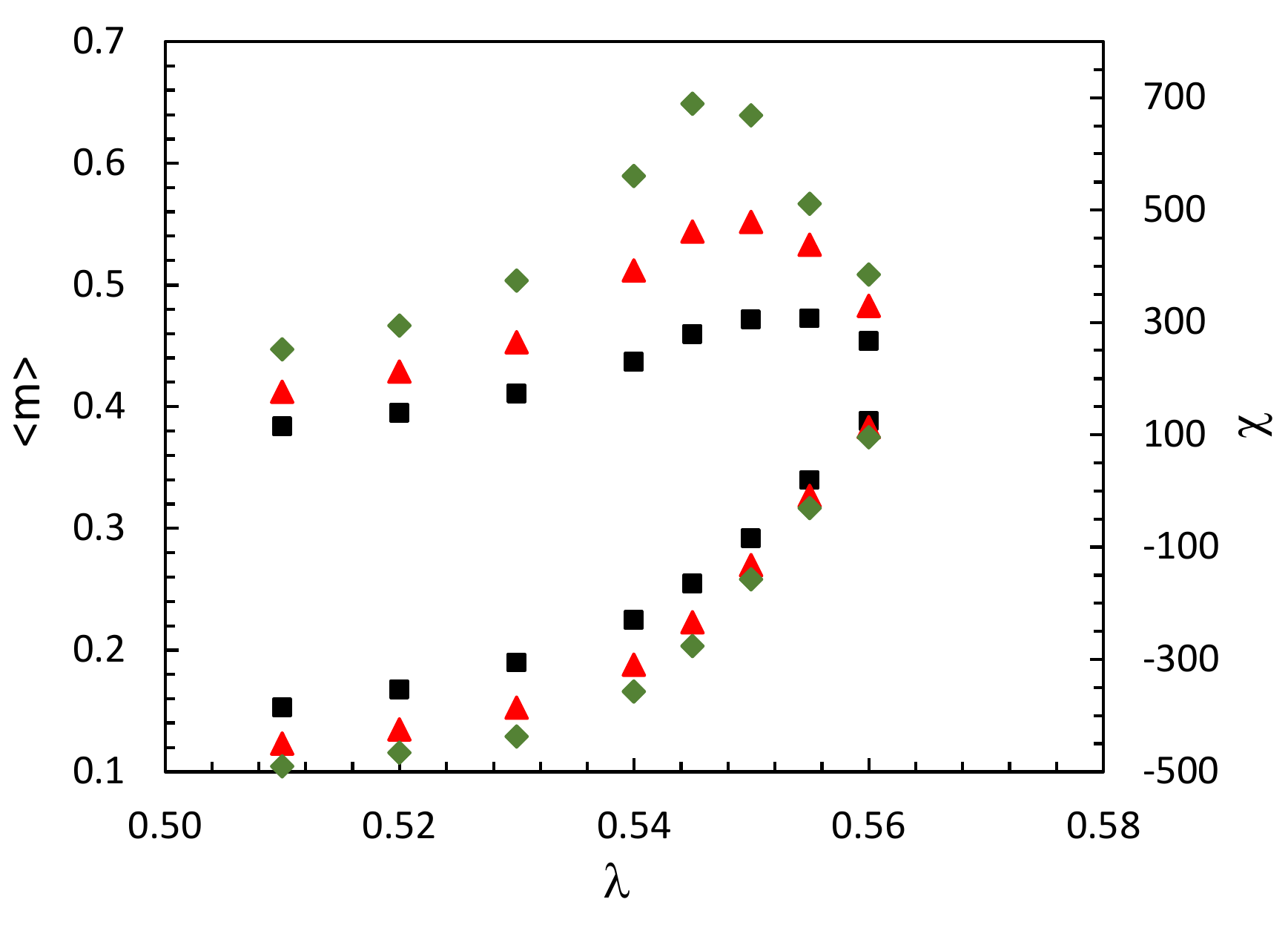}
                                   \caption{Binder cumulant crossing  and $\xi _{\rm{2nd}}/L$ (lower graph - right logarithmic scale)  (a),  and replica magnetization (lower graph) and susceptibility  (b) for $\beta =0.05$. Error ranges are about one-half size of plotted points for susceptibility, and one-tenth for $<m>$. See discussion for others.}
          \label{fig14}
       \end{figure}

Beginning again with the replica-Higgs transition, crossing graphs for both Binder cumulant and $\xi _{\rm{2nd}}/L$ are shown in Fig.~14a.  Each run on $30^3$, $40^3$, and $50^3$ lattices was again $10^ 6$ sweeps measured once every 200 sweeps. A crossing around 0.54 is seen in both quantities.  Above the transition, crossing is verified by $33\sigma$ for $\xi _{\rm{2nd}}/L$ and $27 \sigma$ for $U$. Below the crossing point, separation is up to $7\sigma$ for $\xi _{\rm{2nd}}/L$ but only about $1.5\sigma$ for $U$.  As mentioned before, it is really the upper part of the crossing confirming long-range order which is crucial in determining the presence of a phase transition, since there is never any doubt of a disordered phase. Fig.~14b shows the susceptibility which peaks at 
$\lambda=0.545$ on the $64^3$ lattice, and higher than this on the smaller lattices, implying that the infinite-lattice critical point lies below this value. Scaling collapse plots for all four quantities are shown in Fig.~15.  Here $\xi _{\rm{2nd}}/L$ give a bit of trouble, as it preferred a somewhat different $\lambda _c$, and also could not fit the $30^3$ data below the transition. These were omitted and $\xi _{\rm{2nd}}/L$ was allowed to choose a different $\lambda _c$.  The overall fit had 42 degrees of freedom with $\chi ^2 /$d.f.=1.1. Results are as follows: $\lambda _c = 0.5358(9)$ (0.5413(10) for $\xi _{\rm{2nd}}/L$), $\nu = 1.39(7)$, $\gamma /\nu = 1.632(26)$, $\beta '/\nu = 0.649(13)$, and $d_{\rm{eff}}=2.93(4)$.
\begin{figure}[bt!]\centering
                    \includegraphics[width=0.48\textwidth,  clip]{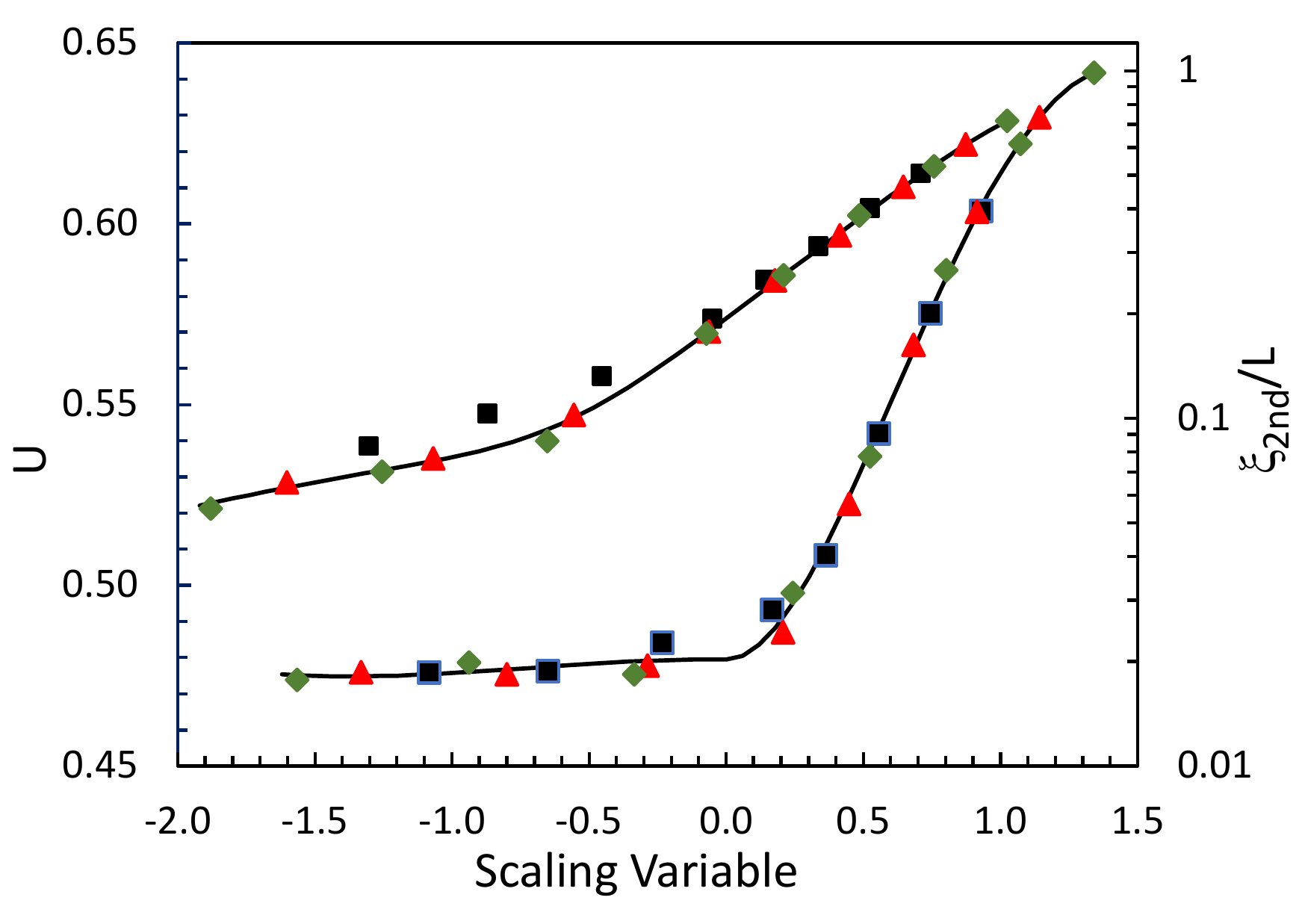}
                         \includegraphics[width=0.48\textwidth,  clip]{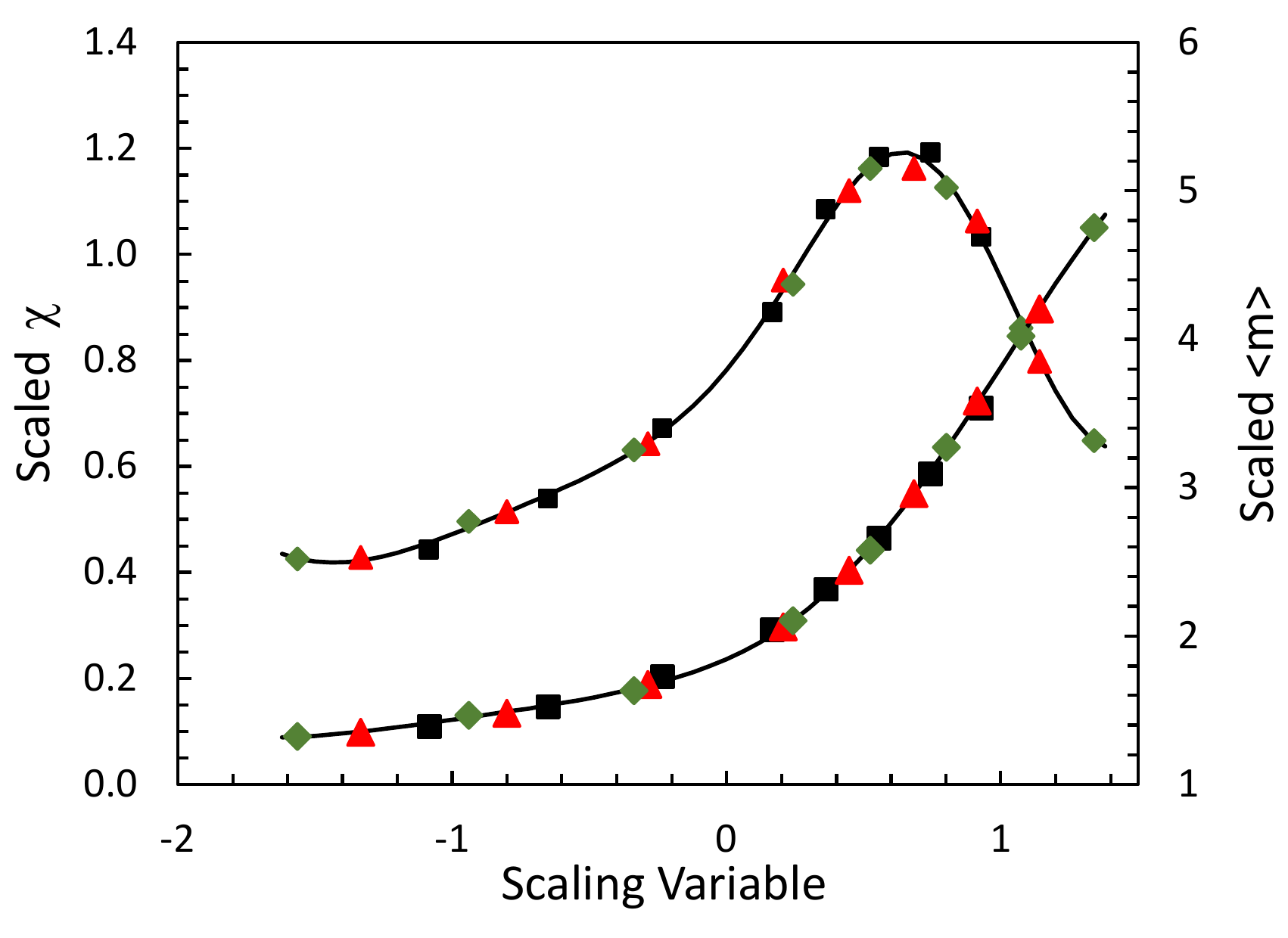}
                                  \caption{Scaling collapse graphs for Binder cumulant (lower graph) and $\xi _{\rm{2nd}}/L$ (a),  susceptibility and replica magnetization (large symbols, right scale) (b) at $\beta =0.05$.}
          \label{fig15}
       \end{figure}

The Coulomb-magnetization transition for $\beta = 0.05$ utilized, for each coupling,  25 million sweeps on the $40^3$ lattice, 20 million on the $50^3$, and 10 million on the $64^3$, each measured once every five sweeps.  Despite higher statistics, the error bars are much larger than at $\beta = 0.5$ because the simulation moved much more slowly through configuration space. This can be seen from the binning of errors, which finally stabilized at around 300,000 measurements per bin.  In all of our other Coulomb studies, 10,000 was more than adequate. In retrospect it was not really sensible here to measure every 5 sweeps, but the Edmonds algorithm is so fast that decreasing the frequency of measurements would not actually save much time.  All of these simulations were also prefaced with $2\times 10^6$ equilibration sweeps to account for the slowness of the Monte Carlo algorithm.

The $U$ crossing graph is shown in Fig.~16a, along with the magnetization which also crosses. At $\lambda = 2.6$ 
 crossings are verified to $4\sigma$ for both $U$ and $<|M|>$ and $3\sigma$ for both at $\lambda = 2.55$.  As mentioned before, the magnetization is not required to cross, but when it does provides strong evidence for long-range order. The susceptibility graph, Fig.~16b,  is typical, showing a broad peak growing with lattice size.   Scaling collapse plots are shown in Fig.~17a,b.  Only the two largest lattices were used and the messy area of the susceptibility at large $\lambda$ was excluded.  The fit had 33 degrees of freedom with
 $\chi^2 /$d.f.=1.25.  The following were determined: $\lambda _c=2.52(3)$, $\nu = 1.47(13)$, $\gamma /\nu = 1.96(4)$, $\beta '/\nu =0.0005(200)$, $d_{\rm{eff}}=1.96(6)$.   So once again an $\alpha$ exponent between $-2$ and $-3$ is predicted. An oddity of this transition is the very low value of the magnetization exponent, $\beta '$.  The rather low precision here is in marked contrast to the extreme precision seen in the $\beta=0.5$ Coulomb transition.  This suggests a study at an intermediate value such as $\beta=0.2$ might be interesting.  One could hope to retain a higher precision with better behavior from the Monte Carlo algorithm, in a place where the transitions are still expected to be widely spread.
\begin{figure}[bth!]\centering
                    \includegraphics[width=0.52\textwidth,  clip]{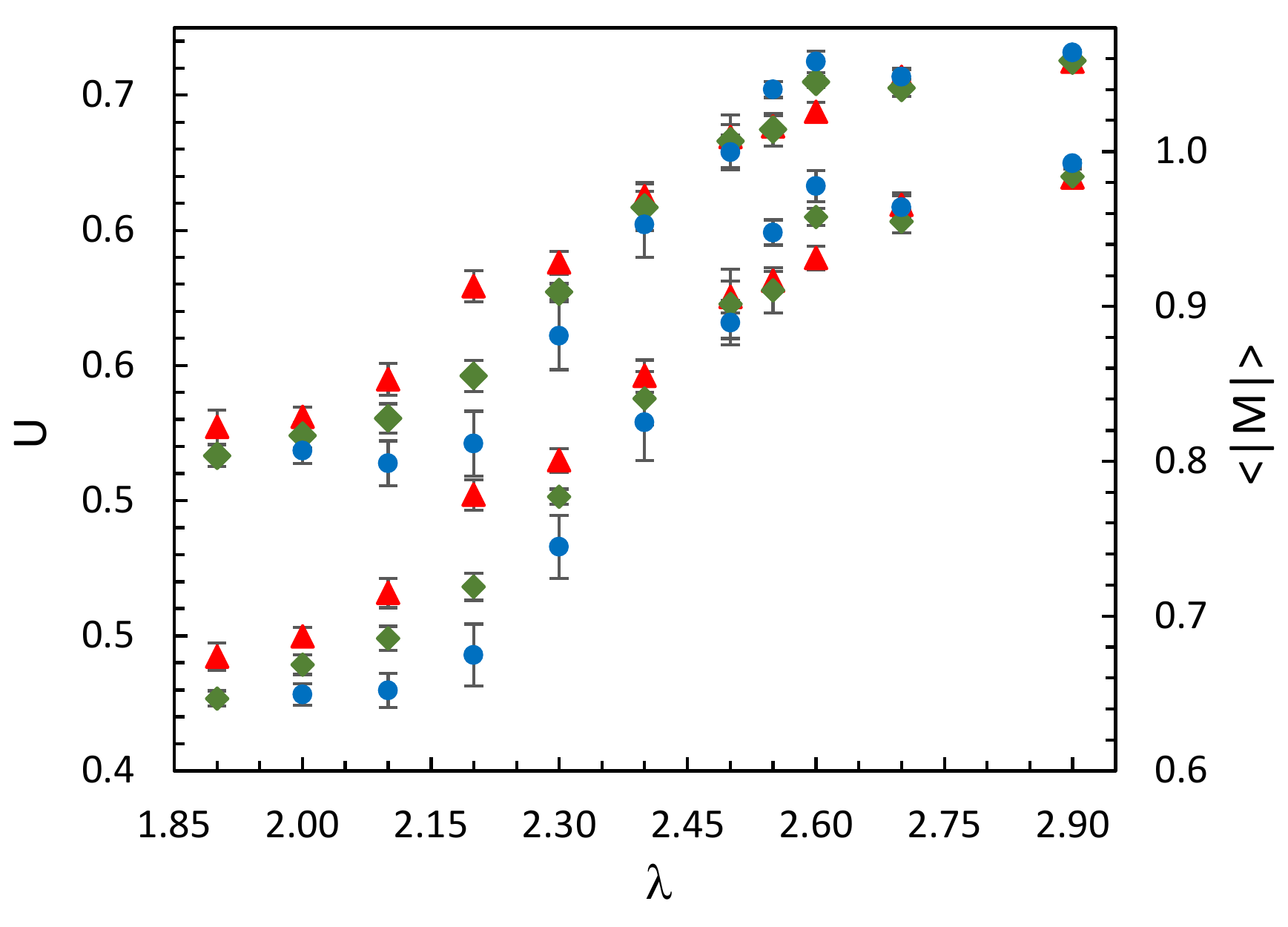} 
 \includegraphics[width=0.47\textwidth,  clip]{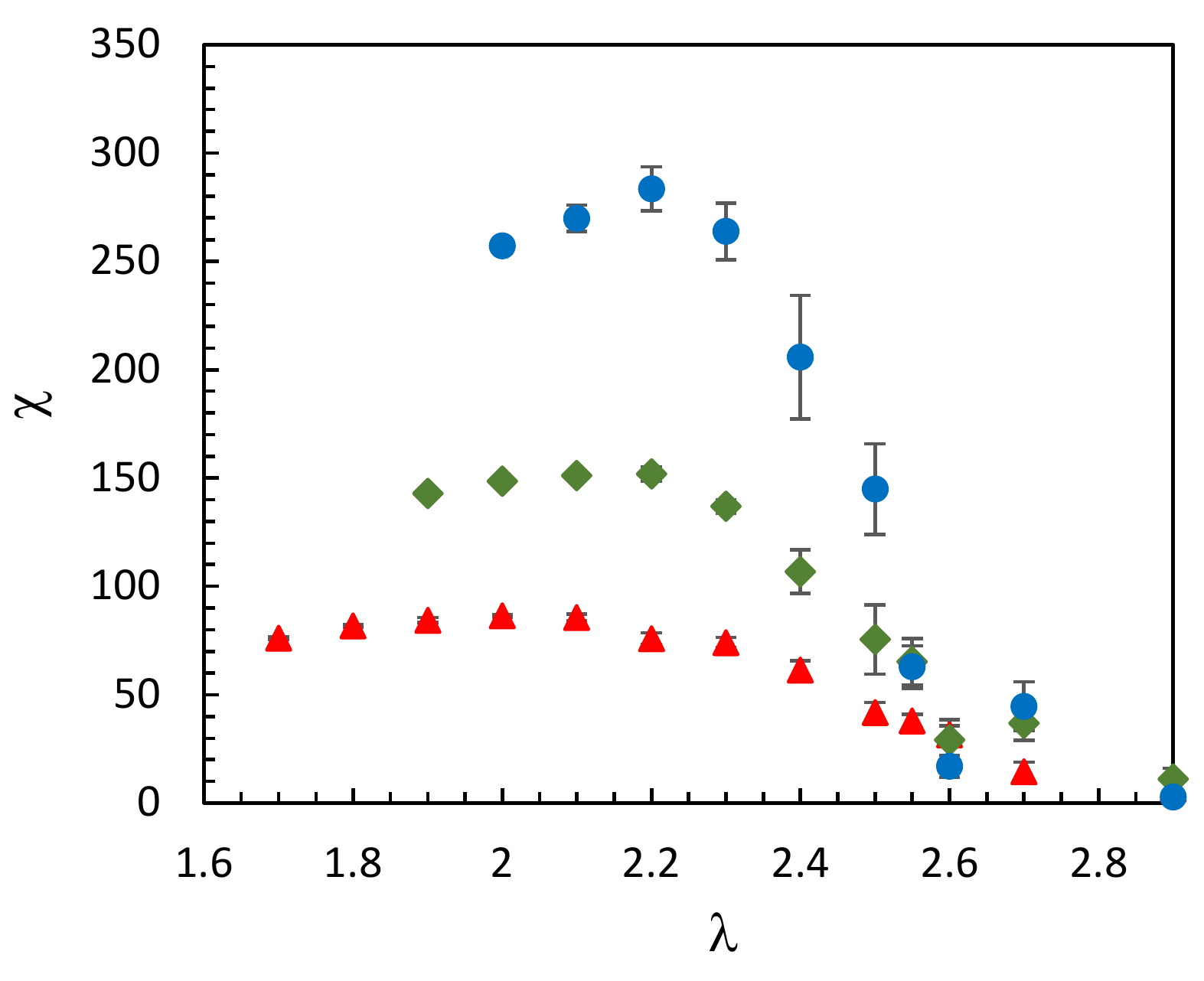}
                     \caption{Binder cumulant(upper plot) and Coulomb magnetization at $\beta =0.05$ (a) and susceptibility (b).}
          \label{fig16}
       \end{figure}
\begin{figure}[bth!]\centering
                    \includegraphics[width=0.46\textwidth,  clip]{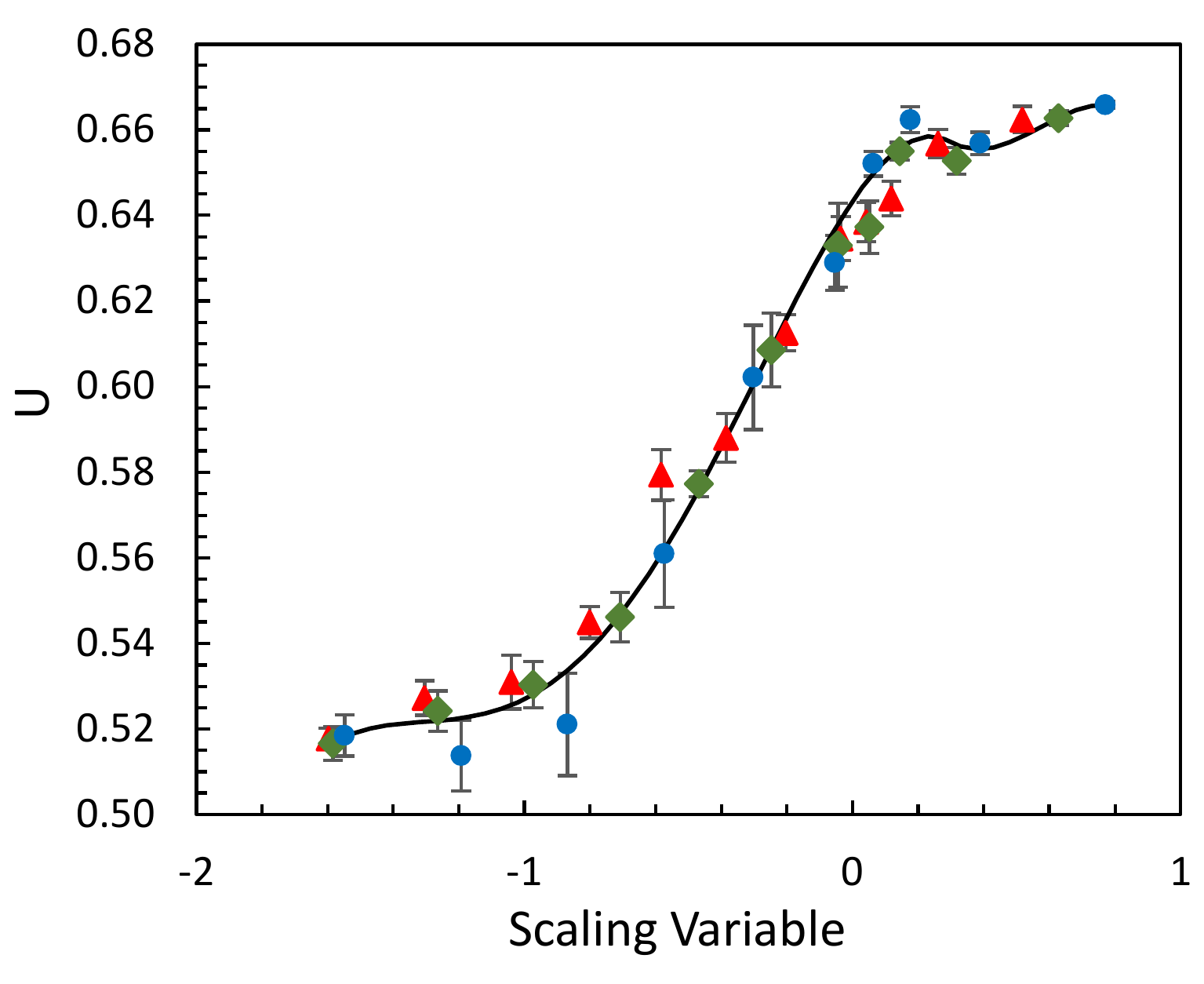}
                         \includegraphics[width=0.53\textwidth,  clip]{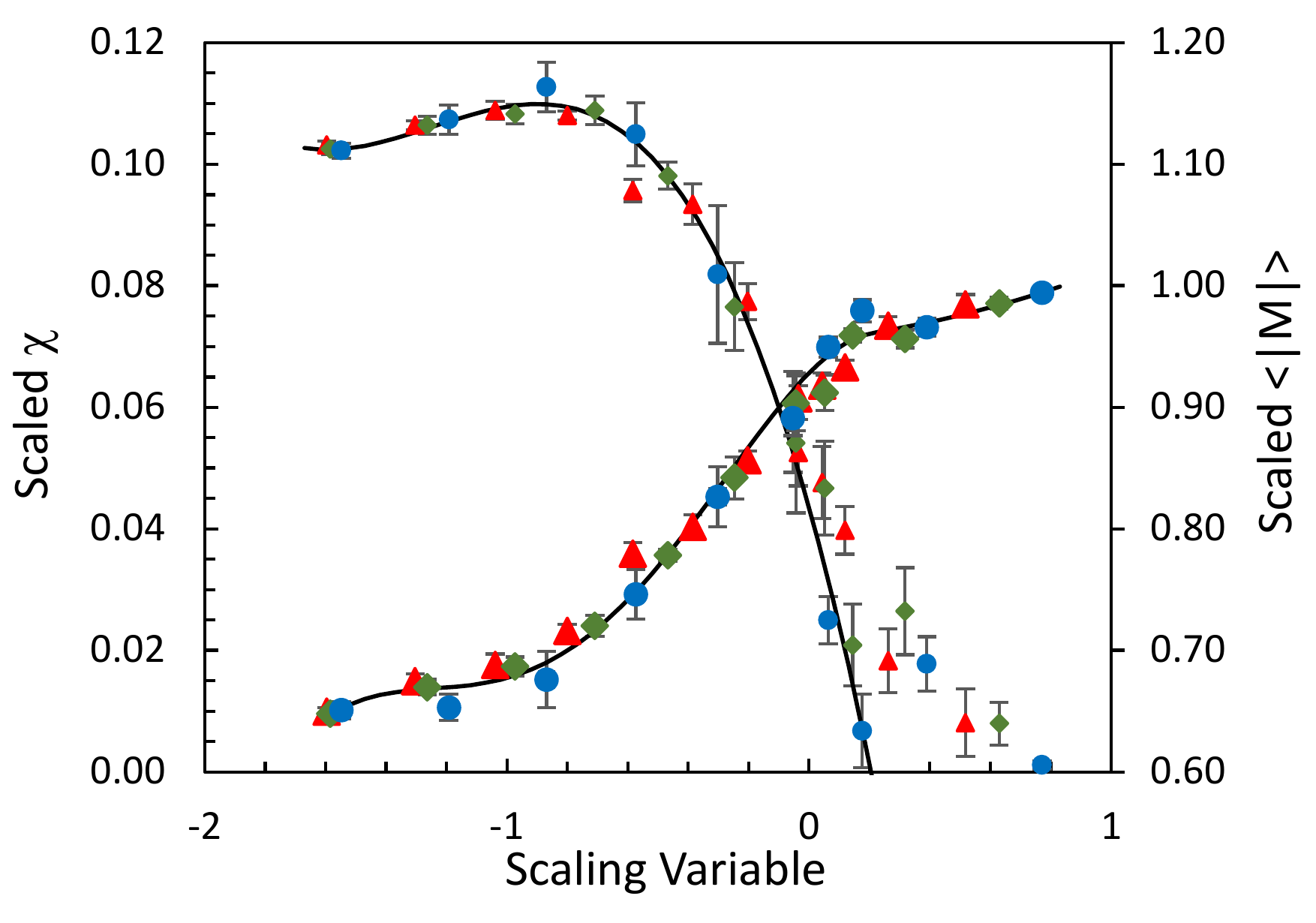}
                                  \caption{Scaling collapse graphs for Coulomb Binder cumulant  (a),  scaled susceptibility and scaled magnetization (large symbols, right scale) (b) at $\beta =0.05$.}
          \label{fig17}
       \end{figure}

\section{Discussion and energy analysis}

To recap, both the Coulomb-gauge magnetization which follows the gauge-deconfinement transition and the replica magnetization which follows the Landau-gauge Higgs transition showed phase transitions to an ordered phase at the three 
$\beta$'s studied: 0.708, 0.5 and 0.05.  The first of these had a clear transition in both order parameter and energy, first-order with periodic boundary conditions but second-order with the OFA boundary conditions used for the Edmonds algorithm. The high-order transitions seen at the latter two gauge couplings had $\nu $ values ranging from 1.39 to 1.54.  A $\nu $ of 1.5  predicts a specific heat exponent $\alpha =-2.5$,
meaning that the first divergent moment of internal energy is the fifth, which is 5th order in the Ehrenfest classification. This certainly explains why no transition has been seen here in energy quantities using numerical simulations.  What is usually done is to look for a specific heat peak growing with lattice size, but that only happens if $\alpha >0$ and here we have a highly negative $\alpha$. We actually did try to look for signs of a divergence in the 5th moment of the plaquette and Higgs energy, but fluctuations were too large to see much.  One difficulty with such an approach is that once you have a divergent moment, the errors in that quantity, which are determined from even higher moments, blow up, because these higher moments have even stronger divergences!  This makes errors grow rapidly with lattice size.  An additional problem with fitting energy quantities to critical behavior that is not present in the order parameter is that the internal energy also has a non-critical part added to it.  This often requires additional terms in fits, which can muddy the extraction of critical behavior.  Higher moments of the energies generally have a smaller non-critical part, since that varies slowly with coupling.  Due to this, we settled on the third moment of both the average plaquette and Higgs energy for careful study. This quantity still has quite low statistical errors in our studies which involved hundreds of millions of sweeps at each datapoint and can, in most cases, be fit without extra terms to handle the non-singular part other than a constant.  Errors were generally higher on the Higgs energy quantities, so only average plaquette results are presented here.  One approach is to test whether the third moment can be fit to the non-infinite singular behavior predicted from the order parameter analysis, i.e. $\propto|\lambda -\lambda _c|^{-\alpha-1}$ where the exponent is of order 1.5.  The singularity here is not only from the fractional power itself, but also the possibility of different coefficients above and below the critical point, which scaling behavior allows.  These fits will be shown to meet expectations in all cases, showing energy behavior in line with critical scaling predictions from the order parameters. In some cases large jumps in the coefficients are seen.   Another approach is looking for cusp-like behavior in the 4th moments which scale with a still positive power, but less than unity.  This is noticeable at the 
$\beta = 0.5$ Landau transition and even more suggestive at the $\beta=0.05$ Coulomb transition, but errors are still fairly large. Finally, one can look at high derivatives of third moments, second moment or even the plaquette or Higgs energy themselves using numerical derivatives, in particular high enough derivatives that infinite singularities are predicted for the infinite lattice.  In one case this produces an interesting result, but mostly our statistics are still not sufficient for this approach.  Numerical derivatives require points spaced closely enough to see the relatively narrow features expected.  Of course, each time the point spacing is halved, the error in the second derivative quadruples, the third by a factor of eight etc. This limits the usefulness of high derivatives.  Also it should be noted that error estimates for second and higher derivatives must be computed directly from the uncorrelated uncertainties of the original measured quantities, rather than chained form the errors of previous derivatives, which are correlated.

All of the energy data in this section are from PBC lattices.  Since the transitions here are clearly higher-order in both order parameter and energy, any differences in boundary conditions on the energy quantities is likely quantitative, rather than qualitative here.  However, for the phase transitions studied with the Coulomb gauge, boundary condition differences could result in additional finite-size differences between energy and order parameter, such as a shift in the finite-lattice $\lambda _c$.

At $\beta=0.5$ the second moments of the average plaquette (i.e. specific heat) and Higgs energy show non-scaling peaks at $\lambda = 0.3917(2)$ and $0.3753(2)$ respectively.  These values were computed from the interpolated zero crossings of third cumulants which is more precise than directly locating the peaks. These values straddle the self-dual point which lies at $0.38597$. The curves are roughly mirror images of each other due to duality, but the relationship is somewhat complicated by the extra term and factor in the duality relation (\ref{duality-relation}).
As expected for a non-divergent peak there is no visible effect on the peak from changing lattice size. The third moment of the plaquette is shown in Fig.~18a.  along with fits to the expected critical behavior.  This is the third central moment (which is the same as the third cumulant).  The quantity plotted has a factor of the square of the lattice volume in order to make it correspond the third derivative of the plaquette with respect to $\beta$.  For higher moments, it is really the moment combinations corresponding to the cumulants which are interesting, since they correspond to derivatives of the energy quantity (i.e. plaquette or Higgs energy).  Although the infinite lattice critical points were determined to be $\lambda _c =0.3371$ (Landau) and $0.3993$ (Coulomb), the susceptibility peaks, which are more relevant to the finite lattices, were at  0.342 for the at the lower critical point and 0.41 at the upper on the largest lattices studied.  Because ``finite-lattice critical points"  can differ somewhat between quantities there is some uncertainty of the critical point to use, so we let the fits themselves choose.  This actually gives a more robust test of the hypothesis that the scaling laws are compatible.  There was not enough data below the lower critical point to attempt a fit there. Above the critical 
point a fit to 
\begin{equation}
c_1+c_2 |\lambda - \lambda _c|^{-\alpha -1}   \label{fittingfunction}
\end{equation}
gives $-\alpha-1$=1.64(6) with $\lambda _c = 0.3326(8)$. This gives $\nu =(-\alpha+2)/3 = 1.55(2)$.  This compares well with the value obtained earlier from the replica overlap order parameter, $1.51(7)$.  The fit had a $\chi ^2$/d.f.$=2.2$. A slight but significant difference between $30^3$ and $40^3$ led us to fit only the latter here.
We also tried fits with the more customary inverse coupling but those had somewhat worse $\chi^2$. These differ only in higher order so the better fit model above was selected.  The $\lambda _c$ value is close to what was found with the order parameter analysis, 0.3371(8), but does differ by 4$\sigma$.  It should be remembered, however, that the scaling fits for the order parameter capture the infinite-lattice critical point, whereas the energy-fit for a single lattice would give a finite-lattice pseudo-critical point.  These quantities could easily exhibit a systematic shift of this size as seen by the susceptibility peaks.

\begin{figure}[t!]\centering
                    \includegraphics[width=0.47\textwidth,  clip]{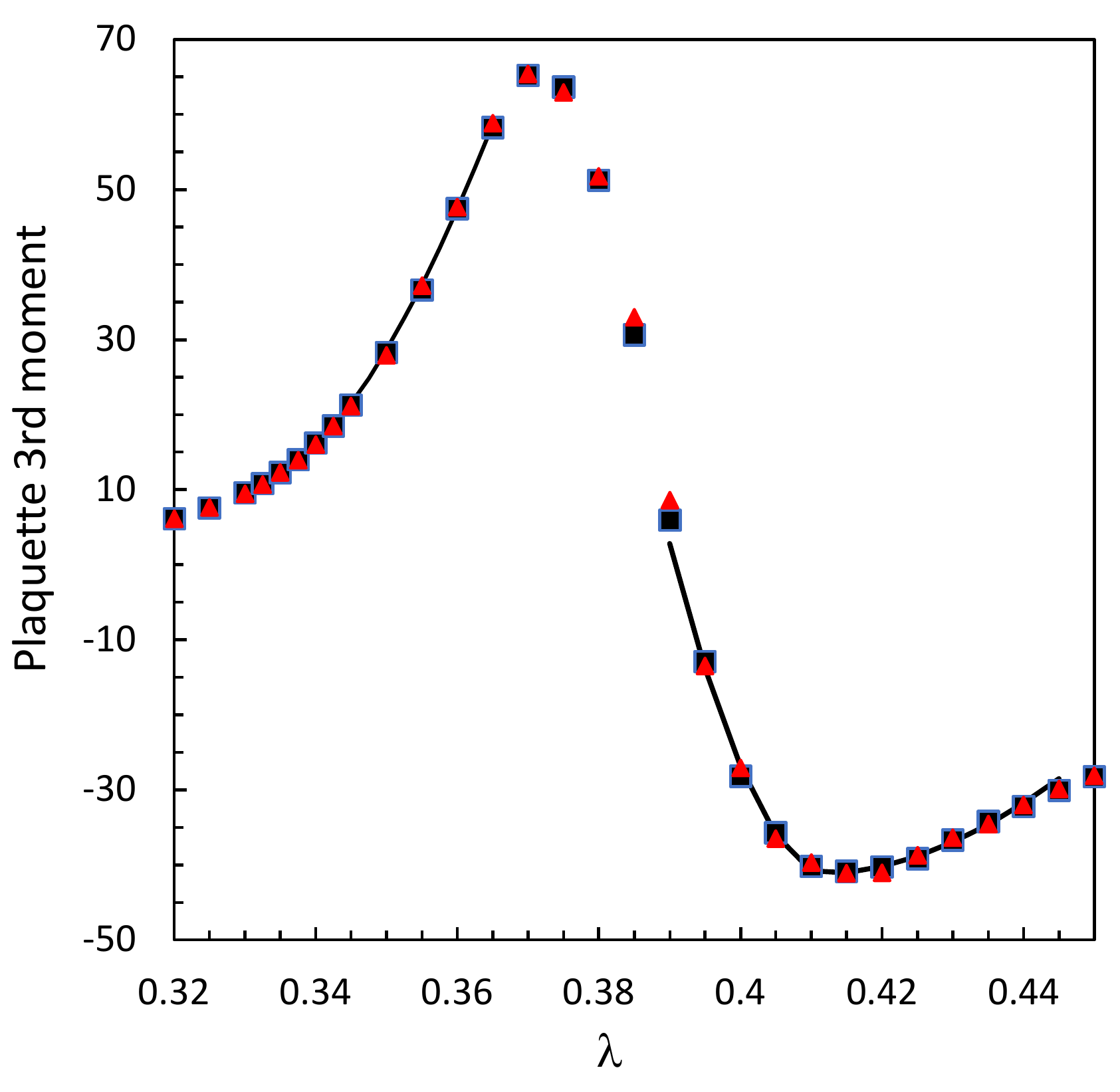}
                      \includegraphics[width=0.51\textwidth,  clip]{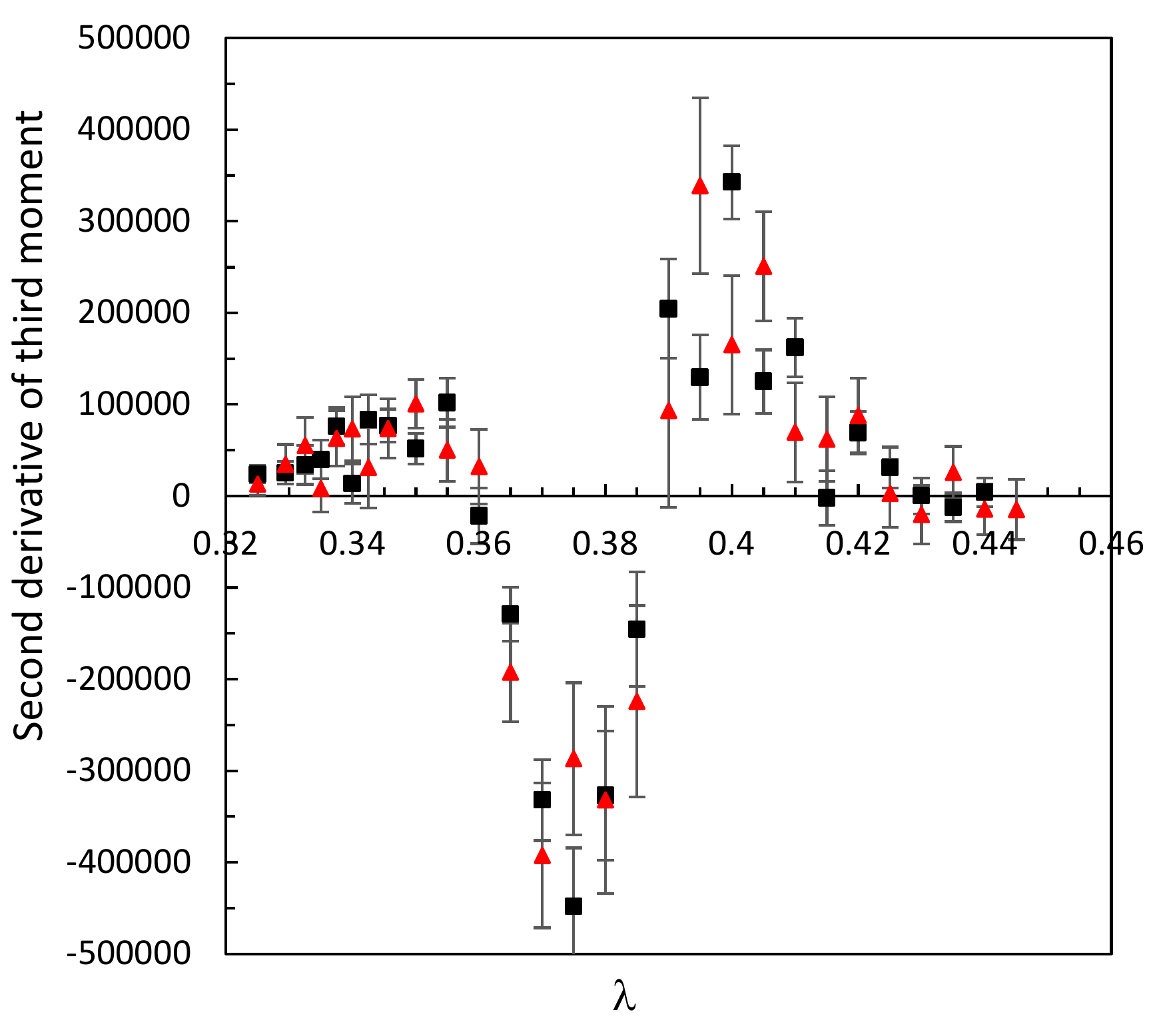}
                     \includegraphics[width=0.52\textwidth,  clip]{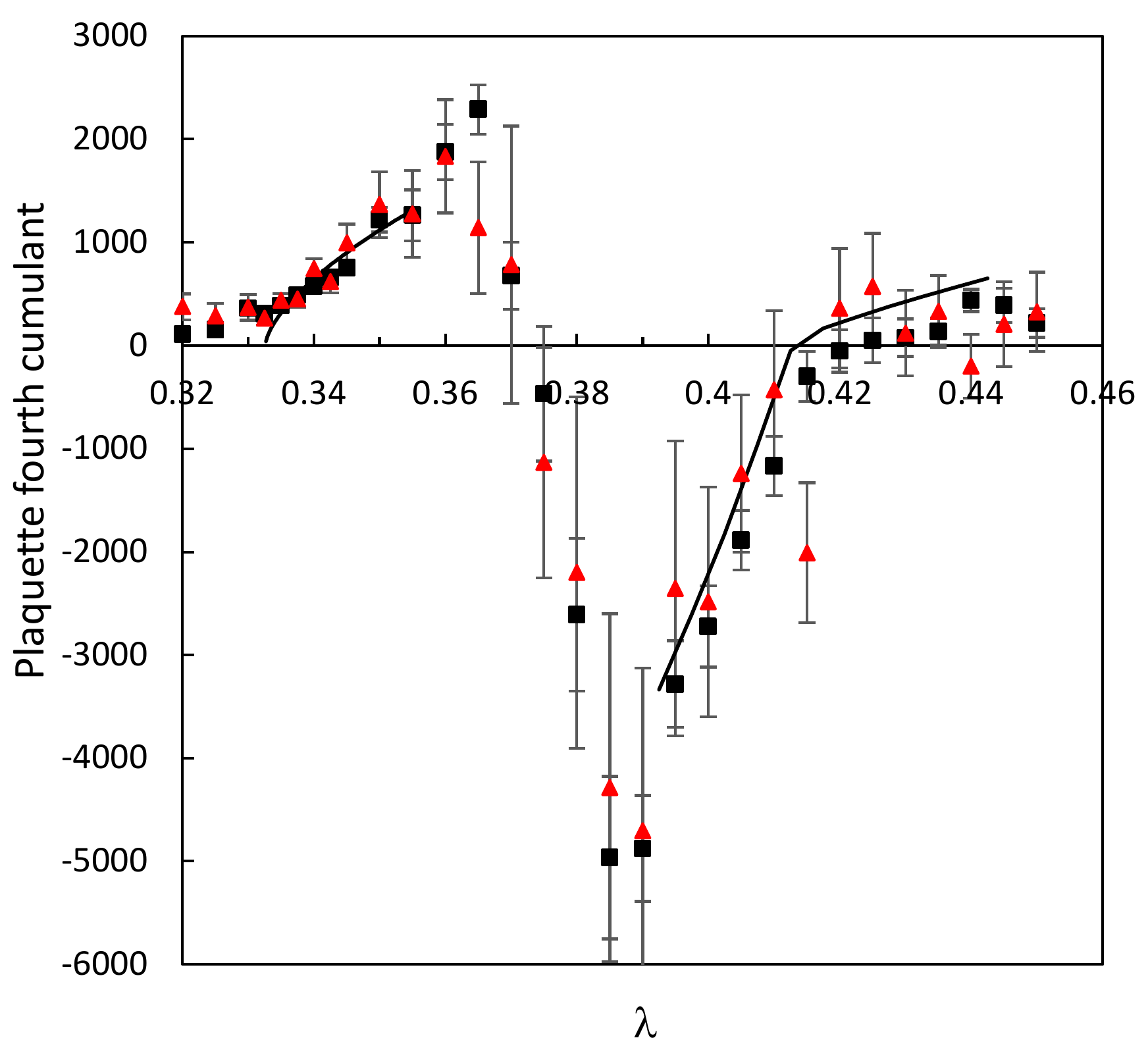}
                                    \caption{Third moment of plaquette (a), and second numerical derivative of same (b) around $\beta = 0.5$ Landau transition.  Error ranges are 5-15 times smaller than datapoint size in former. Fourth cumulant (c) is also shown.  Lines in (c) are scaled derivatives of third moment fits shown in (a).}
          \label{fig18}
       \end{figure}
A fit to both sides of the Coulomb transition is also shown.  The two sides use the same fitting function except for having different coefficients $c_2$, which differed greatly, by  a factor of $7.5(1.2)$.  The fit gave $\lambda _c = 0.412(2)$ and 
$-\alpha -1= 1.82(14)$ with a $\chi^2$/d.f.=1.9.  This value for $\alpha$ 
gives $\nu = 1.61(5)$ which is in reasonable agreement with the value from the Coulomb-gauge order parameter of $1.54(3)$, as is $\lambda _c$, considering the expected finite-lattice shift.  

So for both critical points, behavior of this energy-moment agrees with the critical behavior predicted from the order parameter analysis.  These functions look smooth because it is nearly impossible to see the rapid flattening of curvature singularities in plotted curves, even with excellent data.  Because the error bars are smaller than the plotted points, additional features may emerge from taking numerical derivatives.  One could even hope to see developing divergences from finite-size scaling.  For any derivative that has an infinite singularity on the infinite lattice, one should see differences between results from different lattice sizes.  For instance, for the $q$th derivative of the third moment, the peak height ratio for $40^3$ vs. $30^3$ is expected 
to be $(4/3)^{(\alpha +1+q)/\nu }$, given that $q$ is large enough for the exponent to be positive\cite{fss}. In our case a $q$ of two or larger is needed to have a divergent quantity.  The effect for $q=2$ is small - the $40^3$ peak is predicted to be only 10\% larger than $30^3$. For $q=3$, the peak ratio should be 1.33.  The second derivative of the third moment is shown in Fig.~18b. Note that between $\lambda= 0.33$ and $0.345$, the spacing of points is $0.0025$. This allows one to see peaks much more clearly, but even here one is not sure of hitting them exactly.  Elsewhere in the diagram the spacing is $0.005$ which is less adequate at finding peaks.  Halving the spacing multiplies the error of the second derivative by a factor of four.  We have partially compensated for this with about three times higher statistics in the narrowly-spaced region ($10^9$ sweeps here vs. $3\times 10^8$ elsewhere).  An expected cusp in the first derivative, leads to a down and an up peak in the second, two downs and an up in the third etc., where the singularity is slightly spread on a finite lattice.  For both  $30^3$ and $40^3$ a dip is seen, $2.4\sigma $ at $\lambda = 0.34$ for $30^3$, and $1.5 \sigma$ at $\lambda =0.335$ for $40^3$.  Unfortunately much better statistics would be needed to verify the small finite-size increase in the peak expected.  The $\lambda$ - shift may also be evidence for a finite-size effect. This would require additional lattice sizes to confirm.  Higher resolution is also needed to really confirm peaks.  Ferrenberg-Swendsen reweighting was used as a guide, but its usefulness is very limited because the errors blow up so rapidly away from the base point.  Points are spread too far to use multi-histogram techniques.  

For the Coulomb transition around $\lambda = 0.4$, a simple upward peak is expected, since the derivative of the third moment changes sign here.   An $8.5\sigma$ peak is indeed present for $30^3$ with consistent data from $40^3$.  Note the nearly mirror-reflection peak at $\lambda = 0.37$.  This could be the dual-reflection of the Coulomb transition (the self-dual line is at $\lambda=0.38597$).
 We show this figure mostly to illustrate that it is conceivable that evidence could be found for infinite singularities in high derivatives or high moments, even though demonstration of finite-size scaling has not been achieved yet.  Finally we note that this quantity is proportional to the fourth derivative of the plaquette.  If the plaquette were varying smoothly it is a bit hard to imagine the fourth derivative being so large and varying so rapidly.  To investigate further,
we tried fitting various 12-point segments of the average plaquette itself to an eighth degree polynomial.  Despite using so many parameters, these fits were very far from successful, with $\chi^2/$d.f. in the hundreds or worse.  This could be seen as an additional indirect indication of non-smooth behavior here.

In Fig.~18c, we look at the plaquette fourth cumulant, $(<p^{4}>-3<p^{2}>^{2})V^3 $ where $p$ is the average plaquette and $V$ is the lattice volume. It is this quantity which should equal the third derivative of the plaquette with respect to $\beta$. From the third-cumulant fits, this is expected to have a downward cusp at the Landau critical point and an infinite slope at the Coulomb.  Also shown as a guide are derivatives of the fits to the third moment.  The latter are scaled by an arbitrary factor to compensate for the fact that the fourth cumulant should match the derivative of the third with respect to $\beta$ but we are varying $\lambda$ instead.  Both lattices show  a $1.4\sigma$ dip at $\lambda = 0.3325$, near where a cusp is expected.  That the dip is seen in both lattices near the expected location makes it more significant.  Observing it to go all the way to zero may be unrealistic given how narrow the feature is expected to be. It is subject to finite-size rounding, and there is also always the possibility of a small non-singular part.  The expected rapid change in slope around 0.41 is also borne out in this data.

\begin{figure}[bth!]\centering
                    \includegraphics[width=0.495\textwidth,  clip]{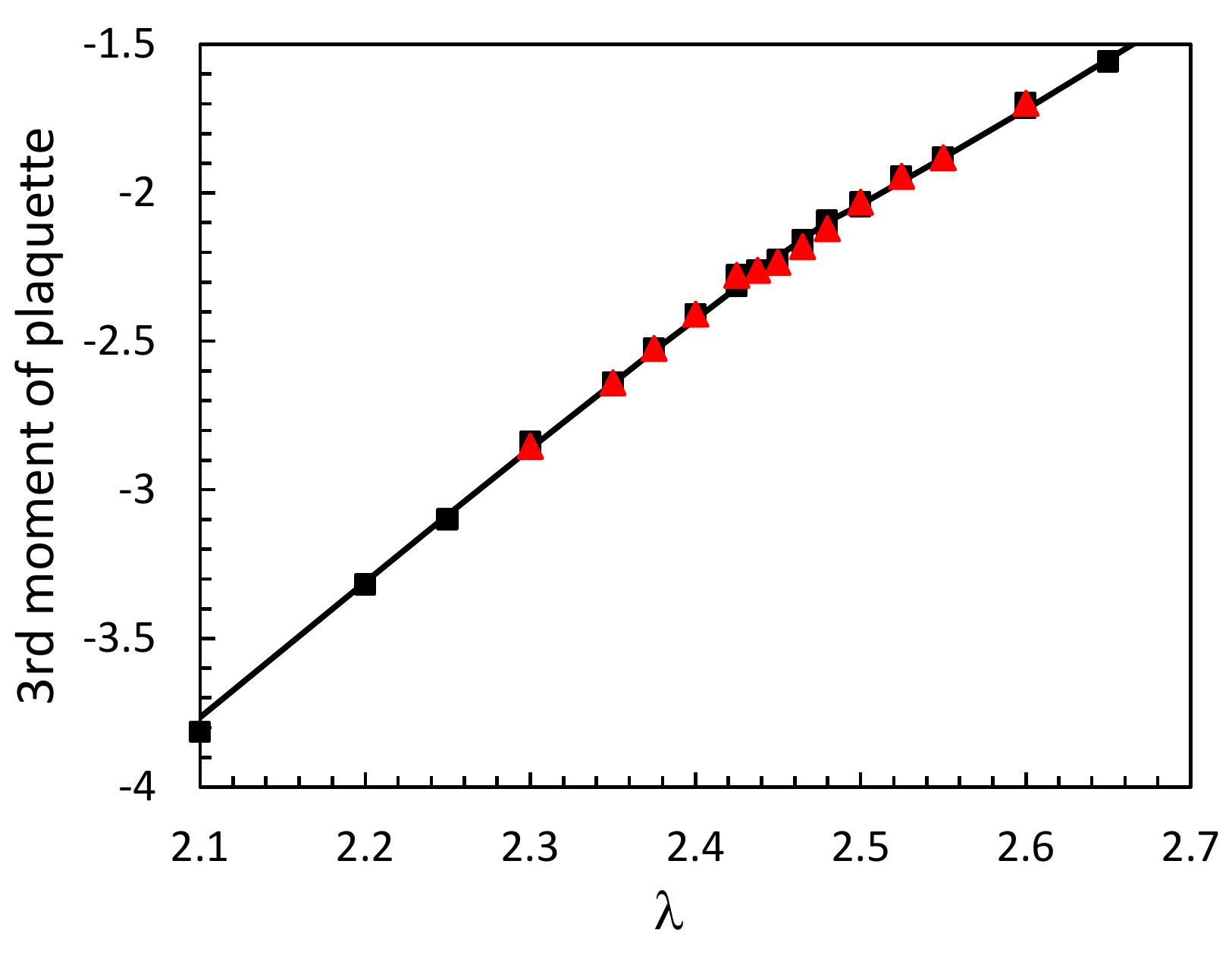}
                      \includegraphics[width=0.495\textwidth,  clip]{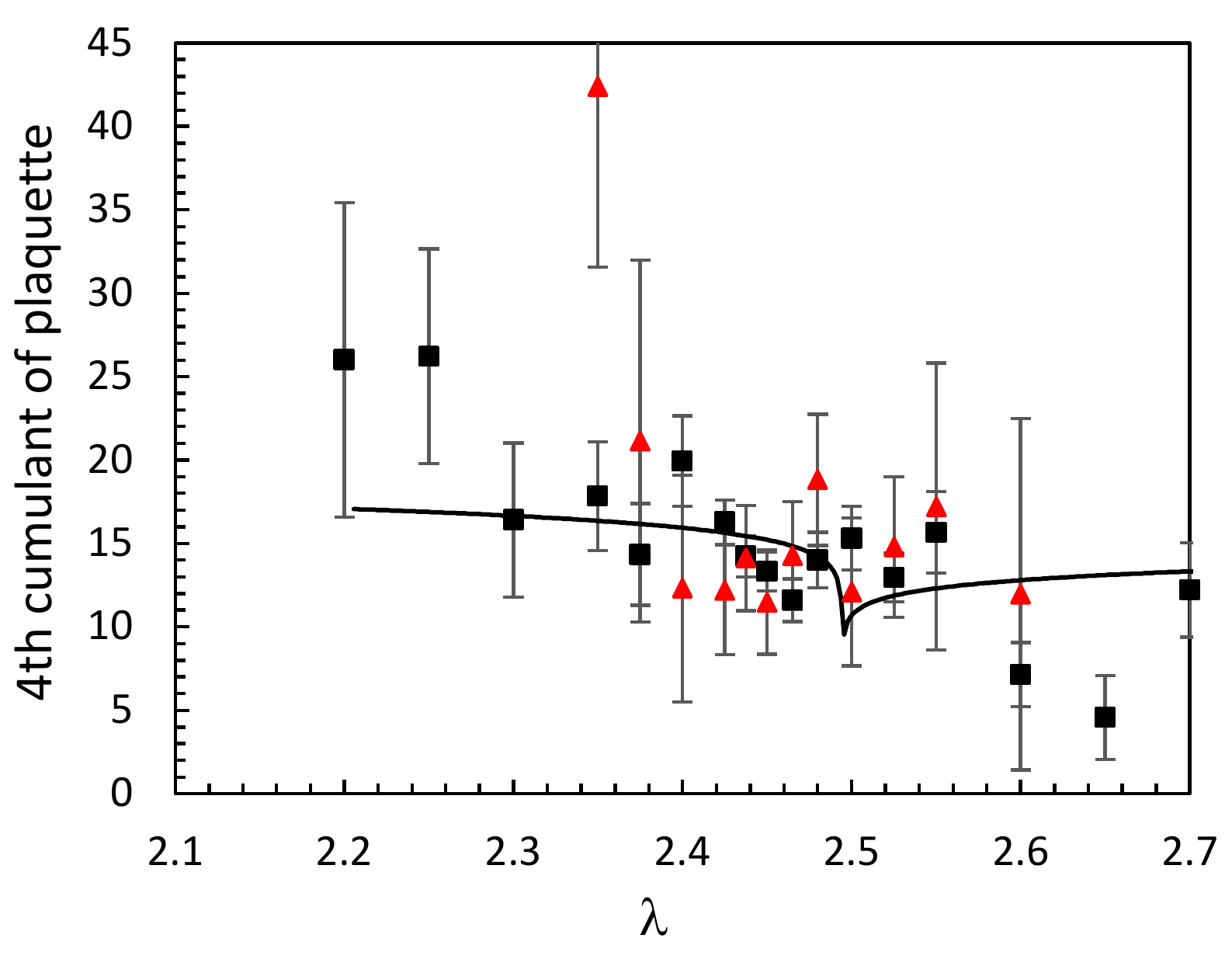}
                                    \caption{Third (a) and fourth (b) cumulant  of plaquette around $\beta = 0.05$ Coulomb transition.  Error ranges are 3-4 times smaller than datapoint size point size in former. Curve in (a) is a fit to critical scaling, and curve in (b) is a numerical derivative of that curve (see text).}
          \label{fig19}
       \end{figure}
At $\beta = 0.05$, the transitions were too widely spread to show both on the same graph.  
First we look at the Coulomb-gauge transition region. The infinite lattice transition is around $\beta=2.5$, but the susceptibility peak for the $40^3$ lattice was around $2.0$ and the $50^3$ was $2.1$ (recall that these have effective lattice sizes 10 less due to the ignored boundary region). So there is a fairly broad region where signals of growing non-analytic behavior could occur here. As mentioned above, energy data were studied on PBC lattices but the order parameter was obtained using our 
OFA-boundary, so this could also introduce finite-size differences in the critical point. For this reason, a fairly wide net was used for this case.  The third moment is shown in Fig.~19a, with fits to fractional-power scaling. Both $30^3$ and $40^3$ were included.  The result had $\lambda _c = 2.495(8)$ with $-\alpha -1 = 1.062(14)$  and a $\chi ^2$/d.f.=1.15. The exponent is close to unity so the data look fairly linear, but the slopes above and below the critical point are significantly different with a ratio of 1.26(4). It is this rather sudden change in slope which is the main visible feature here. There is also a slight flattening around $\lambda = 2.48$. This value of $\alpha$ yields $\nu = 1.35(1)$, agreeing with our previous  determination from the Coulomb order parameter of $1.47(14)$.  The $\lambda _c$ value also agrees with that obtained from the order parameter, 2.52(3).  Note that this data is inconsistent with the possibility of a non-thermal transition, which would require a linear fit to the third moment (giving $\alpha = -2$) with no jump in coefficient(see Appendix A). Such a fit has a $p$-value of $10^{-50}$.  Even if the critical region is halved, the fit is still far from satisfactory ($p=10^{-13}$).

Fig.~19b shows the plaquette fourth cumulant. The fit to the third moment predicts the fourth cumulant should have a rapid cusp with a low but positive exponent of $0.06$.  Indeed, the $30^3$ data is quite suggestive of a cusp at a slightly shifted $\lambda$ of 2.465.  There is a nearly $3\sigma$ dip from the established level followed by about a $2\sigma$ recovery.  Assuming there is no nonsingular contribution a true cusp should go to zero, however it is unrealistic to expect to observe this.  To illustrate we took numerical derivatives of the fit to the third moment on a grid of points eight times finer than our datapoints.  Because this is a derivative with respect to $\lambda$ rather than $\beta$ the result must be arbitrarily scaled to make it comparable.  This is the line drawn on the figure.  One sees a clear cusp but it also does not go anywhere near zero. This is because the cusp is so narrow.  Such a feature is likely to be smoothed by the finite lattice, so it is not just a matter of getting lucky with coupling choice.  Ferrenberg-Swendsen reweighting from $\lambda = 2.465$ indicates a true minimum close to and of a similar value to the one observed, but errors grow so rapidly that it is frankly of little use here.  It is conceivable that a future study with higher statistics and more lattice sizes could see finite-size scaling operating here as a narrowing and deepening cusp.  Unfortunately, even with a similar number of sweeps (approximately $2\times 10^9$ in the best measured region), errors for the $40^3$ lattice were almost double those of the $30^3$.
\begin{figure}[t]\centering
                    \includegraphics[width=0.50\textwidth,  clip]{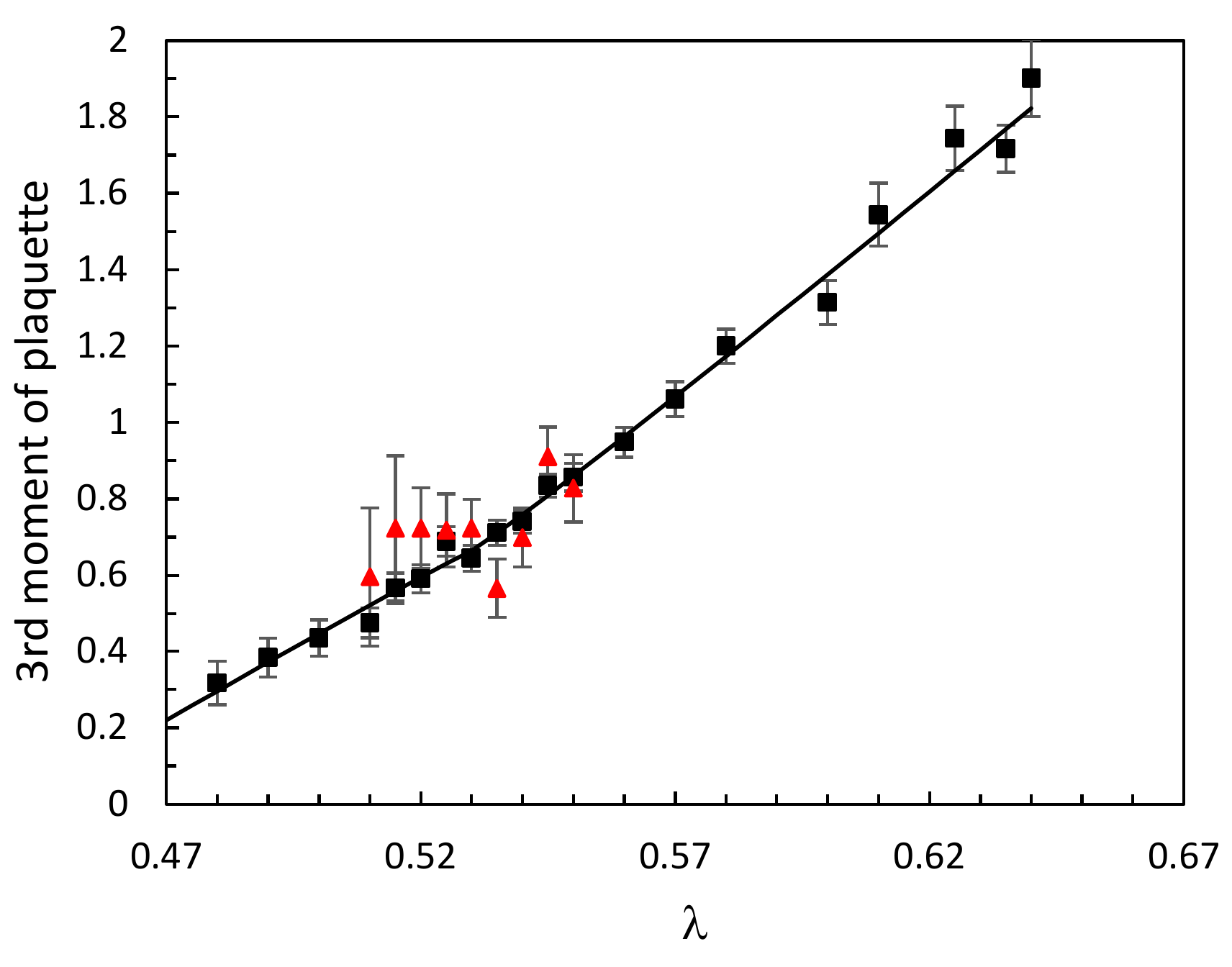}
                                                    \caption{Third moment of the 
plaquette near the Landau transition at $\beta = 0.05$, with fit to critical scaling.}
          \label{fig20}
       \end{figure}

Finally we consider the Higgs (Landau) transition near 
$\lambda = 0.54$.  This is a very noisy place and a big contrast to the Coulomb transition region.  The noise of strong coupling made this the most challenging study.  Up to $1.5\times 10^9$ sweeps were taken per datapoint for $30^3$ and $6\times 10^8$ for $40^3$.   This was more than 200 days of computing on a PC per datapoint. 
Again, the plaquette third moment is shown in Fig.~20 along with a fit to critical scaling behavior. 
Fitting $30^3$ data above and below the critical point, the results were $\lambda _c = 0.530(8)$ and $-\alpha -1= 1.04(6)$ which corresponds to $\nu = 1.35(2)$.  This compares to the order parameter prediction of 1.39(7) with a $\lambda _c$ of $0.536(1)$.  The $\chi ^{2}/\rm{d.f.}$ was 0.69. Again the main feature is a jump in the coefficient above and below the critical point.  The coefficient ratio was 1.39(13). The independently determined values of $\nu$ and $\lambda _c$ are again remarkably close to the predicted values from the order parameter fits.  Direct measurement of the fourth and higher moments were unsuccessful, given the large fluctuations here and the $40^3$ data also did not add much information.  Once again the consistency of energy quantities with the critical scaling predicted by the symmetry-breaking order parameter is demonstrated.  However, for this case a linear fit through the critical region cannot be completely ruled out, since the coefficient ratio was only $3\sigma $ from unity and the exponent is consistent with linear.  Such a fit would be a necessary, though not sufficient, condition for a non-thermal transition.

\section{Updated Phase Diagram}
\begin{figure}[bt]\centering
                    \includegraphics[width=3.2in,  clip]{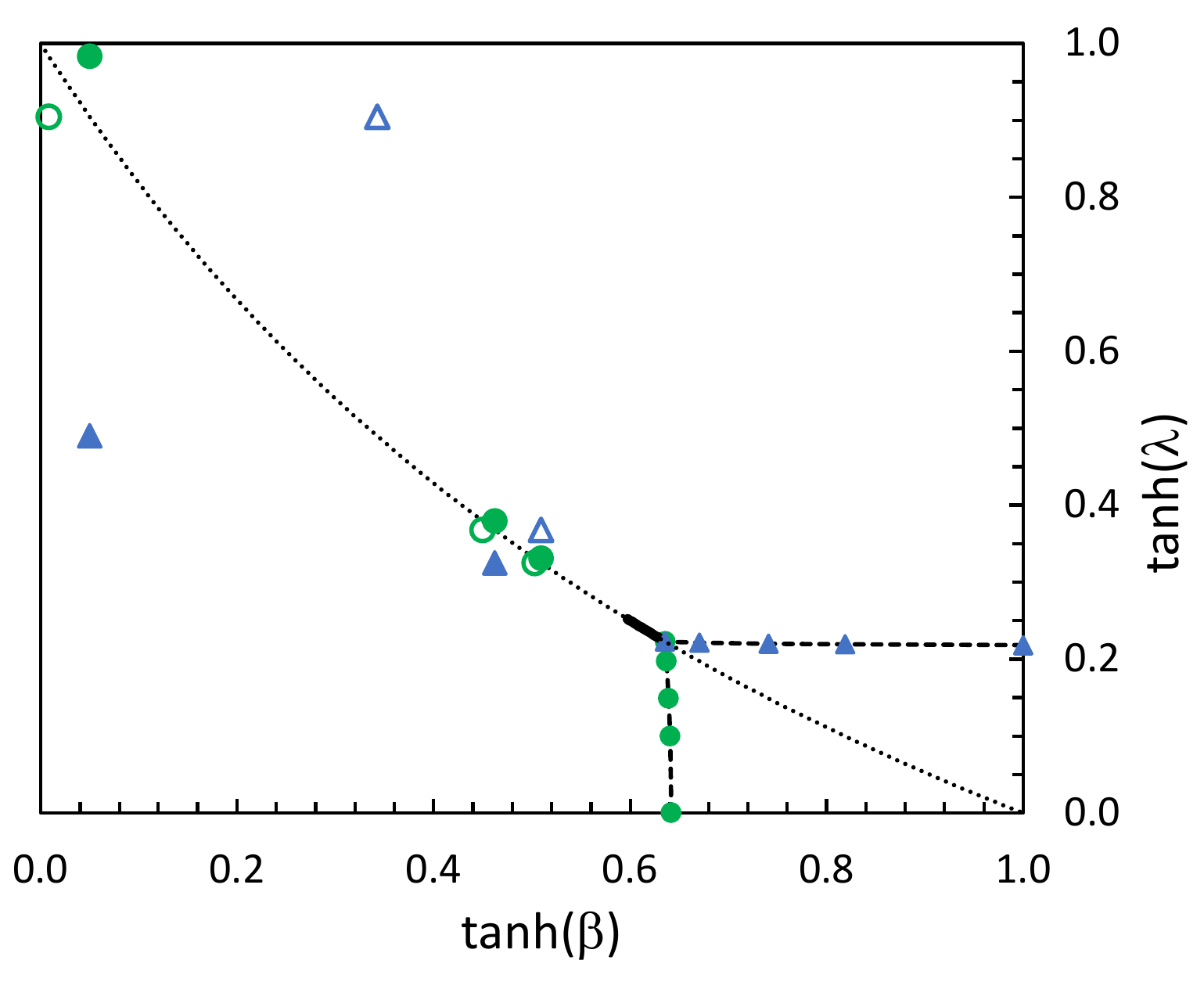}
                                  \caption{Updated phase diagram for the 3D Ising gauge-Higgs theory as seen in the current study. The three large filled circles are higher-order transitions observed with the Coulomb order parameter, two large filled triangles those with the Higgs-replica order parameter.  Small symbols are previously observed second-order transitions using specific heat\cite{tkps}. Open symbols are transitions imputed from duality.  Thick solid line is first-order region.  Dotted line is self-dual line.}
          \label{fig-final}
       \end{figure}

In Fig.~\ref{fig-final} the transitions seen here for $\beta=0.5$ and $0.05$ are added to the previously known phase diagram (Fig.~1).  Open points show the dual-reflections of the observed transitions, where transitions must also exist.  Note that for $\beta = 0$ there are no possible transitions for finite $\lambda $, as explained below, but there is a singularity at $\lambda = \infty$.  So this must be the ultimate destination of all  transition lines shown.  Also of interest is the dual reflection of the Coulomb transition at $\beta = 0.05$. This lies at ($\beta =0.0084$, $\lambda = 1.498$), showing a singularity at an even stronger gauge coupling than $\beta = 0.05$. Any singularity off the self-dual line is required to have a twin dual singularity.

\section{Fradkin-Shenker theorem revisited}
The upper left corner of the phase diagram at $\beta =0$, $\lambda = \infty$ has a demonstrable singularity\cite{me}. 
The $\beta =0$
theory is exactly solvable.  In axial gauge, no interactions perpendicular
to the gauge-fixed direction remain, resulting in a set of 1-d Ising models. 
These are
disordered for all finite $\lambda$ but are ordered at $\lambda =\infty$.  The 1-d Ising model
has a phase transition at $T=0$, accompanied by an essential singularity, in our notation at  $1/\lambda =0$.
Thus a singularity
exists at the heart of the FS analyticity region (FSAR). However, it is possible this is an isolated singular
point. The FS theorem is evaded simply because the zeroth-order coefficient in the cluster expansion is itself singular
here. Nevertheless, it gives one pause and provides a handy endpoint for the continued lines of phase transitions if they exist. 

The FS theorem relies on proving
the convergence of the cluster expansion. It uses unitary gauge which removes the Higgs field entirely.
The Higgs action becomes
\begin{equation}
S_H =-\lambda \sum _{\vec{r},\mu}  U_\mu (\vec{r}) .
\end{equation}
Consider the expectation value of a local operator $\cal{F}$ such as the average plaquette.  The cluster expansion rewrites
the individual plaquette 
Boltzmann factor 
\begin{equation}
\exp (-\beta S_{p}) = 1+p_{p}
\end{equation}
For small $\beta$, $p_p$ are small. The Higgs  action, now local on links,
is absorbed into the link measure $dU$.  The cluster expansion for $<\cal{F}>$ is given by
\begin{equation}
 <\mathcal{F}>= \sum_{Q_{1}(Q_{0})}\int dU \mathcal{F}\prod _{p\in Q_{1}}p_{p}\frac{Z(\overline{Q_{1} \cup Q_{0}})}{Z} .
\end{equation}
Here $Q_0$ is the set of plaquettes connected to $\cal{F}$. The expansion is over all connected sets of plaquettes $Q_1$
connected with $Q_0$. $Z$ is the partition function and $Z(\overline{Q_1 \cup Q_0})$ is the partition function
missing all plaquettes in $Q_1 \cup Q_0$ and any touching its boundary. For details see FS\cite{fs} and OS\cite{os}.
As $Q_1$ grows terms have a larger number of the small factors $p_{p}$ which for small $\beta$ form
a convergent series.  The ratio of the two partition functions sums the disconnected diagrams and does not
spoil the series convergence for small enough $\beta $ or large enough $\lambda $. Small $\beta$ and large $\lambda$ both 
aid convergence
which explains the slope of the FSAR.  Quoting FS, ``Analyticity of $<\cal{F}>$ in (the couplings) (follows),
because the series converges uniformly and the terms are each analytic."  There seems little doubt the series 
converges in the region claimed.
However, the second condition that the individual terms are analytic is not addressed in either FS or OS. Presumably it
was thought too obvious to require proof. Recall that non-analyticity of an individual term is the loophole that
allows a singularity at the upper corner. Could this problem be more widespread? The suspicious factor is the 
ratio of the two partition functions, one missing some of its plaquettes. Since a partition function is the sum of an infinite
number of terms in the thermodynamic limit, here we have a finite ratio of two infinite quantities. It is precisely
such a ratio that gives rise to thermodynamic singularities. It is not at all apparent that such a factor is 
singularity free.  Consider the simplest case of $Z$(missing a single plaquette)$/Z$ which can be written $<\exp (\beta S_p)>$ 
where $S_p$ is the single plaquette action.  If we expand the exponential this contains a term of $<S_p >$ the expectation 
value of the average
plaquette itself.  So in the cluster expansion for the average plaquette, some of the terms on the RHS
{\em also} include factors of the average plaquette and other more complex expectation values.  
Thus if the average plaquette has
a singularity, there are singularities on both sides of the equation which seems perfectly consistent regardless of
convergence. Therefore, a convergent cluster expansion is not sufficient to prove analyticity as a function of temperature or coupling.
It is consistent
with expectation values either being singular or not.  One must also 
prove that each ratio of partition functions on the RHS is itself analytic. This flaw is common to both FS and the second part of OS which concerns the Higgs mechanism.  

Seiler has also developed an alternative proof of analyticity based on a polymer expansion, which has a 
different structure and 
does not have the
suspicious infinite-lattice factors\cite{seilernew,poly}. However, this proof appears to make somewhat more 
restricted claims on analyticity than FS does.  
It
claims analyticity either if both $\beta $ and $\lambda $ are small or if $\lambda $ itself is large enough, 
given $\beta$ (see page 59 of \cite{poly}). This 
seems to leave open the possibility of a phase transition threading through at intermediate $\lambda $. A critique of our arguments is given in \cite{seilernew}.

It seems that sufficient doubt exists in the completeness of all of these analyticity arguments, that one can
entertain the possibility of an unbroken line of phase transitions, as is required by the other ``analyticity argument" of Landau -- the one
requiring symmetry-broken and unbroken phases to be separated by phase transitions due to non-analyticity of the order parameter at the point of symmetry-change.

\section{Conclusion}
The 3D Ising gauge-Higgs model is the simplest of the gauge-Higgs theories.  It certainly makes sense to understand it very well, as a step toward understanding the more complex theories relevant to particle physics.  This simple, but not too simple, theory is a good place to test fundamental ideas with high-precision computing.  The main question we are asking here is whether  symmetry-breaking transitions exist in gauge theories. As earlier emphasized, if the observed phase transitions are symmetry-breaking then they cannot end mid-diagram as has been previously assumed.  They must continue to the edge of the phase diagram.  However, this would contradict the widely-cited Fradkin-Shenker theorem which purports to prove the existence of an analyticity region which would not allow phase transitions to continue.  For the theorem to be true, either the transitions are not symmetry breaking, or they are a new class of transitions dubbed non-thermal. For these, the transitions would still have symmetry breaking order parameters but these would have no effect on the free energy.  This latter suggestion is probably impossible for this system as detailed in the appendix below.  Thus, given the Fradkin-Shenker theorem, it must be concluded that the phase transitions are not symmetry-breaking within the diagram (we know they are symmetry-breaking on the edge).

In this paper we have presented the opposite case, that these phase transitions are indeed symmetry breaking, do affect the energy, and continue all the way to the edge of the phase diagram.  This requires something to be wrong with the Fradkin-Shenker theorem.  We have suggested that although the cluster expansion may indeed converge, there is a largely unexplored possibility that the coefficients in this highly complex expansion could themselves be non-analytic functions of the couplings.  Our main effort, however, is simply to present the rather voluminous data that trace the phase transitions, in order to uncover many previously unexplored facets of this fascinating system.  Experiment is the final arbiter for physical theories of real systems. Precision numerical studies play a similar role for statistical models.

The theoretical justification for there being no symmetry-breaking phase transitions in gauge theories is Elitzur's theorem, which states that local symmetries cannot be spontaneously broken.  Since the global symmetry operation can be obtained from a compilation of an infinite number of local symmetry operations, one might think that is also included in this prohibition. However, one must be extremely careful when adding an infinite number of anything.  Elitzur's theorem relies on the symmetry operation changing only a finite number of degrees of freedom, so it does {\em not} apply to an underlying global symmetry.   Therefore, it is possible that the gauge symmetry simply hides the underlying symmetry breaking, by erasing the order parameter. Such a symmetry breaking can be uncovered through gauge fixing to a gauge which leaves a global remnant symmetry unfixed, such as Landau or Coulomb gauge, which have different remnant symmetries and order parameters. These can break either simultaneously or separately.  A very clear example of a gauge symmetry effectively hiding the established symmetry-breaking transition of the 2D Ising model is given in Appendix B.  Another example is the 3D Ising gauge theory, which is dual to the 3D Ising model. It must have a phase transition identical to that of the 3D Ising model, due to duality, but there is no order parameter for this in the unfixed gauge theory.

Our studies rely on Coulomb and Landau gauge-fixing.  However, these are difficult to implement in practice, and have difficult to control systematic errors arising from imperfect gauge-fixing algorithms (usually simulated annealing).  An important ingredient to our study is the use of two new techniques, which avoid simulated annealing entirely.  For the Coulomb case, an exact gauge-fixing is accomplished using Edmunds' minimum-weight graph-matching algorithm. Unfortunately it only applies to this specific model, but allows very high statistics with zero systematic error, the only disadvantage being giving up periodic boundary conditions.  For the Landau gauge, we use instead a gauge-invariant substitute using a real replica Higgs field which was introduced earlier.  We verified this traces the same phase transition that Landau gauge-fixing does, by running some simulated-annealing simulations.  The real replica involves an equilibration step similar to that at the beginning of any Monte Carlo simulation.  This has a systematic error, but one that is easily characterized and falls exponentially with equilibration time.

We first verified that both of these symmetry-breaking order parameters show very clear transitions along the known first-order line. In particular the replica order-parameter shows jumps which are highly correlated with jumps in average plaquette and Higgs energy.  This result alone establishes our hypothesis that phase transitions in the interior of the phase diagram are symmetry breaking.  A first-order transition which is not symmetry-breaking, such as the liquid-gas transition, can end at a critical point.  However, a symmetry-breaking first-order transition cannot end.  It can only change into one or more higher-order transitions at a tricritical point.  Looking beyond the first-order endpoint, first at $\beta = 0.5$ and then at $\beta = 0.05$, we find the Landau and Coulomb transitions continuing as high-order transitions, but split from each other.  These have high values of correlation length exponent $\nu \sim 1.5$, which result in highly negative specific heat exponents $\alpha \sim -2.5$. The specific heat does not have an infinite singularity, nor does its first two derivatives.  This makes observing the phase transition from energy quantities extremely challenging.  However, by using of order $10^9$ sweeps per datapoint, evidence of coincident transitions in energy quantities is found. This includes fits to non-infinite but still singular behavior of the third moment of the plaquette, which yields critical exponents and critical points that agree in all cases with what is obtained from order parameter scaling fits.  It also includes evidence for expected cusps in fourth moments and indications of possible developing infinite singularities in the second derivative of the third moment.

In the future we plan to add more lattice sizes to be able to include higher-order corrections to scaling in the order-parameter analysis. This will allow critical exponents to be determined with much higher precision.  Another goal is to observe finite-size scaling of high derivatives or high moments of energy quantities as well as coupling-shift.   Another area that needs further investigation is the fairly strong evidence already collected that seems to show four transition lines rather than two, due to dual reflections falling far from the established lines.  One's initial reaction is ``Who ordered that?'' Experience has taught, however, that one must follow numerical data to its own logical conclusion, regardless of one's theoretical expectations.  It will also be interesting to explore the physical properties of the intermediate phase or phases. 

\section*{Appendix A: Does thermodynamics allow for a non-thermal transition?}
The idea behind a non-thermal transition is the possibility that the Landau or Coulomb-gauge magnetizations could 
be in some sense non-physical in that they rely on gauge fixing. This worry is somewhat relieved by our use of the gauge-invariant replica overlap to replace the Landau-gauge magnetization, but our Coulomb-gauge results still rely on gauge fixing.   One could imagine
gauge-fixed objects undergoing a phase transition without feeding back to physical
quantities related to energy, such as the specific heat, as suggested by Caudy and Greensite\cite{cg}.  However, this ignores the essential
reason behind the occurrence any phase transition which is ultimately thermal.  
If one separates 
gauge configurations by whether the order parameter is spontaneously broken or not, 
these generally have different dependence of entropy on temperature. 
If a phase transition occurs the system shifts from the configurations making an
important contribution to the partition function
being overwhelmingly in one category to overwhelmingly in the other.
The
shift in the entropy function from the symmetry-unbroken one to the symmetry-broken one leads to a singularity in entropy as a 
function in temperature because not all derivatives of the two entropy functions will match.  
If this did not happen then there would be no reason 
for the sudden change in order parameter, from a zero to non-zero value (because both types of configurations would be present).  
The reason behind any phase transition is 
essentially thermal since it is the entropy which is in control, and it would seem impossible for energy quantities not to pick
up a singularity through the entropy.  Another way of saying this is that 
correlations go both ways. If the magnetization depends on temperature, then 
$<mU>\neq 0$ since this correlator is proportional to $\partial m / \partial \beta$. (Here $m$ is the magnetization, $U$ is the
internal energy and $\beta$ is the inverse temperature.).
But the same correlator is also proportional to $\partial U /\partial H$, where $H$ is the external magnetic field
coupled to $m$.   If energy
quantities do not care about this magnetization it doesn't seem consistent that
the internal energy depends on the external magnetic field coupled to this
magnetization. Clearly 
the fact that $m$ and $U$ are correlated shows that this $m$ {\em is} 
relevant to the energy, just as the energy and temperature 
are relevant to the magnetization (i.e. they are correlated).  
 
Basic thermodynamics of the simple magnetic system 
can be used to precisely demonstrate the 
linkage between magnetic and thermal quantities as far as their
critical behavior is concerned. 
Following Stanley\cite{stanley} and Fisher\cite{fisher}
\begin{equation}
C_H -C_M = T \frac{\alpha _{H}^2}{\chi _T},
\end{equation}
where $\alpha _H = \left( \frac{\partial M}{\partial T}\right) _H$ and
$\chi _T = \left(\frac{\partial M}{\partial H}\right) _T$.  This follows from definitions
and multivariate calculus only, along with the fact that such a system has
two independent variables. On the left hand side are thermal quantities and on
the right are magnetic quantities.  If the combined magnetic quantities follow a non-analytic
critical behavior, so must the thermal quantity on the LHS.  The critical behavior of
$\alpha _H$ is $t^{\beta '-1}$ and that of $\chi _T$ is $t^{-\gamma }$ where $\beta '$ and
$\gamma$ are the standard magnetization and susceptibility critical exponents and $t$ is
the reduced temperature $t=|T-T_c |/T_c $.  The RHS, therefore, has the dependence 
$t^{2\beta '+ \gamma -2}$.  $C_H$ at $H=0$ is the specific heat of the 
original system without any reference to the magnetization or magnetic field, so it
can be thought of as a pure thermal quantity. $C_M$ is still a specific heat, but holding
$M$ constant does mix in the magnetization, so one is really interested in the critical
behavior of $C_H$ alone.  The above can be seen as a derivation of the well-known scaling relation\cite{pa}
by which $2\beta '+\gamma -2$, for systems obeying correlation length scaling and below the 
upper critical dimension, is equal to the negative of the specific heat exponent, $-\alpha$.  
If this is negative
leading to a divergence of the RHS as $t\rightarrow 0$ then $C_H$ itself must diverge,
because $C_H \geq C_M$ and both are positive. Thus $C_H$ {\em cannot avoid} inheriting
the singularity of the magnetic quantities.

If the exponent $2\beta '+ \gamma -2 >0$ then the situation is more nuanced, because
the specific heat remains finite in such a case but generally has some 
derivative that is infinite. An example is the liquid helium lambda transition which has a small negative $\alpha$. One can 
solve for $C_H$ from the above, using the relation $C_H / C_M = \chi _T / \chi _S$ where $\chi _S =
\left( \partial M /\partial H \right) _S$. This gives 
\begin{equation}
C_H^{-1}=\frac{ (\chi _T - \chi _S)}{T\alpha _H ^2}
\end{equation}
In order to give a finite result at $t=0$, in this case the leading divergences of 
$\chi _T$ and $\chi _S$ must exactly cancel, with the next to leading divergence cancelling
that of $\alpha _H ^2$. The behavior for $T<T_C$ is then controlled 
by corrections to scaling, i.e.
the next to leading
behavior of $\alpha _H$ and second sub-leading behavior of the $\chi$ 's.  In almost
all cases these will also be fractional powers leading to a non-divergent but still 
non-analytic behavior (some derivative diverges).  It is extremely rare for any critical exponent to be an exact
integer, and logarithmic corrections would likely be present in such a case.  So even for a 
finite specific heat, it is still almost always the case that the specific heat 
picks up non-analytic behavior from
its necessary relationship to magnetic quantities.  
Thus it seems that any magnetization defined from local fields, 
displaying an exact global symmetry which undergoes 
spontaneously symmetry breaking,
will necessarily cause singularities in thermal quantities which scale according to
the correlation length scaling laws generally associated with such transitions. Only in the extremely rare case that $\alpha$
is a non-positive even integer would it be possible for the energy quantities to evade a singularity when the order parameter has one.

However, there is one interesting case, perhaps the only one, in which this actually occurs. It is the {\em  two}-dimensional Ising-gauge theory.  The 2D Ising model has $\alpha = 0$, so it meets the necessary requirement.  In this case, though, it still has a logarithmic singularity in the specific heat.  The interior of the phase diagram for this system, however, is known rigorously to have an analytic free energy, and therefore specific heat.  This is due to the dual of the theory being the 2D Ising model in an external magnetic field (the gauge part maps into the magnetic field).  The free energy for that theory is known from the Lee-Yang theorem and extensions to be analytic for any non-zero magnetic field\cite{lee-yang,fisher}.  Therefore the free energy of the original dual-related theory must be analytic for any non-zero gauge coupling.  Nevertheless, the gauge-Higgs theory in Landau gauge still has an exact remnant symmetry just as it does in three dimensions.  This can be seen to break spontaneously using the 
replica Higgs order parameter used in this paper, on a line connecting to the Ising transition, using the Binder cumulant crossing method.  We performed a replica-Higgs Monte Carlo study at $\beta= 2.3$, which corresponds to an external field of about 0.01 in the dual theory. Here we find the specific heat does not show any finite-size scaling, whereas for the pure Ising model finite-size scaling from the logarithmic singularity is clearly visible.  Our data for the third moment of the Higgs energy at $\beta=2.3$ on a $100^2$ lattice is consistent with a linear fit with no jump in coefficient.  The second moment shows a quadratic peak near the suspected critical point, $\lambda _c =0.466$.  These are both consistent with a non-thermal transition based on $\alpha = -2$. Order parameter scaling from runs on $100^2$, $150^2$, and $256^2$ lattices gives a $\nu$ value consistent with $2$, which is what is needed in two dimensions to get $\alpha = -2$. This is not inconsistent with the Ising value of $\alpha = 0$, since in the absence of a logarithmic singularity, $\alpha = 0$ predicts merely a constant behavior.  The $\alpha = -2$ fit could therefore be interpreted as the sub-leading behavior, which in this case is also not singular.   So it appears that this is a genuine case of a non-thermal transition, one for which the order parameter is singular but the specific heat and free energy are not.

\section*{Appendix B: Hiding the symmetry-breaking transition in the 2D Ising model} Here we explore the idea of hiding symmetry in the system that possesses the most well-established symmetry-breaking
phase transition, the venerable 2D Ising model.  This will show that explicit symmetry is a convenient but unnecessary 
luxury. The phase transition will exist with all necessary features even if there is no longer an identifiable order parameter.
If one extends the Ising model to the 2D Ising gauge-Higgs theory, then the Ising model is 
obtained in the $\beta \rightarrow \infty$ limit. As in the 3D theory, with no gauge fixing the order parameter vanishes even in the ordered phase, due to the local gauge symmetry.  In Landau gauge all links become unity in this limit, and one is
just left with the standard Ising model action $S=-\sum \lambda \phi _i \phi _j$ with a global Z2 symmetry.  In unitary 
gauge, however, all of the gauge freedom is used to set the Higgs field to unity.  The gauge-Higgs model  has action
\begin{equation}
S=-\beta \sum U_p - \lambda \sum U_{ij}.
\end{equation}
Here there is no Higgs field and no apparent symmetry left.  If one imagines first starting from Landau gauge (all links unity at $\beta = \infty$) and then
transforms to unitary gauge, one must apply a (-1) gauge transformation at each negative Higgs site. This completely erases the domain structure
of the Higgs field (since they are now all positive), but introduces negative U links along boundaries of the former Higgs domains.  So the boundaries of the domain structure are
preserved
in the gauge field.  It is the percolation of the boundaries that can be thought of as inducing the phase transition - one does not 
really need the domains themselves nor equal-energy configurations related by a symmetry operation.  To solve the Ising model in
the unitary gauge, one must simply sum $\exp (\lambda \sum U_{ij})$ for all $\{U_{ij}\}$ which satisfy the plaquette constraints (all plaquettes unity).
The easiest way to implement the plaquette constraints is to write the gauge fields $U_{ij}=\eta_i \eta_j$ where the $\eta$'s are Z2
valued auxiliary fields on sites.  This gives all of the allowed U fields at $\beta = \infty$.  This then results in an action
\begin{equation}
S=-\lambda \sum \eta _i \eta _j.
\end{equation}
The Higgs field is back! In this gauge it is actually the gauge field degrees of freedom in disguise, but the result is obviously
no different from the Landau gauge result in which the Higgs fields are the only survivors. The symmetry is 
back only because the constraint solution chosen had an accidental double counting. This could even be seen as an artificially introduced symmetry, but there is actually nothing wrong with introducing such a symmetry by multiplying the degrees of freedom, so long as the overcounting is uniform.
One
could insist on eliminating the double counting by only summing over positive magnetization configurations, but this would not 
change the phase transition in any important way.
Nevertheless, the order parameter definable in the formulation with an explicit symmetry (Landau gauge) certainly makes
studying the phase transition a lot easier.  The equivalence of the two gauges means that any phase transition found when
using Landau gauge must also be present in unitary gauge, or without fixing any gauge, regardless of the possible 
hiding of the symmetry.  This also shows that the concept of symmetry-breaking phase transitions applies to systems with no explicit symmetry, so long as it is possible to introduce a symmetry through a uniform expansion of degrees of freedom.  This simple example also shows that symmetries can easily be hidden in two different ways: either by evenly reducing the degrees of freedom responsible for the symmetry, as in unitary gauge, or by vastly expanding the degrees of freedom as in full local gauge symmetry.


\begin{thebibliography}{99}
\bibitem{ll}L.D. Landau and E.M. Lifshitz, {\em Statistical Physics - Vol. 5 of the Course of Theoretical Physics}, Pergamon Press, London, 1958, p452.
\bibitem{elitzur}S. Elitzur, Impossibility of spontaneously breaking local symmetries, Phys. Rev. {\bf D12} (1975) 
3978-3982 . 
\bibitem{megh}M. Grady, Solution to the gauge-Higgs analyticity paradox, arXiv:1502.04362 (2015).
\bibitem{ggrt}L. Genovese, F. Gliozzi, A. Rago, and C. Torrero, The phase diagram of the three-dimensional $Z_2$ gauge-Higgs system at zero and finite temperature, Nucl. Phys. B Proc. Suppl. {\bf 119} (2003) 894-899.
\bibitem{cl}P.M. Chaikin and T.C. Lubensky, {\em Principles of Condensed Matter Physics}, Cambridge University Press, Cambridge, 1995.
\bibitem{cg}W. Caudy and J. Greensite, On the ambiguity of spontaneously broken gauge symmetry, Phys. Rev. D {\bf 78} (2008) 025018.
\bibitem{fs}E. Fradkin and S.H. Shenker, Phase diagrams of lattice gauge theories with Higgs fields, Phys. Rev. {\bf D19} (1979) 3682-3697.
\bibitem{os}K. Osterwalder and E. Seiler, Gauge field theories on a lattice, Ann. Phys. (NY) {\bf 110} (1978) 440-471.
\bibitem{gjs}J. Glimm, A. Jaffe, and T. Spencer, The particle structure of the weakly coupled $P(\phi )_2$ model and other applications of high temperature expansions part II: the cluster expansion,  in: {\em Lecture Notes in Physics Vol. 25},
G. Velo and A. Wightman, eds., Springer-Verlag, Berlin, 1973, pp. 199-242.
\bibitem{poly}E. Seiler, {\em Gauge theories as a problem of constructive quantum field theory and statistical mechanics}, Lecture Notes in Physics Vol. 159, J. Ehlers et. al. eds., Springer-Verlag, Berlin, 1982 (see p. 59).
\bibitem{seilernew}E. Seiler, On the Higgs-confinement complementarity, arXiv:1506.00862 (2015).
\bibitem{stack} G.A. Jongeward, J.D. Stack, and C. Jayaprakash, Monte Carlo calculations on $Z_2$ gauge-Higgs theories, Phys. Rev. D {\bf 21} (1980) 3360-3368.
\bibitem{tkps}I.S. Tupitsyn, A. Kitaev, N.V. Prokof'ev, and P.C.E Stamp, Topological multicritical point in the phase diagram of the toric code model and the three-dimensional lattice gauge Higgs model, Phys. Rev. B {\bf 82} (2010) 085114.
\bibitem{2rr} K. Binder and W. Kob, {\em Glassy Materials and Disordered Solids}, World Scientific, New Jersey, 2005, pp. 248, 261.
\bibitem{zw}J. Greensite, S. Olejn\'{i}k, and D. Zwanziger, Coulomb energy, remnant symmetry, and phases of non-Abelian gauge theories, Phys. Rev. D {\bf 69} (2004) 074506; D. Zwanziger, No confinement without Coulomb confinement, Phys. Rev. Lett. {\bf 90} (2003) 102001.
\bibitem{edmonds}J. Edmonds, Paths, trees and flowers, Canadian Journal of Mathematics {\bf 17} (1965) 449-467; 
Maximum matching and a polyhedron of 0,1-vertices, Journal of Research of the National Bureau of Standards {\bf 69B} (1965) 125-130.
\bibitem{bieche}J. Bieche, R. Maynard, R. Rammal, and J.P. Uhry, On the ground states of the frustration model of a spin glass by a matching method of graph theory, J. Phys A: Math. Gen. {\bf 13} (1980) 2553-2576.
\bibitem{burkard-derigs}R.E. Burkard and U. Derigs, {\em Assignment and Matching Problems:
Solution Methods with FORTRAN-Programs}, Lecture Notes in Economic and Mathematical Systems 184,
Springer-Verlag, Berlin, 1980.
\bibitem{otheralgs}W. Cook and A. Rohe, Computing minimum-weight perfect matchings,  INFORMS Journal on Computing {\bf 11} (1999) 138-148;
V. Kolmogorov, Blossom V: A new implementation of a minimum cost perfect matching algorithm, Mathematical Programming Computation {\bf 1} (2009) 43-67.
\bibitem{fl} A.M. Ferrenberg and D.P. Landau, Critical behavior of the three-dimensional Ising model: A high-resolution Monte Carlo study, Phys. Rev. B {\bf 44} (1991) 5081-5091.
\bibitem{mh}M. Hasenbusch, A finite size scaling study of models in the three-dimensional Ising universality class, Phys. Rev. B {\bf 82} (2010) 174433. 
\bibitem{cooper}F. Cooper, B. Freedman, and D. Preston, Solving $\phi _{1,2}^4$ field theory with Monte Carlo, Nucl. Phys. B {\bf 210} (1982) 210-228; D.J. Amit and V. Mart\'{i}n-Mayor, {\em Field Theory, the Renormalization Group, and Critical Phenomena: Graphs to Computers},  3rd ed., World Scientific, Singapore, 2005.
\bibitem{rougheningising}K.K. Mon, S. Wansleben, D.P. Landau and K. Binder, Anisotropic surface tension, step free energy, and interface roughening in the three-dimensional Ising model, Phys. Rev. Lett. {\bf 60} (1988) 708-711, Erratum {\bf 61} (1988) 902;
K.K Mon, D.P. Landau, and D. Stauffer, Interface roughening in the three-dimensional Ising model, Phys. Rev. B {\bf 42} (1990) 545-547.
\bibitem{savit} R. Savit, Duality in field theory and statistical systems,  Rev. Mod. Phys. {\bf 52} (1980) 453-487.
\bibitem{fss}M.N. Barber, Finite-size scaling, in: {\em Phase Transitions and Critical Phenomena  Vol. 8}, C. Domb and J.L. Lebowitz eds., Academic Press, NY, 1983, pp. 146-266.
\bibitem{me}M. Grady, Reconsidering gauge-Higgs continuity, Phys. Lett. B, {\bf 626} (2005) 161-166.
\bibitem{stanley} H.E. Stanley, {\em Introduction to Phase Transitions and Critical Phenomena}, Oxford Univ. Press, NY, 1971.
\bibitem{fisher}M.E. Fisher, The nature of critical points,  in: {\em Lectures in Theoretical Physics Vol. VIIC - Statistical physics, weak interactions, field theory},
W.E. Brittin, ed., University of Colorado Press, Boulder, 1965, pp. 1-159.
\bibitem{pa}R.K. Pathria, {\em Statistical Mechanics}, 2nd ed., Elsevier, Amsterdam, 1996, pp. 344-347.
\bibitem{lee-yang}T.D. Lee and C.N. Yang, Statistical theory of equations of state and phase transitions. II. Lattice gas and Ising model, Phys. Rev. {\bf 87} (1952) 410-419.
\end{thebibliography}
\end{document}